\documentclass[usenatbib,useAMS]{mnras}
\usepackage{multirow}
\usepackage{xcolor}
\usepackage{graphicx}
\usepackage{amsmath}
\usepackage{amssymb}	
\usepackage{multicol}   
\usepackage{bm}		
\usepackage[normalem]{ulem} 
\usepackage{caption} 
\usepackage{url}
\usepackage{longtable}
\usepackage{pdfpages}
\usepackage{subcaption}
\usepackage{hyperref}
\usepackage{pdflscape}
\usepackage{rotating}
\usepackage{tablefootnote}
\title[Multiwavelength variability of the blazar AO\,0235$+$164]
{Multiwavelength variability of the blazar AO\,0235$+$164}

\author[Vlasyuk et al.]{%
V.~V.~Vlasyuk,$^{1}$\thanks{E-mail: vvlassao@gmail.com}
Yu.~V.~Sotnikova,$^{1,2}$
A.~E.~Volvach,$^{3}$
T.~V.~Mufakharov,$^{1,2}$
Yu.~A.~Kovalev,$^{4,5}$
\newauthor
O.~I.~Spiridonova,$^{1}$
M.~L.~Khabibullina,$^{1}$
Yu.~Yu.~Kovalev,$^{6}$
A.~G.~Mikhailov,$^{1}$
V.~A.~Stolyarov,$^{1,7}$
\newauthor
D.~O.~Kudryavtsev,$^{1}$
M.~G.~Mingaliev,$^{1,2,8}$
S.~Razzaque,$^{9}$  
T.~A.~Semenova,$^{1}$
A.~K.~Kudryashova,$^{1}$
\newauthor
N.~N.~Bursov,$^{1}$
S.~A.~Trushkin,$^{1}$
A.~V.~Popkov,$^{10,4}$
A.~K.~Erkenov,$^{1}$
I.~A.~Rakhimov,$^{8}$
\newauthor
M.~A.~Kharinov,$^{8}$
M.~A.~Gurwell,$^{11}$
P.~G.~Tsybulev,$^{1}$
A.~S.~Moskvitin,$^1$
T.~A.~Fatkhullin,$^1$
\newauthor
E.~V.~Emelianov,$^1$
A.~Arshinova,$^{12}$
K.~V.~Iuzhanina,$^{1,2}$
T.~S.~Andreeva,$^{8}$
L.~N.~Volvach,$^{3}$
A.~Ghosh$^{9}$
\\
$^{1}$ Special Astrophysical Observatory of
the Russian Academy of Sciences,
Nizhny Arkhyz, 369167, Russia\\
$^{2}$ Kazan Federal University, 18 Kremlyovskaya St, Kazan 420008, Russia\\
$^{3}$ Crimean Astrophysical Observatory of the Russian Academy of Sciences, 298409, Nauchny, Russia \\
$^{4}$ Astro Space Center, Lebedev Physical Institute, Russian Academy of Sciences, 117997, Moscow, Russia \\
$^{5}$ Institute for Nuclear Research, Russian Academy of Sciences, 60th October Anniversary Prospect 7a, Moscow 117312, Russia\\
$^{6}$ Max-Planck-Institut f\"ur Radioastronomie, Auf dem H\"ugel 69, Bonn 53121, Germany\\
$^{7}$ Astrophysics Group, Cavendish Laboratory, University of Cambridge, Cambridge, CB3 0HE, UK \\
$^{8}$ Institute of Applied Astronomy of the Russian Academy of Sciences, Kutuzova Embankment 10, St. Petersburg 191187, Russia\\
$^{9}$ Centre for Astro-Particle Physics and Department of Physics, University of Johannesburg, Auckland Park, 2006, South Africa\\
$^{10}$ Moscow Institute of Physics and Technology, Institutsky per. 9, Dolgoprudny 141700, Russia\\
$^{11}$ Center for Astrophysics, Harvard \& Smithsonian, 60 Garden Street, Cambridge, MA 02138, USA\\
$^{12}$ St. Petersburg State University, 7/9 Universitetskaya Emb., St Petersburg 199034, Russia\\
}
\date{Accepted 2024 October 30. Received 2024 October 27; in original form 2024 July 9}
\pubyear{2024}

\begin{document}
\label{firstpage}
\pagerange{\pageref{firstpage}--\pageref{lastpage}}
\maketitle

\begin{abstract}
We present a study of the multiwavelength (MW) variability of the blazar AO\,0235$+$164 based on the radio-to-$\gamma$-ray data covering a long time period from 1997 to 2023. The radio data are represented by the 1--22 GHz measurements from the SAO RAS RATAN-600 radio telescope, the 5 and 8 GHz data from the IAA RAS RT-32 telescopes, and the 37 GHz data from the RT-22 telescope of CrAO RAS. The optical measurements in the $R$-band were collected with the SAO RAS 1-m Zeiss-1000 and 0.5-m AS-500/2 telescopes. Additionally we used the archive data at 230~GHz from the Submillimeter Array (SMA) and the $\gamma$-ray data in the 0.1--100 GeV band from the Fermi-LAT point source \mbox{4FGL-DR2} catalogue. The variability properties during four epochs containing major flares and one epoch of relatively low activity were analysed using the fractional variability indices, discrete correlation functions, Lomb--Scargle periodograms and structure functions. A significant correlation ($\geq\!2\sigma$) between the radio, optical, and $\gamma$-ray bands is found for all these periods with time delays from 0 to 1.7 yrs. The relation between time delay and frequency is described by a linear law with a negative slope of $-10$ day GHz$^{-1}$. The discovered properties of MW variability for the low activity period and for flaring states suggest that the mechanisms dominating the radio--$\gamma$-ray variations are not substantially different. The detected quasi-periodic oscillations of about 6 and 2 years are tentative, as the time span of the observations includes fewer than 4 full cycles for the radio and optical data and only about 3 cycles for the Fermi-LAT data. These results should be interpreted with caution, given the limited number of observed cycles and the influence of red noise. We used cluster analysis to reliably separate the high and low activity states and determined statistical differences in the main properties of AO\,0235$+$164 non-thermal emission. The physical parameters of the radio jet were obtained using the Hedgehog model applied to the average radio spectrum of AO\,0235$+$164 in the range 0.1--300 GHz. The effectiveness of replacing electrons with protons in the synchrotron radio emission of relativistic jets is shown for describing the nature of blazars and the generation of high energy neutrinos.
\end{abstract}
\begin{keywords}
    galaxies: active --
    galaxies: BL Lacertae objects: individual: AO\,0235$+$164 --
    galaxies: jets --
  radio continuum: galaxies
\end{keywords}

\maketitle

\section{Introduction}

Blazars are a specific type of active galactic nuclei (AGNs) known for their relativistic jets aligned with the observer's line of sight, which causes the radiation to be strongly Doppler boosted \citep{1995PASP..107..803U}.
Blazars encompass two categories: BL Lacertae objects (BL\,Lacs) and flat spectrum radio quasars (FSRQs). BL\,Lacs display a continuous, featureless emission or weak narrow emission lines in their optical/UV spectra with an equivalent width of $\leqslant 5$\r{A} (e.g., \citealt{1991ApJ...374..431S,1996MNRAS.281..425M}), while FSRQs exhibit prominent broad emission lines. The radiation of blazars has a bimodal spectral energy distribution (SED). It has been reliably established that the low-frequency peak is caused by the synchrotron emission of electrons moving in the magnetic field of the jet. The low-frequency peak is most often located in the region from the infrared to the optical part of the spectrum, but for some objects of the BL Lac type it is in the X-ray part. The high-frequency peak extends into the $\gamma$-energy region up to the TeV region and is caused by the inverse Compton scattering of thermal or synchrotron photons. Based on the location of the synchrotron peak ($\nu_{\rm peak}$) in a SED, blazars are classified as high, intermediate, and low synchrotron-peaked types, i.e., HSP, ISP, and LSP. According to the common criteria, HSP blazars have $\nu_{\rm peak}>10^{15}$ GHz, the LSP have $\nu_{\rm peak}<10^{14}$ GHz, and ISP class blazars are with interim values of the peak frequency \citep{2010ApJ...716...30A}. The flux, polarization, and spectra of blazars are highly variable across the electromagnetic (EM) spectrum from radio to $\gamma$-rays (e.g., \citealt{2017MNRAS.472..788G} and references therein). The flux variability for blazars is found on different time-scales, ranging from minutes to several decades \citep{1989Natur.337..627M,1993MNRAS.262..963G,1995ARA&A..33..163W,2004A&A...422..505G}. The jet, presumably originating from accretion onto a supermassive rotating black hole (SMBH) surrounded by an accretion disk, contains relativistic electrons, which produce soft photons from radio up to UV (or even soft \mbox{X-rays}) through synchrotron emission and high-energy photons up to GeV and TeV energies via the inverse Compton process which involves scattering of synchrotron photons as well as externally produced soft photons.

One of the effective ways to understand relativistic jet emission is studying the changes in the blazar physical properties that cause the observed multiwavelength (MW) light-curve variations. Investigation of the time-scales of flare events and their correlation across different frequency bands makes it possible to study  a connection between spatially different emission regions in the AGN central parts \citep{2019MNRAS.490.5300D,2020MNRAS.492.3829L,2023ApJS..266...37A,2024MNRAS.527.6970K}.

The BL\,Lac blazar AO\,0235$+$1644 at z=0.94 \citep{1987ApJ...318..577C} is a good candidate for a MW study of variability processes in AGNs, exhibiting extreme variability of non-thermal radiation across all EM spectrum. It has been extensively studied during the last four decades, across radio to  $\gamma$-rays, displaying variability at time-scales from less than an hour to several years \citep{{2000AJ....120...41W, 1997A&A...326...77R, 2000A&A...360L..47R, 2004ApJ...610..151P, 2001A&A...377..396R, 2005A&A...438...39R, 2008A&A...480..339R, 2008ApJ...672...40H, 2012ApJ...751..159A, 2014Ap&SS.351..281W, 2015ARep...59..145V, 2017ApJ...837...45F}}.

Ground-based VLBI interferometric observations reveal its extreme compactness on submilliarcsecond scales (\mbox{$\leq\!0.5$}~mas) with superluminal apparent speeds and indications of a broad projected jet opening angle \citep{2001ApJS..134..181J, 2006ApJ...640..196P, 2017ApJ...846...98J}. The RadioAstron space VLBI mission suggests the presence of two scales in the inner compact structure of AO\,0235$+$164: the first is the ``core\!'', which is resolved by ground-based VLBI and equals to $\sim\!350$ $\mu$as, the second is ultra-compact, less than about 10 $\mu$as or 0.1 pc, and remains unresolved even with the longest ground--space baselines
\citep{2018MNRAS.475.4994K}.

The compact nature and extreme flaring along with a small viewing angle ($\theta=1.7^{\circ}$) suggest a favorable geometry for observing the source's jet activity \citep{{2009A&A...494..527H}}. AO\,0235$+$16 has undergone significant changes in the direction of its jet, which distinguishes it from most other blazars \citep{2018ARep...62..103H}.

The first attempt to investigate the time-domain relationship between radio and $\gamma$-ray emission in order to localize the $\gamma$-ray emission site in blazars was done by \cite{2014MNRAS.445..428M}. If both types of emission are triggered by shocks propagating along a relativistic jet, the time delay between flares in the two bands depends on their separation. The \mbox{$\gamma$-ray} flare becomes observable after crossing the surface of unit \mbox{$\gamma$-ray} opacity. In the same way, the radio flare becomes observable upon crossing the surface of unit radio opacity (hereafter ``radio core''). The time lag between these wavebands provides an estimate of the interval between the emergence of $\gamma$-ray and radio radiation. The blazar AO\,0235$+$164 was the only case with a significance $\ge3\sigma$ amongst the 86 objects in the study of \cite{2014MNRAS.445..428M}; a correlation at a delay of $150$~days was demonstrated. Later \cite{2020ApJ...902...41W} defined this delay as $45$~days at a 3$\sigma$ significance level. In \cite{2024MNRAS.527..882C} the 2$\sigma$ $\gamma$-ray--radio correlation was detected with time lag about 50 days for the period 2013--2019.

The analysis of the correlation between the optical and \mbox{$\gamma$-ray} flux variations in AO\,0235$+$164 highlights their close relationship, emphasizing the interplay between different emission components \citep{2021MNRAS.504.1772R}. \cite{2020ApJ...902...41W} show that the $\gamma$-ray and optical emitting regions are roughly the same and are located $\sim\!6.6$~pc upstream of the core region of 15~GHz. The estimation of
the radio core size predicts that the optical and $\gamma$-ray emitting regions are far away from the broad-line region (BLR).

For AO\,0235$+$164 the quasi-periodic oscillations (QPO) at 3-6 yr time-scale on base of data taken between 1975 and 2000 in the radio and optical light curves were found \citep{{2001A&A...377..396R,  2002A&A...381....1F}}, indicating a possible binary black hole system at its centre \citep{{2003ChJAA...3..513R, 2004A&A...419..913O}}. The analysis of radio data at 37 GHz proposed a model of close binary SMBH with components of similar masses around $10^{10} M_\odot$ \citep{2015ARep...59..145V}. \cite{2022MNRAS.513.5238R} present a comprehensive analysis of a long-term optical light curve from 1982 to 2019, revealing a periodicity of $\sim\!8$ yrs. The five major flares were followed by minor flares roughly after $\sim\!2$ yrs. This double-peaked periodicity is consistent with  the hypothesis of a binary supermassive black hole system.

The investigations of \cite{2023MNRAS.518.5788O} suggest a potential  QPO in the $R$-band light curve with a period of $\sim\!8.2$ yrs, in accordance with the previous reports of \cite{{2006A&A...459..731R}} and \cite{2022MNRAS.513.5238R}. But the authors noted that the temporal coverage of the $\gamma$-ray and radio data does not allow one to determine the existence of this period reliably.

This work presents a systematic study of the multiband variability of AO\,0235$+$164. We used the $\gamma$-ray, optical, and radio observations covering the period of \mbox{1997--2023}. The primary objective of our collaboration is to achieve dense data sampling to identify potential correlations between the blazar emission in different spectral ranges in order to determine the variability time-scale and possible quasi-periodicity. Such a vast database for this specific source provides an opportunity to investigate variability characteristics on both temporal and spectral domains.

The paper is organised as follows: the details of multiband observations are outlined in Section 2. The MW light curves and some features of the most prominent flares are presented in Section 3; and in Section 4 the results of the used statistical methods, namely the structure function, discrete correlation function, and Lomb--Scargle periodogram, are discussed. The intraweek variability of AO\,0235$+$164, studied from the RATAN-600 radio data obtained on a daily basis, is discussed in Section~5. The average radio spectrum in the range between 0.1 and 300 GHz is constructed in Section 6 with the aim to model it by combining the low and high frequency components. A set of quasi-simultaneous broadband radio spectra allows us to discriminate the high and low activity states of the blazar in Section 7. Finally, the general results of our work are discussed in Sections 8 (Discussion) and 9 (Summary).

\section{Multiwavelength data}
\label{sec:radio_studies}

We have collected MW datasets from four radio and two optical instruments and the $\gamma$-ray Fermi-LAT telescope (Table~\ref{tab:faci}). Part of the data are from public archives, and some recent observations are published for the first time, see the description of the instruments and data reduction below. At 230 GHz (1.3 mm) we utilized measurements from the Submillimeter Array\footnote{\url{http://sma1.sma.hawaii.edu/callist/callist.html}} (SMA; \citealt{2007ASPC..375..234G}), obtained in the period from October 2002 to January 2024.

\begin{table}
\caption{The instruments whose data have been used in this paper}
\label{tab:faci}
\centering
\begin{tabular}{cccc}
\hline
Telescope & Institution & Period & Passband \\
\hline
 RATAN-600 &  SAO RAS  & 1997--2023 & 1-22 GHz \\
 RT-32     & IAA RAS  & 2020--2023 & 5, 8 GHz\\
 RT-22     & CrAO RAS  & 2002--2023 & 37 GHz \\
 SMA &  SAO \& ASIAA & 2002--2023 & 230 GHz \\
 Fermi-LAT & NASA  & 2008--2023 & $\gamma$-rays \\ 
 Zeiss-1000 & SAO RAS  & 2002--2023 & optical R \\
 AS-500/2 & SAO RAS  & 2021--2023 & optical R \\
\hline
\end{tabular}
\end{table}

\subsection{RATAN-600}

Flux densities at 1--22 GHz were obtained with the \mbox{RATAN-600} radio telescope in 1997--2023. Broad-band radio spectra at six frequencies, 0.96/1.2, 2.3, 4.7, 7.7/8.2, 11.2, and 21.7/22.3 GHz were measured quasi-simultaneously within \mbox{3--5}~minutes (secondary mirrors N1 and N2). The \mbox{RATAN-600} antenna and radiometer parameters are given in Table \ref{tab:radiometers}. We use the word ``instantaneous'' to such RATAN-600 spectra in order to differ them from the ``quasi-simultaneous'' spectra, measured often during months at different frequencies. Detailed descriptions of the RATAN-600 antenna, radiometers, and data reduction are given in \cite{1993IAPM...35....7P,1997ASPC..125...46V,1999A&AS..139..545K,2011AstBu..66..109T,2016AstBu..71..496U,2018AstBu..73..494T,2020gbar.conf...32S}. The flux densities at 1-22~GHz $S_{\rm \nu}$, their errors $\sigma$, and average observing epochs (MJD and yyyy.mm.dd) are presented in Table~\ref{TableA2_part}.

Additional measurements of AO\,0235$+$164 at 2.3 and 4.7~GHz were conducted with daily cadence from May 2021 to June 2022 in the radio survey mode on the RATAN-600 Western Sector (secondary mirror N5). The parameters of four four-channel radiometers with a central frequency of 4.7 GHz and a radiometer at 2.3 GHz are given in Table~\ref{tab:radiometers}. The measurements from the four 4.7~GHz radiometers were averaged. The total data set consists of  370 observing epochs. The averaged flux density errors are equal to 7 and 3 per cent at 2.3 and 4.7~GHz, respectively. The average observing epochs (MJD and yyyy.mm.dd), flux densities and their errors are presented in Table~\ref{table4}. Details of the observations and data processing are described in \cite{2023AstBu..78..429M,2024AstBu..79...36K}.

\begin{table}
\caption{RATAN-600 continuum radiometer and antenna parameters for
secondary mirrors 1, 2, and 5: the central frequency $f_0$, bandwidth $\Delta f_0$ and detection limit for point sources per transit $\Delta F$. ${\rm FWHM}_{\rm {RA} \times \rm {Dec.}}$ is the angular resolution along RA and Dec., calculated for the average angles.}
\label{tab:radiometers}
\centering
\begin{tabular}{cccr@{$\,\times\,$}l}
\hline
$f_{0}$ & $\Delta f_{0}$ & $\Delta F$ &  \multicolumn{2}{c}{FWHM$_{\rm {RA} \times \rm{Dec.}}$}\\
GHz    &   GHz           &  mJy beam$^{-1}$   &   \multicolumn{2}{c}{}  \\
\hline
\multicolumn{5}{c}{Secondary mirrors 1, 2} \\
\hline
 $22.3$ & $2.5$  &  $50$ & $0\farcm17$ & $1\farcm6$  \\
 $11.2$ & $1.4$  &  $15$ & $0\farcm34$ & $3\farcm2$ \\
 $8.2$  & $1.0$  &  $10$ & $0\farcm47$ & $4\farcm4$   \\
 $4.7$  & $0.6$  &  $8$  & $0\farcm81$ & $7\farcm6$   \\
 $2.25$  & $0.08$  &  $40$ & $1\farcm7$ & $16\arcmin$  \\
 $1.25$  & $0.08$ &  $200$ & $3\farcm1$ & $27\arcmin$ \\
\hline
\multicolumn{5}{c}{Secondary mirror 5} \\
\hline

$4.40-4.55$  & $0.15$  &  $10$  & $1\farcm5$ & $35\arcmin$ \\
$4.55-4.70$  & $0.15$  &  $10$  & $1\farcm5$ & $35\arcmin$ \\
$4.70-4.85$  & $0.15$  &  $10$  & $1\farcm5$ & $35\arcmin$ \\
$4.85-5.00$  & $0.15$  &  $10$  & $1\farcm5$ & $35\arcmin$ \\
$2.21-2.29$  & $0.80$  &  $40$  & $3\farcm0$ & $70\arcmin$\\
\hline
\end{tabular}
\end{table}

\subsection{RT-32}

Flux densities at frequencies  of  4.8 and 8.6~GHz were measured at several observing epochs using three RT-32 radio telescopes: Svetloe~(Sv), Zelenchukskaya~(Zc) and Badary~(Bd) \citep{2019..VLBI..Quasar}. All the antennas and receivers have similar parameters: bandwidth $\Delta f_{0}=900$~MHz for both receivers with the central frequencies specified above; beam width at half-power level $HPBW=7\farcm0$ and $3\farcm9$, respectively; flux density limit $\Delta F$ reaching about 20~mJy per scan with a time constant of 1~s for both frequencies under optimal observing conditions. Observations were performed in the elevation (Bd, Zc) or azimuth (Sv) drift scan mode.

During the first observing epoch in 2015--2017, the intraday variability (IDV) monitoring was carried out at 4.8 and 8.6~GHz with the Zc and Bd telescopes---about two 24-hour experiments per month, previously published in \cite{2020ARep...64..350K}. In this paper we use flux densities averaged over a day. The results for AO\,0235$+$164 show weak flux density variations at levels of about 2--3~Jy on a year-long time-scale. For the calibration source PKS~0316+152 we used accepted flux densities from the ``CATalogs supporting System'' (CATS, \citealt{2005AstBu..58..118V}): 3.09 and 1.72~Jy at 4.8 and 8.6~GHz, respectively.

The second observing epoch consists of a single experiment on Aug~4, 2019 at 8.6~GHz with the Zc telescope: eight measurements during 00:17-07:14~UTC with an average flux density of 0.34~Jy for AO\,0235$+$164. The source J0521$+$1638 with an accepted flux density of 2.49~Jy was used as a reference point.

On the third observing epoch, from January 2020 to October 2023, the Sv RT-32 had been performing weekly observations at both available frequencies. The source 3C295 was used as a reference with the flux densities from \cite{2017ApJS..230....7P}. Each data point represents an average over about a 30-min exposure ($\sim\!30$ drift scans).

The flux densities $S_{\nu}$, their errors $\sigma$, and average observing epochs (MJD and yyyy.mm.dd) are presented in Table~\ref{TableA2_part}.

Since the observing frequencies of the RT-32 (5.05, 8.63 GHz) and RATAN-600 (4.7, 8.2) are close, we use their rounded values, 8 and 5 GHz, in the further analysis. The frequencies 21.7/22.3, 11.2, 2.3, and 0.96/1.2 GHz are also used in rounded form: 22, 11, 2, and 1 GHz.

\begin{table*}
\caption{\label{TableA2_part}
The RATAN-600 measurements in 1997--2023 and the RT-32 data in 2015--2023: MJD (Col.~1), yyyy.mm.dd (Col.~2), flux densities at 22, 11, 8, 5, 2, and 1 GHz and their errors in Jy (Cols.~3--8), and the name of the telescope (Col.~9). The full version is available as online supplementary material.}
\begin{tabular}{|l|c|c|c|c|c|c|c|l|}
\hline
MJD  &  yyyy.mm.dd & $S_{22}$, $\sigma$, Jy & $S_{11}$, $\sigma$, Jy & $S_{8}$, $\sigma$, Jy & $S_{5}$, $\sigma$, Jy & $S_{2}$, $\sigma$, Jy & $S_{1}$, $\sigma$, Jy & Telescope \\
1 & 2 & 3 & 4 & 5 & 6 & 7 & 8 & 9  \\
\hline

57746 & 2016.12.23 & $1.52\pm0.37$   & $1.76\pm0.11$ & $1.98\pm0.02$ & $2.03\pm0.04$ & $1.87\pm0.07$ & $1.57\pm0.24$ & RATAN-600 \\
57777 & 2017.01.23 & $1.11\pm0.10$   & $1.34\pm0.10$ & $1.54\pm0.10$ & $1.77\pm0.10$ & $1.67\pm0.10$ & $0.69\pm0.10$ & RATAN-600 \\
57787 & 2017.02.02 & $1.35\pm0.09$   & $1.43\pm0.02$ & $1.75\pm0.01$ & $1.96\pm0.06$ & $1.89\pm0.36$ & --            & RATAN-600 \\
57817 & 2017.03.04 & $1.33\pm0.06$   & $1.48\pm0.03$ & $1.74\pm0.01$ & $1.93\pm0.02$ & $1.76\pm0.06$ & --            & RATAN-600 \\
59671 & 2022.04.01 & --              & --            & $1.86\pm0.03$ & --            & --            & --            & RT-32     \\
59672 & 2022.04.02 & --              & --            & $1.82\pm0.02$ & --            & --            & --            & RT-32     \\
59686 & 2022.04.16 & --              & --            & --            & $1.74\pm0.03$ & --            & --            & RT-32     \\
59686 & 2022.04.16 & --              & --            & --            & $1.80\pm0.03$ & --            & --            & RT-32     \\
\hline
\end{tabular}
\end{table*}

\begin{table}
\caption{The RATAN-600 daily measurements in 2021--2022: MJD (Col.~1), yyyy.mm.dd (Col.~2), flux densities at 5 and 2 GHz and their errors, Jy (Cols.~3--4). The full version is available as online supplementary material.}
\label{table4}
\centering
\begin{tabular}{cccc}
\hline
MJD  &  yyyy.mm.dd &  $S_{5}$, $\sigma$, Jy & $S_{2}$, $\sigma$, Jy \\
1 & 2 & 3 & 4 \\
\hline
59364 & 2021.05.29 & 2.49$\pm$0.05 & 1.27$\pm$0.12 \\
59365 & 2021.05.30 & 2.50$\pm$0.05 & 1.77$\pm$0.13 \\
59366 & 2021.05.31 & 2.60$\pm$0.06 & 1.53$\pm$0.09 \\
59367 & 2021.06.01 & 2.58$\pm$0.06 & 1.38$\pm$0.08 \\
59368 & 2021.06.02 & 2.50$\pm$0.04 & 1.43$\pm$0.08 \\
\hline
\end{tabular}
\end{table}

\subsection{RT-22}
The data on the flux densities at the 36.8 GHz frequency (further 37 GHz) for AO\,0235$+$164 were obtained with the 22-metre RT-22 radio telescope. The beam-modulated receivers were used to acquire the data. The antenna temperature from the source was measured as a difference between the signals from the radiometer output in two antenna positions, when the radio telescope was pointed at the source alternately with one or the other receiving horns (``on-on observation'' method). Observations of the blazar consisted of 5--20 such measurements to achieve a required signal-to-noise ratio. Details of the observations and data reduction are presented in
\cite{2023Galax..11...96V,2022AstBu..77..246S}.

The total period of observations with the RT-22 spans the time interval since February 2002 till October 2023. The flux densities, their errors and average observing epochs are presented in Table~\ref{RT22_obs}. The data up to 2012 have previously been used and published in \cite{2015ARep...59..145V}.

\begin{table}
\caption{The RT-22 measurements in 2003--2023:
MJD (Col.~1), yyyy.mm.dd (Col.~2), flux densities at 37 GHz and their errors, Jy (Col.~3). The full version is available as online supplementary material.}
\label{RT22_obs}
\centering
\begin{tabular}{ccccc}
\hline
MJD &  yyyy.mm.dd &  $S_{37}$, $\sigma$, Jy \\
1 & 2 & 3 \\
\hline
52679 & 2003.02.08 & $1.42\pm0.06$ \\
52681 & 2003.02.09 & $1.32\pm0.08$ \\
52693 & 2003.02.22 & $1.49\pm0.15$ \\
52695 & 2003.02.24 & $1.43\pm0.16$ \\
52697 & 2003.02.26 & $1.55\pm0.12$ \\
\hline
\end{tabular}
\end{table}

\subsection{Zeiss-1000 and AS-500/2 reflectors}
\label{sec:opt_studies}

The optical study (mainly in the $R$-band) was carried out with the 1-metre Zeiss-1000 (August 2002 -- December 2023) and 0.5-metre AS-500/2 (January 2021 -- December 2023) optical reflectors.

First optical observations of AO\,0235$+$164 in 2002 were mostly performed using the Zeiss-1000 telescope with a CCD photometer  based on a small ($520 \times 580$ pixels) front-illuminated CCD with  rectangular pixels: $18 \times 24$~$\mu$m. The photometer provided a total field of view of about \mbox{$2\farcm4\times 3\farcm7$}. The maximum quantum efficiency of the system was only  55 per cent.

Due to these reasons, further observations with the optical telescope Zeiss-1000 till the end of 2023 were implemented with another CCD photometer in the same Cassegrain focus, equipped with a $2048\times2048$~px back-illuminated E2V~chip CCD\,42-40. Details of the instrumental setup are described in our previous paper \citep{2023AstBu..78..464V}.

The main characteristics of the instrumental complex of an AS-500/2 reflector are presented in \cite{2022Photo...9..950V, 2022AstBu..77..495V}. In order to reduce the system readout noise and significant dark current of the FLI camera used, we installed in July 2023 a smaller but more sensitive detector: the back-illuminated electron-multiplying CCD camera Andor IXon$^{\rm EM}$+897 with high quantum efficiency ($\sim\!90$ per cent in the 450--700 nm range).

This CCD camera with $512 \times 512$ pixels, when mounted in the Cassegrain focus of AS-500/2, provides a $7'$ field of view with $0\farcs82$/pixel data sampling (1~pixel~$=$~16~$\mu$m). The system readout noise is about 6~$e^-$, the CCD operating temperature has been chosen to be $-70^\circ$C to minimize the dark current.

Both CCD photometers are equipped with similar sets of filters, which are close to  the standard broadband Johnson--Cousins system given the sensitivity of both CCDs.

The typical integration time for observations of the blazar was 300~s for Zeiss-1000 and 120~s for AS-500/2. During the period of high activity of the object, the integration time was less: up to 30~s for higher time cadence.

In order to obtain blazar flux estimates, we performed the standard reduction steps, which were described earlier \citep{1993BSAO...36..107V}. We collected the optical data as the instrumental magnitudes of the source and comparison stars in the same field. Magnitude calibration was performed using preferably
stars~8--11 from \cite{2001AJ....122.2055G}.

\begin{figure}
\includegraphics[width=1.0\linewidth]{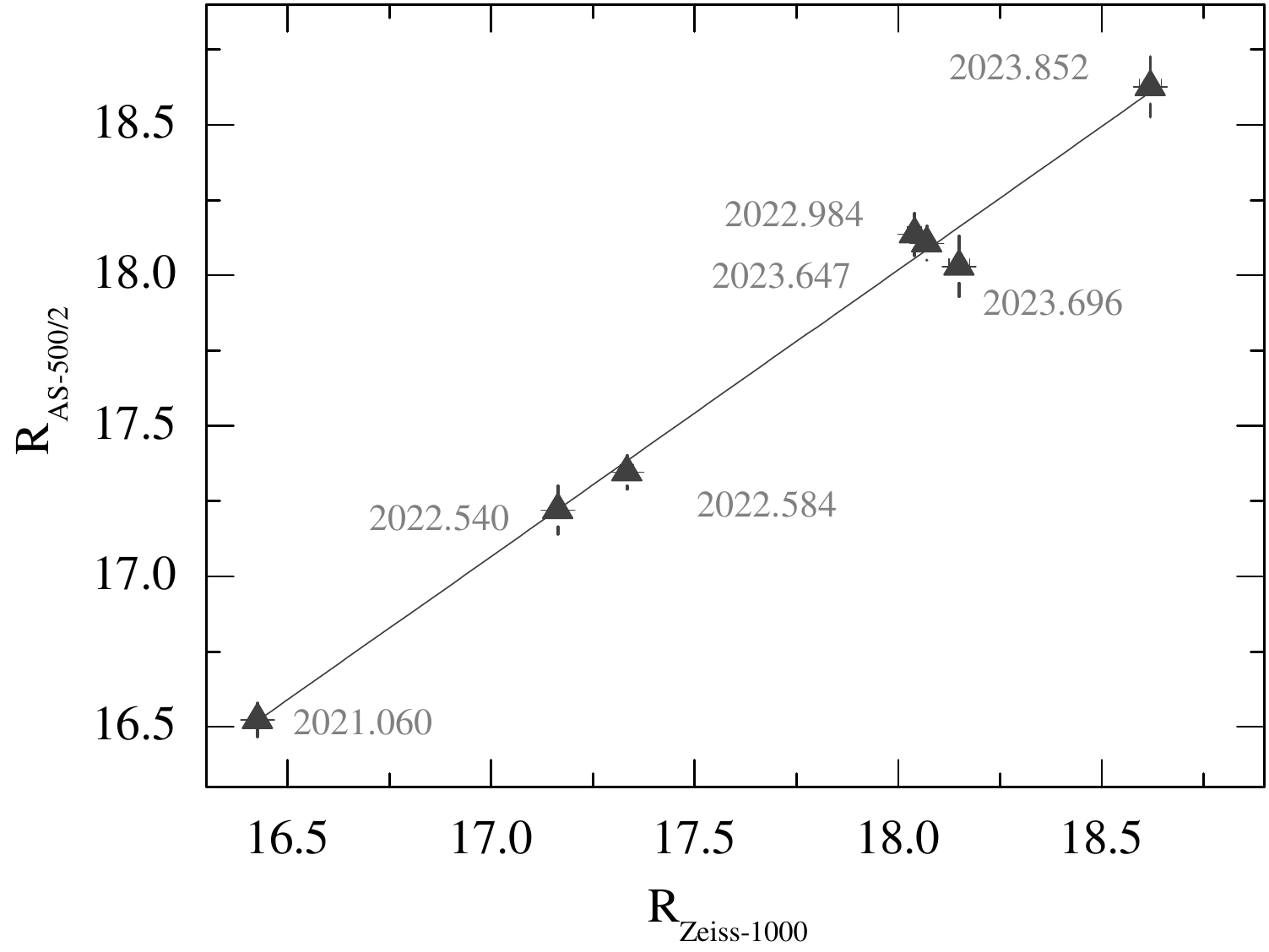}
\caption{Comparison of the AO\,0235$+$164 $R$-magnitudes taken with both optical telescopes ($x$-axis: Zeiss-1000, $y$-axis: AS-500/2).}
\label{fig:r-r}
\end{figure}

A comparison with the data obtained on the same nights with different telescopes has shown general agreement; in the cases where variations have been detected, they are consistent with the usual variations of the source, as it is shown in Fig.~\ref{fig:r-r}. The individual flux estimates are marked by their epochs. The mean deviation of the data from the linear law is $0\fm035$.

In order to provide a joint analysis of the optical and radio data, we averaged the measurements over individual nights and transformed the resulting values into fluxes according to the constant from \cite{1990A&AS...83..183M}. We did not subtract from the optical fluxes the addition from ELISA, the faint southern companion of AO\,0235$+$164 with $R=20\fm50$ \citep {2008A&A...480..339R}, supposing that this addition is insufficient for our following analysis.

The optical data are collected from 568 observing nights spanning between August 2002 and December 2023. The average observing epochs, number of observations for each night, and obtained fluxes with their errors are presented in Table~\ref{optic_obs}. Further in this paper we use the $R$-band fluxes not corrected for the Galactic extinction. 

\begin{table}
\caption{Daily averaged $R$-band measurements of AO\,0235$+$164 in August 2002 -- December 2023: the MJD (Col.~1), yyyy.mm.dd (Col.~2),
number of observations $N_{\mathrm{obs}}$ (Col.~3),
flux densities in the $R$-band and their errors in mJy (Col.~4). The full version is available as online supplementary material.}
\label{optic_obs}
\centering
\begin{tabular}{ccccc}
\hline
MJD &  yyyy.mm.dd & $N_{\mathrm{obs}}$ & $R_{\mathrm{flux}}$,  $\sigma_{\mathrm{R}}$, mJy \\
1 & 2 & 3 & 4 \\
\hline
52518  &  2002.08.31  &   5  &   $0.20\pm0.01$ \\
52529  &  2002.09.11  &   7  &   $0.17\pm0.01$ \\
52606  &  2002.11.27  &   6  &   $0.90\pm0.02$ \\
52609  &  2002.11.30  &   5  &   $0.63\pm0.01$ \\
52610  &  2002.12.01  &   5  &   $0.66\pm0.02$ \\
\hline
\end{tabular}
\end{table}

\subsection{Fermi-LAT}
We have made use of the $\gamma$-ray light curves available in the Fermi Large Area Telescope (Fermi-LAT) public Light Curve Repository (LCR)\footnote{\url{https://fermi.gsfc.nasa.gov/ssc/data/access/lat/LightCurveRepository/about.html}} \citep{2023ApJS..265...31A}. The LCR is a database of multicadence flux-calibrated light curves for over 1500 sources considered as variable in the 10-yr Fermi-LAT point source (4FGL-DR2) catalog \citep{2020arXiv200511208B}. The analysis has been performed with the standard Fermi-LAT science tools (version \verb|v11r5p3|), which use maximum likelihood analysis \citep{2009ApJS..183...46A}, considering a region with a radius of $12^{\circ}$ centered on the location of the source of interest and the $\gamma$-rays energy range between 100 MeV and 100 GeV. A limit of Test Statistics (TS) ${\rm TS}\geq 4$ (approximately $2\sigma$)  was applied to compute the fluxes for each bin of the light curve, while only the measurements with ${\rm TS}\geq 9$ were selected for further MW analysis. We adopted the monthly binned light curves to have more evenly-sampled time series with higher time cadence in the period from August 2008 to December 2023.

\begin{figure*}
\includegraphics[width=\linewidth]{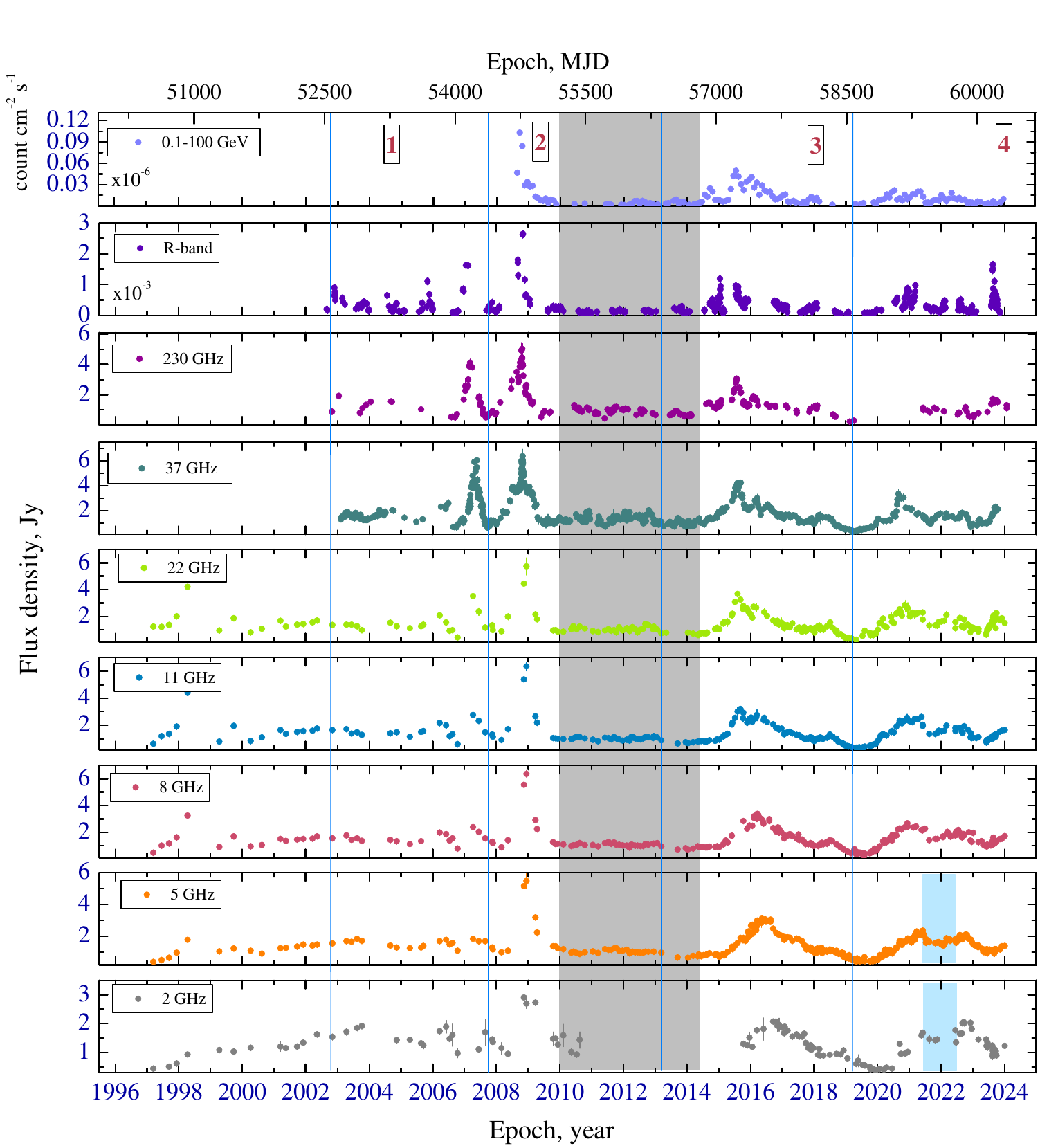}
\caption{Multiband light curves of AO\,0235$+$164 in 1997--2023. Four epochs chosen for analysis are depicted by the straight vertical lines. The historically quiet state between 2009.11--2014.03 is indicated by the grey area and shown in detail in Fig.~\ref{fig:low}. The blue area at 2 and 5 GHz is the time interval of the daily observations (see the description of RATAN-600 observations and Fig.~\ref{west_lc}).
}
\label{fig:l1}
\end{figure*}

\section{MW light curves}

\begin{figure}
\includegraphics[width=1.05\linewidth]{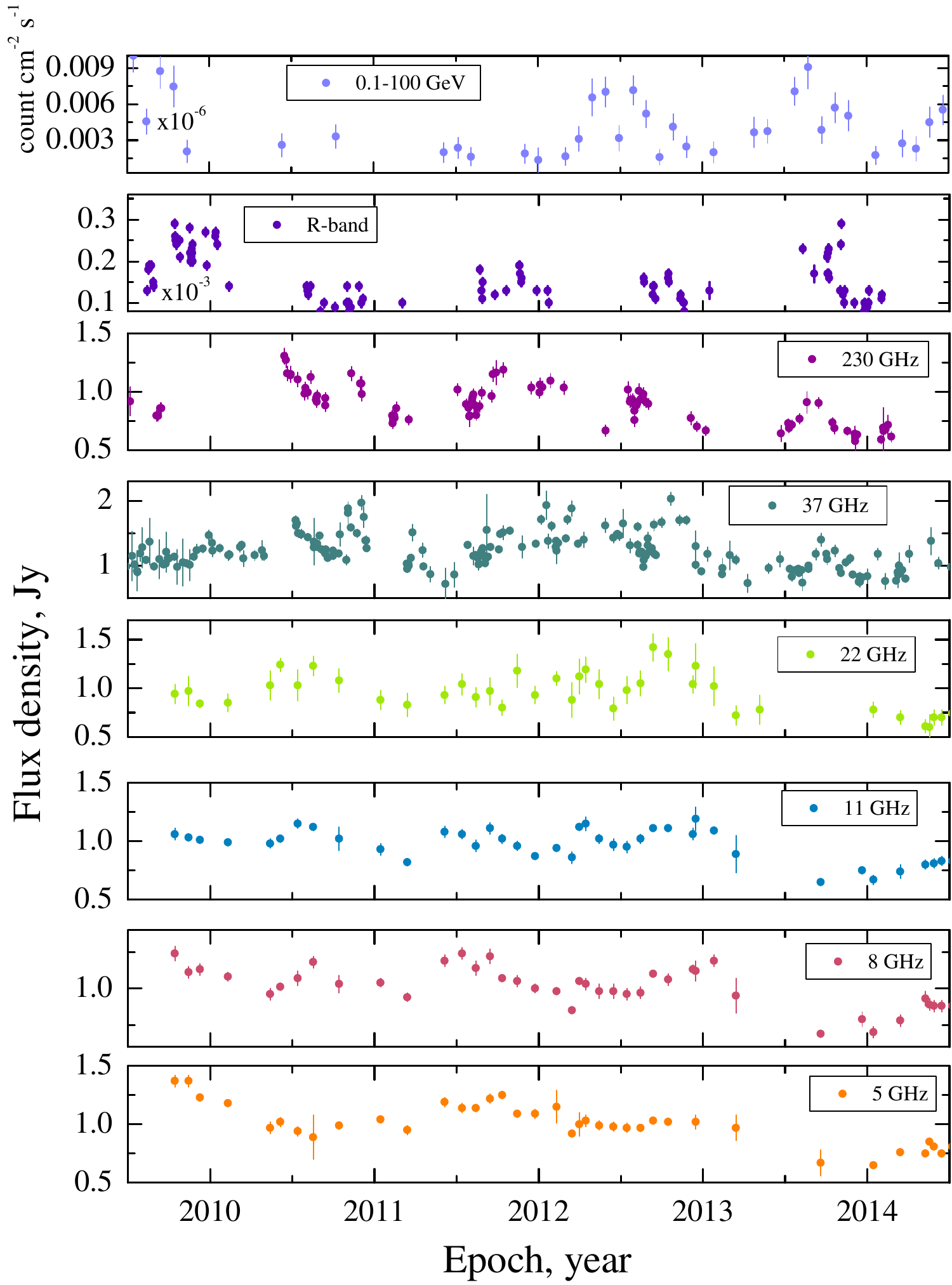}
\caption{The multiband light curves of AO\,0235$+$164 during
the period of low activity between November 2009 and March 2014.}
\label{fig:low}
\end{figure}

Figure~\ref{fig:l1} shows the MW light curves of AO\,0235$+$164
collected from March 1997 until December 2023.
The most complete light curves covering this period are those at radio frequencies of 5--22 GHz. The data for 37 GHz, 230 GHz, and the $R$-band cover a shorter interval: 2002--2023. The low-frequency radio curve at 2 GHz contains extended gaps due to strong radio interference. There are also seasonal gaps in the optical data, caused by the periods when the object is not visible in the spring--summer seasons. During the years 2009--2014, AO\,0235$+$164 had been showing quiescent broadband behaviour with a remarkable feature: very low activity lasting almost 4.5 yrs, from November 2009 to March 2014. We consider this time period as the low state epoch with the minimum variability level ($F_{\rm var}$, see Section~\ref{sec:vari}) across the light curves in all the bands. The light curves for this epoch are presented separately in Fig.~\ref{fig:low}.

We divided the light curves into four epochs (denoted by the blue vertical lines in Fig.~\ref{fig:l1} and listed in Table~\ref{states}) in a way that each epoch includes one of the four major flares. As the major flares, we adopt the most prominent flares ($>\!3$~Jy) at 37 GHz, because for this frequency we have the most uniform and abundant measurements. The first epoch covers the period from August 2002 to October 2007, during which the blazar underwent quasi-periodical flares in the optical range, separated by \mbox{480--620}~days and having amplitudes between $\sim\!0.6$~mJy and $\sim\!1.6$~mJy. The typical duration of these flares lies between 100 and 300 days. This value may be incorrect because the measurements do not cover the total period. In epoch 1, the maximum flux density was observed at 37 GHz, achieving 5.5 Jy in April 2007, ${\rm MJD}\sim54200$.

The second epoch spans the period of October 2007~-- April 2013, when the blazar demonstrated only one strong optical flare with a complex shape, located near epoch ${\rm MJD}=54774$, November 2008, with an $R$-band amplitude near 2.6 mJy. This optical outburst corresponds to powerful flares with flux densities between 5 and 6 Jy in the sub-mm and radio waves at a close time interval. The most powerful event in $\gamma$-rays over all the analysed period, $1.2 \times 10^{-6}$~ph\,cm$^{-2}$\,s$^{-1}$, is also located near this date.

Within $\sim\!1600$ days after this flare, AO\,0235$+$164 had been in a low state with optical fluxes between 0.1 and 0.3 mJy. The radio flux densities had been varying from $\sim\!0.5$~Jy to \mbox{$\sim\!1.3$--$2$}~Jy at 1--22~GHz within this epoch. The $\gamma$-ray emission was also at minimal levels during this time.

The third epoch spans the period of April 2013 -- March 2019, when AO\,0235$+$164 demonstrated four pronounced flares in the optical range with maxima between $R \sim 0.52$~mJy (${\rm MJD}=56952$, October 2014) and 1.2~mJy (${\rm MJD}=57042$, January 2015). These outbursts do not show any significant periodicity, and their duration did not exceed 200 days for the longest flare (${\rm MJD}=57221$, July 2015). The flares at the radio frequencies seem to be more smooth and may be a superposition of few events with amplitudes between 1.5 and 3 Jy at ${\rm MJD} \sim 57250$--57500, August 2015~-- April 2016.

The $\gamma$-ray emission is represented here by serial flares. The most powerful one with an amplitude of \mbox{$\sim\!7 \times 10^{-7}$~ph~cm$^{-2}$~s$^{-1}$} is also located near ${\rm MJD}=57230$, which is close to one of the brightest optical events.

The last epoch covers the observations between March~2019 and December 2023. The optical light curve is represented here by a set of $\sim\!15$ fast flares with amplitudes from 0.29 mJy (${\rm MJD}=59938$, December 2022) to 1.64 mJy (${\rm MJD}=60179$, August 2023). The longest outburst had a duration of about 70 days (the peak near ${\rm MJD}=59197$, December 2020), other flares are much shorter: the most intensive flare mentioned above had a duration of about 10 days. The fastest flares span about 6--8 days (the outbursts near ${\rm MJD}=59630$, February 2022, and ${\rm MJD}=59938$). Unfortunately, the real durations of some flares are unknown due to the absence of optical data within the periods when the blazar was unobservable (e.g., the flare with a maximum at ${\rm MJD}=59287$, March 2021). The most powerful (and much slower) two or three radio flares with amplitudes between 2--2.5~Jy are located near epochs ${\rm MJD}\sim 59200$, December 2020, and ${\rm MJD}\sim 59600$, January 2022, demonstrating slight delays as the frequency decreases. The $\gamma$-ray emission in this period seems to be 2--3 times fainter than in the previous epochs, approaching levels of
about $2 \times 10^{-7}$~ph~cm$^{-2}$~s$^{-1}$.

\begin{table}
\caption{Four epochs with flaring activity and a period with low activity}
\label{epochs}
\label{states}
\centering
\begin{tabular}{ccccc}
\hline
Epoch &  Period, yyyy.mm  & MJD \\
1 & 2 & 3 \\
\hline
1  &  2002.08--2007.10  & 52518--54373 \\
2  &  2007.10--2013.04  & 54374--56367 \\
3  &  2013.04--2019.03 & 56368--58529 \\
4  &  2019.03--2024.01 & 58530--60237 \\
\hline
\hline
low state & 2009.11 --2014.03 & 55151--56731 \\ 
\hline
\end{tabular}
\end{table}

\subsection{Parameters of the most prominent flares}

At high radio frequencies above 10 GHz, flux density variations in AGNs can be well modeled by a superposition of flare components \citep{1999ApJS..120...95V,2009A&A...494..527H}. According to these papers, the ascending and descending wing of each flare may be modeled by an exponential law with the same amplitude and epoch of the flare and different time-scales. We used independent time-scales for the rise and decay of a flare in the approximation procedure, whilst the authors of the papers cited above supposed a close connection between them.

\begin{figure*}

\centerline{\includegraphics[height=5cm]{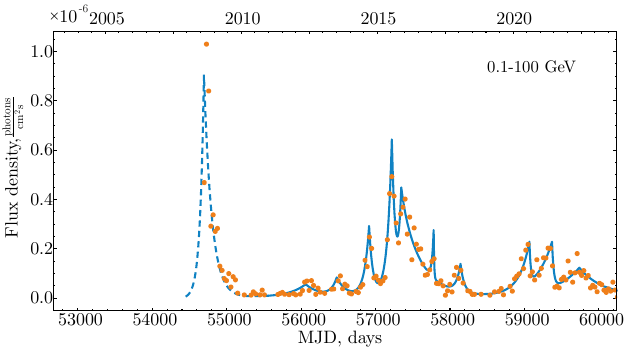}\includegraphics[height=5cm]{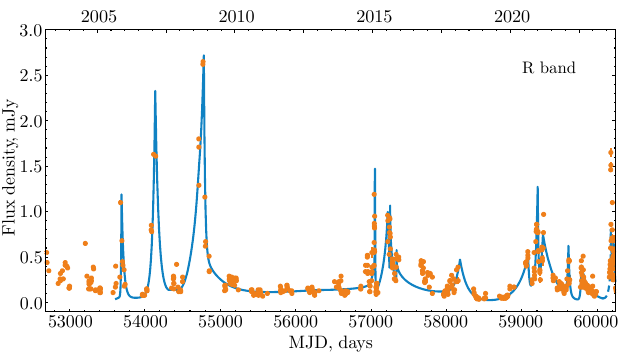}}
\centerline{\includegraphics[height=5cm]{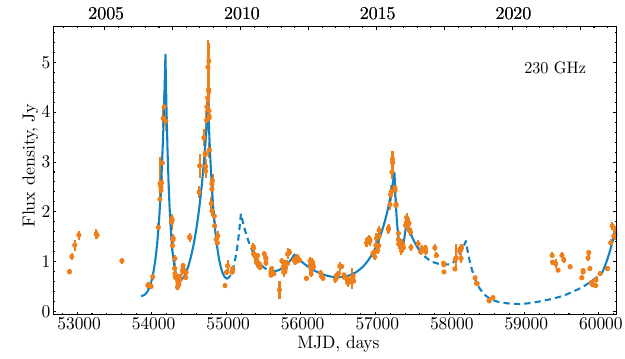}\includegraphics[height=5cm]{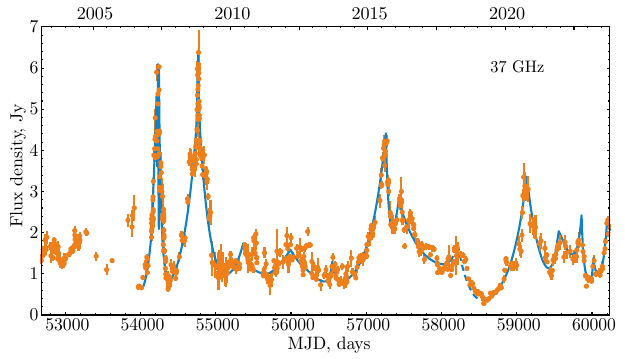}}
\centerline{\includegraphics[height=5cm]{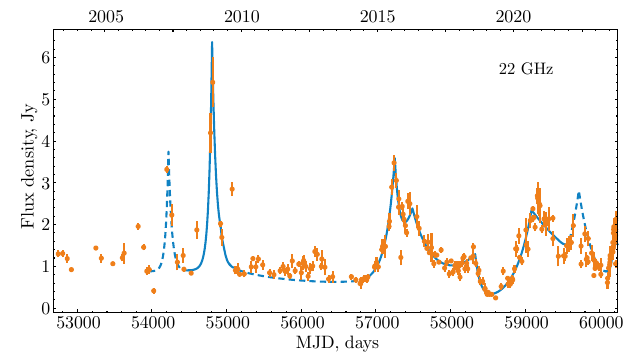}\includegraphics[height=5cm]{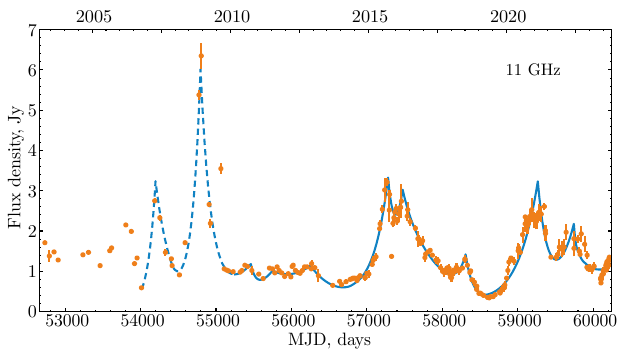}}
\centerline{\includegraphics[height=5cm]{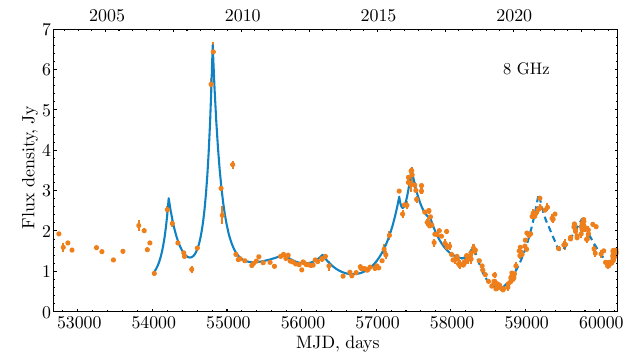}\includegraphics[height=5cm]{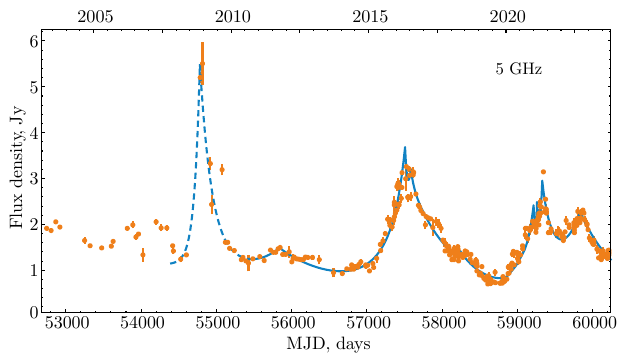}}
\caption{MW light curves of AO 0235$+$164 for $\gamma$-rays,
optical $R$-band and radio emission (230, 37, 22, 11, 8 and 5 GHz) with the most prominent flares modeled by exponential laws (the blue curves). The dotted lines indicate events that are poorly described by this method due to incomplete data.}
\label{fig:exp}
\end{figure*}

To preprocess the light curves, we used cubic Hermite interpolation. The flares were then detected using the Bayesian block method described in \citet{2013ApJ...764..167S} and implemented in Python by \citet{2022icrc.confE.868W}. As a result, for all the light curves we obtained a picture of decomposition into exponential components (Fig.~\ref{fig:exp}).

The behaviour of the flares along the spectral ranges shows systematic variation: they become slower and more smooth as the frequency decreases, depicting less energetic processes. The most abundant ranges are the optical band and $\gamma$-rays: they demonstrate the presence of tens of strong and faint flares, both fast and slow. With emission energy falling down, these flares become wider, exhibiting a time delay with respect to the optical/$\gamma$-ray flares, the greater as the wavelength grows.

As a result, for the light curves at frequencies of 5 and 8~GHz, we can see only few strongest of them: the fainter events that leave only slight smooth waves in the total curve. For example, the delay of the most powerful outburst (${\rm MJD} \sim 54730$, September 2008 for $\gamma$-ray and optical emission) increases from 30 days for radio frequencies of 37 and 230 GHz to 70 days at 5 and 8 GHz.

The amplitudes of the most powerful flares vary through radio bands within 20--30 per cent of the mean values with only one exception: the leftmost outburst in Fig.~\ref{fig:exp} (MJD near 54200, April 2007). Its amplitude, being near 5--6 Jy at high frequencies (230 and 37 GHz), becomes 3 times lower at 8~GHz and is only slightly noticeable at 5 GHz. A similar behaviour of flare components may be noticed for the group of outbursts at epoch ${\rm MJD} \sim 57000$--$57300$, January--September 2015. A complete analysis of event delays over different wavebands is presented in Section~\ref{sec:dcf}.

The behavior of the flare in epoch 1 differs from the flares in epochs~2--4 by a steeper decrease of its amplitude with frequency decrease. As a result, the flare practically does not appear at 2 and 5 GHz in epoch 1.   This behavior corresponds to an optically thick emission area, where the spectrum is inverted at 2--22 GHz ($\alpha_{2-5}=+0.59$ and $\alpha_{5-22}=+0.21$ in the epoch around ${\rm MJD}= 54267$).

Within the framework of the ``generalized shock model'' of \cite{1992A&A...254...71V}, which supposes the presence of growing and decaying shocks in relativistic jets, we can attribute this outburst as ``high-peaked,\!'' and the others should be attributed as ``low-peaked'' flares. Among  possible explanations, there are higher magnetic fields or a smaller size of the emitting region at the moment of ``high-peaked'' flare initiation.

Also in the case of this flare in epoch 1, the relevant explanation is the possibility of higher absorption of low-frequency radio emission because of a larger region where it originates compared to the flares at other epochs, which also appears as a more stretched variability time-scale. The analysis of AO 0235$+$164 long-term variability in  \cite{2018MNRAS.475.4994K} found a similar behavior of some flares observed in 1980--2012, which corresponds to extreme compactness of the blazar core, although the flare in epoch~1 stands out between them as possessing the steepest spectral index.

\section{Analysis of long-term variability}

\subsection{Fractional variability}
\label{sec:vari}

Quantification of the degree of variability in each band can provide
information about the particle population that dominates in
the emission in a certain energy band. For this purpose, we used
fractional variability $F_{\rm var}$ \citep{2003MNRAS.345.1271V}:
\begin{equation}
\label{eq:frac}
F_{\rm var}=\sqrt{\frac{V^2-\bar\sigma^{2}_{\rm err}}{\bar S^2}}
\end{equation}
where $V^2$ is variance, $\bar S$ is
the mean flux density, and $\sigma_{\rm err}$ is the root mean square error. The uncertainty of $F_{\rm var}$ is determined as
\begin{equation}
\label{eq:frerr}
\bigtriangleup F_{var}=\sqrt{\left(\sqrt{\frac{1}{2N}}\frac{\bar\sigma^{2}_{\rm err}}{F_{var}*\bar S^2}\right)^2 + \left(\sqrt{\frac{\bar \sigma^{2}_{\rm err}}{N}}\frac{1}{\bar S}\right)^2}
\end{equation}

\begin{figure}
\includegraphics[width=\columnwidth]{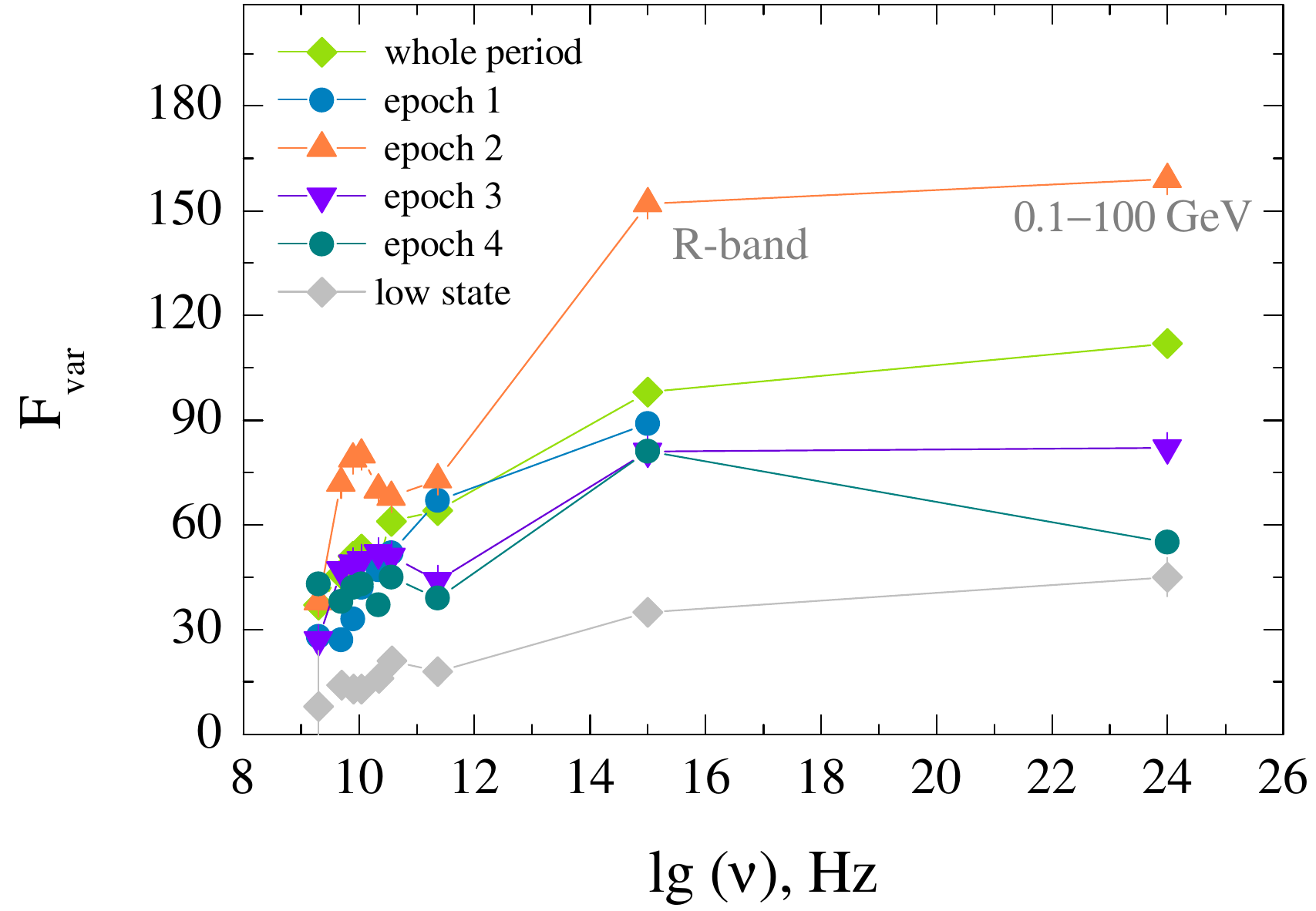}
\caption{Fractional variability of AO\,0235$+$164. Different markers represent observations during the whole period, low-state epoch, and
four activity epochs.}
\label{fig:var-band}
\end{figure}

Fractional variability was computed for the light curves shown in Fig.~\ref{fig:l1} for the whole period as well as for each of four flaring states and the low-state period. The $F_{\rm var}$--$\log \nu$ plot is given in Fig.~\ref{fig:var-band}. The $F_{\rm var}$ errors are from 0.2 to 6 per cent in different wavelength bands. Fractional variability increases with increasing frequency for the whole time period and for each selected epoch.

$F_{\rm var}$ reaches maximum levels in all bands in epoch 2 (above 150 per cent for the optical and $\gamma$-ray bands). In the low-state epoch fractional variability is from 7 to 20 per cent at frequencies of 2--230 GHz and reaches 35 and 52 per cent in the optical and $\gamma$-ray bands.

\subsection{Structure function analysis}
\label{sec:SF}

The structure function (SF) is a method of searching for typical time-scales and periodicities in non-stationary processes \citep{1985ApJ...296...46S,1992ApJ...396..469H}. SF analysis provides time variability quantification and gives information on the nature of the process that caused the variations. A characteristic time-scale in a light curve, defined as the time interval between a maximum and an adjacent minimum or vice versa, is indicated by a maximum of the SF, whereas periodicity in the light curve causes a minimum \citep{1996A&A...305...42H}.

The SF of the first order normalized to the variance of the signal $\sigma^{2}$ is usually determined as:
\begin{equation}
D_{1}(\tau) = \langle\{[f(t)-f(t+\tau)]\}^2\rangle,
\end{equation}
where $f(t)$ is the signal at a time $t$, and ${\tau}$ is the lag. The slope of the power part of the curve is estimated as:
\begin{equation}
b=d\log{D_{1}}/d\log{\tau}.
\end{equation}

An ideal SF consists of two plateaus and an upward-sloping curve with a slope $b$ between them. One of the important characteristics of the SF is $\tau$,  the point where the SF reaches its higher plateau with an amplitude equal to 2$\sigma^{2}$. This time-scale gives the maximum time-scale of correlated signals or, equivalently, the minimum time-scale of uncorrelated behaviour.
The value of the slope $b$ determines the nature of the variable process. If the light curve can be modeled as a combination of white (flicker) and red (shot) noises, then the slope value is between 0 and 1 \citep{1992ApJ...396..469H}. For a single dominating outburst in the light curve, the slope is usually steeper than 1. If the slope $b \sim 2$, then there is a strong linear trend or a strong periodic oscillation \citep{1992ApJ...386..473H}.

In this work, for the non-uniform and finite data series $f(i)$ with $i=1,2,...,N$, obtained from the original temporally non-uniform observed data series by partitioning into intervals with the values averaged over the interval, the first-order SF is calculated as:
\begin{equation}
D_{1}(k) = \frac{1}{N_{1}(k)}\sum_{i=1}^{N}{w(i)w(i+k)[f(i+k )-f(i)]^2},
\end{equation}
where $N_{1}(k)=\sum{w(i)w(i+k)}$ and the weight factor $w(i)$ is equal to $1$ if the observations exist in the interval $i$ and is $0$ if there are no observations.

For $k=1,2,...,L$, we construct our own set of intervals. The initial time interval $k=1$ is chosen in such a way that it is equal to or slightly larger than the average time interval between observations (ignoring very long gaps in them). For radio frequencies of 2--22 GHz, intervals between 10 (for the low-state epoch) and 50 (epochs 1--4) days were taken as the initial ones. The corresponding intervals for 37 GHz, 230~GHz, and optical data lie between 4 and 20 days, whereas $\gamma$-ray data are described by an SF with an initial time interval of 30 days. The final interval $k=L$ is calculated based on the length of the time-scale of the original sequence so that all the values of the original series are divided into two intervals.

\begin{table}
\centering
\caption{\label{tab:sf} The time-scales of the SF for the $\gamma$-rays, $R$-band, and radio emission at 2, 5, 8, 11, 22, and 37 GHz, calculated for epochs~1--4 and
the quiet state of AO\,0235$+$164}
\begin{tabular}{|c|cccccc|}
\hline
Band & \multicolumn{5}{c}{$\tau$, years}\\
\hline
 &  ep 1 & ep 2 & ep 3 & ep 4 & epoch$_{\rm low}$\\
\hline
2 GHz   & 1.1 & 0.9 & 0.9  & 1.7 & --  \\
5 GHz   & --  & 0.5 & 1.1 & 1.7 & 0.5 \\
8 GHz   & 0.9 & --  & 0.9 & 1.7 & 0.5 \\
11 GHz  & 0.9 & --  & 1.1 & 1.4 & 0.5 \\
22 GHz  & 0.5 & 0.7 & 0.9 & 1.4 & 0.5 \\
37 GHz  & 0.4 & 0.7 & $\geq$0.9  & 0.9 & $\geq$0.5 \\
230 GHz & 0.3 & 0.7 & 1.7 & 1.1 & 0.5 \\
R       & 0.3 & --  & --  & 0.4 & 0.4 \\
$\gamma$& --  & 0.4  & 0.4 & 0.4 & 0.5 \\
\hline
\end{tabular}
\end{table}

We estimated the uncertainties on the SF through an independent bootstrap technique. A model light curve was generated for each source and frequency by applying a 2-day boxcar smoothing and a subsequent averaging process to the original light curve. The resulting 2-day model was then subtracted from each light curve, yielding a set of residuals. To preserve the time sampling of the original data, a new simulated light curve was created by adding a randomly selected residual to each point in the model. This process ensured that each residual was utilized once. Repetition of this procedure 1000~times allowed us to generate multiple simulated light curves, allowing the SF to be recalculated each time. The confidence limits were then determined, such as the 99 per cent confidence limit on a given time-scale, which was identified as the point where only 5 data points deviated from the value. This approach does not mandate symmetric confidence limits.

The computed time-scale values for the four epochs and for the low-state epoch are given in Table~\ref{tab:sf} and in \mbox{Figs.~\ref{fig:SF1}--\ref{fig:SF}}. The dashes in Table~\ref{tab:sf} for some frequencies denotes the absence of a clear structure in the obtained SFs. The absence of the variability scale at 37 GHz for the low state epoch is due to the impossibility to determine its value reliably. We have indicated its lower limit ($\geq0.5$~yrs) as well as for epoch~3 at 37 GHz ($\geq0.9$~yrs).

As it was mentioned by \cite{2010MNRAS.404..931E}, results of SF analysis must be interpreted with some caution. Strong SF breaks frequently occur in data sets without any sort of characteristic time-scales. The position of these artificial breaks depends on the length of the observations and the nature of the variation process. On the other side, intensive modeling shows that SFs derived from typical blazar observations resemble those coming from the shot-noise model. However, in the frequency domain the same model does not describe correctly the observations and thus may not be physically realistic.

Also it was emphasised in the paper cited above that the data gaps affect severely the SF estimates in an unpredictable way, introducing systematic deviations. Even the bootstrap method cannot yield statistically meaningful errors depicting the deviations between the gappy and continuous SFs. This effect as well as the presence of some powerful outbursts may be responsible for difficulties with SF estimates in the optical data, where SF was successfully built only at the quiet epoch between 2009 and 2014.

The SF analysis reveals different variability time-scales: for epochs 1 and 2 the $\tau$ value is between 0.3 and 1 yr; for epochs 3 and 4 $\tau$ is almost two times higher, between 0.9 and 1.7 yrs, respectively. The variability time-scale grows steadily from $\gamma$-rays (about 0.4--0.5 yrs) to longer radio waves (between 0.9 and 1.7 yrs for different epochs), but significantly differs from one epoch to another. The variability time-scales for the quiet epoch are similar at all wavelengths: $\tau=0.4$--$0.5$ yrs.

The given trends of $\tau$ are consistent with the findings of \citep{2001A&A...377..396R, 2002A&A...381....1F}, which also indicate time-scales of up to a year that are related to accretion processes and jet activity.

\subsection{Multiband correlations}
\label{sec:dcf}

Cross-correlation functions (CCFs) are commonly used in the multifrequency analysis of light curves observed over approximately the same period. The CCF analysis described in many
digital signal processing texts is usually applied to the regular time series, which is not always possible in the case of astronomical observations. To solve this problem, discrete correlation functions (DCFs) for irregularly sampled time series began to be used several decades ago to find the time lags between the light curves at different frequencies \citep{1988ApJ...333..646E}.

In this paper the DCF method was applied in the same way as in  our earlier work \citep{2023AstBu..78..464V}. The Python-based software described in \cite{2015MNRAS.453.3455R} was used to calculate cross correlation functions between all the frequency bands mentioned above.
Bin widths $DT=10$--$40$ days were taken.

To estimate the confidence level of a resulted DCF, we utilized Monte-Carlo simulations. The method is outlined in \cite{2013MNRAS.433..907E}, where related software is also described. The point is to simulate a large number of artificial light curves that have the same statistical properties (probability density function, power spectrum density) as the real light curve has. To find the probability of getting a given DCF value purely by chance, one can calculate cross-correlation functions for an ensemble of paired artificial light curves and analyse the DCF distribution for each lag.

\begin{table*}
\centering
\caption{\label{tab:ccf-lags} Pairwise DCF results for all
the AO\,0235$+$164 light curves. A positive lag for a DCF means that the variations of light curve 2 lag with respect to those of light curve 1 for the relation ``frequency 1 vs frequency 2''}
\begin{tabular}{c|cccccccccc}
\hline
band  & $lag$, days & $DCF$ & $lag$, days  & $DCF$  & $lag$, days & $DCF$ & $lag$, days & $DCF$ \\
\hline
  & \multicolumn{2}{c}{epoch 1} &  \multicolumn{2}{c}{epoch 2} & \multicolumn{2}{c}{epoch 3} & \multicolumn{2}{c}{epoch 4}\\
\hline
$\gamma$ vs 230 GHz & -- & -- &  -- & -- &  10 $\pm$ 10 & 0.84 $\pm$ 0.21 &     --      &     --      \\
$\gamma$ vs 37 GHz  & -- & -- &  -- & -- &  30 $\pm$ 10 & 0.87 $\pm$ 0.17 &     --      &     --      \\
$\gamma$ vs 22 GHz  & -- & -- &  120$\pm$20 & 0.73$\pm$0.31 &  30 $\pm$ 10 & 0.92 $\pm$ 0.22 & 10 $\pm$ 10 & 0.78 $\pm$ 0.24   \\
$\gamma$ vs 11 GHz  & -- & -- &  -- & -- &  190 $\pm$ 10 & 0.87 $\pm$ 0.22 & 10 $\pm$ 10 & 0.89 $\pm$ 0.24   \\
$\gamma$ vs 8 GHz   & -- & -- &  -- & -- &  230 $\pm$ 10 & 0.88 $\pm$ 0.16 & 150$\pm$ 10 & 0.83 $\pm$ 0.19   \\
$\gamma$ vs 5 GHz   & -- & -- &  -- & -- &  190 $\pm$ 10 & 0.86 $\pm$ 0.12 & 130$\pm$ 10 & 0.78 $\pm$ 0.10 \\
$\gamma$ vs 2 GHz   & -- & -- &  -- & -- &  310 $\pm$ 10 & 0.86 $\pm$ 0.21 & 243$\pm$ 15 & 0.72 $\pm$ 0.18 \\
\hline
R vs 230 GHz & 0 $\pm$ 20  & 0.93 $\pm$ 0.12 & 210$\pm$10 & 0.73$\pm$0.10 & 30 $\pm$ 10 & 0.85 $\pm$ 0.14 & 40 $\pm$ 20 & 0.66 $\pm$ 0.07  \\
R vs 37 GHz  & 0 $\pm$ 20  & 0.87 $\pm$ 0.08 & -- & -- &  65 $\pm$ 5  & 0.87 $\pm$ 0.15 &      --      &     --   \\
R vs 22 GHz  & 40 $\pm$ 20 & 0.98 $\pm$ 0.56 & -- & -- &  10 $\pm$ 10 & 0.82 $\pm$ 0.24 & 70 $\pm$ 10 & 0.79 $\pm$ 0.11  \\
R vs 11 GHz  & 40 $\pm$ 20 & 0.96 $\pm$ 0.57 & -- & -- &  10 $\pm$ 10 & 0.83 $\pm$ 0.21 & 70 $\pm$ 10 & 0.88 $\pm$ 0.11  \\
R vs 8 GHz   & 40 $\pm$ 20 & 0.96 $\pm$ 0.59 & -- & -- & --  & --              &110 $\pm$ 10 & 0.74 $\pm$ 0.11  \\
R vs 5 GHz   & 40 $\pm$ 20 & 0.94 $\pm$ 0.59 & -- & -- &  -- & --              &145 $\pm$ 5 & 0.82 $\pm$ 0.12  \\
R vs 2 GHz   & 40 $\pm$ 20 & 0.90 $\pm$ 0.55 & -- & -- &  -- & --              &     --      &     --    \\
\hline
230 GHz vs 37 GHz  & 0  $\pm$ 20 & 0.86 $\pm$ 0.03 & 0$\pm$20 & 0.75$\pm$0.06 & 10 $\pm$ 10 & 0.93 $\pm$ 0.11 & 0  $\pm$ 20 & 0.86 $\pm$ 0.16   \\
230 GHz vs 22 GHz  & 40 $\pm$ 20 & 0.95 $\pm$ 0.13 & -- & -- & 10 $\pm$ 10 & 0.90 $\pm$ 0.20 & 40 $\pm$ 20 & 0.76 $\pm$ 0.16   \\
230 GHz vs 11 GHz  & 40 $\pm$ 20 & 0.92 $\pm$ 0.14 & -- & -- & 10 $\pm$ 10 & 0.79 $\pm$ 0.16 &200 $\pm$ 20 & 0.90 $\pm$ 0.19  \\
230 GHz vs 8 GHz   & 40 $\pm$ 20 & 0.90 $\pm$ 0.14 & -- & -- & 250 $\pm$ 10 & 0.80 $\pm$ 0.12 &120 $\pm$ 20 & 0.83 $\pm$ 0.21   \\
230 GHz vs 5 GHz   & 40 $\pm$ 20 & 0.87 $\pm$ 0.14 & -- & -- & 390 $\pm$ 10 & 0.77 $\pm$ 0.12 &160 $\pm$ 20 & 0.74 $\pm$ 0.14   \\
230 GHz vs 2 GHz   & 40 $\pm$ 20 & 0.89 $\pm$ 0.13 & -- & -- & 450 $\pm$ 10 & 0.79 $\pm$ 0.21 &440 $\pm$ 20 & 0.88 $\pm$ 0.25   \\
\hline
37 GHz vs 22 GHz  & 0  $\pm$ 20 & 0.83 $\pm$ 0.08 & -- & -- & 10$\pm$ 10 & 0.95 $\pm$ 0.14 & 30$\pm$ 10 & 0.87 $\pm$ 0.14   \\
37 GHz vs 11 GHz  & 40 $\pm$ 20 & 0.81 $\pm$ 0.09 & -- & -- & 30$\pm$ 10 & 0.90 $\pm$ 0.12 & 30$\pm$ 10 & 0.90 $\pm$ 0.14   \\
37 GHz vs 8 GHz   & 40 $\pm$ 20 & 0.80 $\pm$ 0.09 & -- & -- & 230$\pm$ 10 & 0.91 $\pm$ 0.07 & 90$\pm$ 10 & 0.92 $\pm$ 0.13   \\
37 GHz vs 5 GHz   & 40 $\pm$ 20 & 0.79 $\pm$ 0.09 & -- & -- & 250$\pm$ 10 & 0.83 $\pm$ 0.08 &190$\pm$ 10 & 0.84 $\pm$ 0.10   \\
37 GHz vs 2 GHz   &160 $\pm$ 20 & 0.71 $\pm$ 0.11 & -- & -- & 370$\pm$ 10 & 0.85 $\pm$ 0.12 &230$\pm$ 10 & 0.85 $\pm$ 0.15  \\
\hline
22 GHz vs 11 GHz  & 0  $\pm$ 20 & 0.98 $\pm$ 0.48  & 0$\pm$20 & 1.01$\pm$0.32 & 10  $\pm$ 10 & 0.96 $\pm$ 0.22 & 50  $\pm$ 10 & 1.00 $\pm$ 0.18 \\
22 GHz vs 8 GHz   & 0  $\pm$ 20 & 0.94 $\pm$ 0.48  & -- & -- & 10  $\pm$ 10 & 0.95 $\pm$ 0.17 & 50  $\pm$ 10 & 1.02 $\pm$ 0.18 \\
22 GHz vs 5 GHz   & 0 $\pm$ 20  & 0.87 $\pm$ 0.47  & -- & -- & 250 $\pm$ 10 & 0.91 $\pm$ 0.12 &190  $\pm$ 10 & 0.94 $\pm$ 0.09 \\
22 GHz vs 2 GHz   & 120 $\pm$ 20 & 0.84 $\pm$ 0.44 & -- & -- & 250 $\pm$ 10 & 0.93 $\pm$ 0.29 &210  $\pm$ 10 & 0.80 $\pm$ 0.21 \\
\hline
11 GHz vs 8 GHz   & 0  $\pm$ 20 & 0.99 $\pm$ 0.5 & 0$\pm$20& 0.82$\pm$0.30 & 10  $\pm$ 10 & 0.99 $\pm$ 0.12 & 10  $\pm$ 10 & 0.98 $\pm$ 0.13 \\
11 GHz vs 5 GHz   & 0  $\pm$ 20 & 0.94 $\pm$ 0.48 & -- & -- & 10  $\pm$ 10 & 0.95 $\pm$ 0.12 & 130 $\pm$ 10 & 0.93 $\pm$ 0.11 \\
11 GHz vs 2 GHz   & 120 $\pm$ 20 & 0.83 $\pm$ 0.42 & -- & -- & 250  $\pm$ 10 & 0.99 $\pm$ 0.21 & 210 $\pm$ 10 & 0.80 $\pm$ 0.20 \\
\hline
8 GHz vs 5 GHz   & 0  $\pm$ 20 & 0.97 $\pm$ 0.48 & 0$\pm$20 & 0.82$\pm$0.34 & 10  $\pm$ 10 & 0.98 $\pm$ 0.11 & 70  $\pm$ 10 & 0.94 $\pm$ 0.07 \\
8 GHz vs 2 GHz   & 120 $\pm$ 20 & 0.85 $\pm$ 0.42 & -- & -- & 90  $\pm$ 10 & 0.96 $\pm$ 0.16 & 190 $\pm$ 10 & 0.94 $\pm$ 0.11 \\
5 GHz vs 2 GHz  & 160 $\pm$ 20 & 0.89 $\pm$ 0.59 & -- & -- & 90  $\pm$ 10 & 0.99 $\pm$ 0.16 & 130 $\pm$ 10 & 0.93 $\pm$ 0.11 \\
\hline
\end{tabular}
\end{table*}

Since the quality of the data was quite different at various frequencies and in different epochs, sometimes a DCF had very large errors that made impossible to assert confidently its values, such cases were skipped. The DCF examples are presented in Figs.~\ref{fig:dcf1}-\ref{fig:dcf4}.

The $2\sigma$ confidence level DCFs between the pairs of frequencies and the corresponding time lags over the whole period and for epochs 1, 2, 3, and 4 are displayed in Table~\ref{tab:ccf-lags}. For this analysis we changed the definition of the epochs due to the poor number of data points for the first and second flares and complicated DCF calculation. Therefore, for epoch 1 we combined the first two flares, the second epoch is the low state, the third and forth epochs are the same.

The analysis of the DCF data shows different
time delays and behaviour  of multiband variability for
epochs 1--4. Evidently, all the analysed light curves exhibit flux variations, which are typical for blazars, where the flux changes at
higher frequencies lead those at lower frequencies, as it was noted in \cite{2001A&A...377..396R},\cite{2015A&A...582A.103G}, and references therein. For AO\,0235$+$164 the value of such lags differs for epochs under analysis. A minimum delay, 0--40~days, is observed in epoch 1 with the two brightest flares for the pairs ``37 GHz -- 2 GHz'' and ``5 GHz -- 2 GHz.\!'' The time lags for epochs 3 and 4 are higher, hundreds of days, and approach 450 days for the pairs ``230 GHz -- 5 GHz'' and ``230~GHz~-- 2~GHz.\!'' This observational result confirms the scenarios where time lags at lower frequencies are due to different opacities of the matter, as greater synchrotron self-absorption could occur at the lower frequencies compared to the higher ones.

\subsection{Search for long-term periodicity}

The Lomb--Scargle (L--S) periodogram method \citep{1976Ap&SS..39..447L,1982ApJ...263..835S} is widely used in astronomy to search for periodicities in unevenly sampled time series, like the light curves. The result of this method is a periodogram similar to the power spectrum estimate in the standard Fourier analysis, where the most prominent peaks show the periods of oscillations in given data.

To find the long-term periodicity in our data, we used the {\tt astropy} L--S implementation\footnote{\url{https://docs.astropy.org/en/stable/timeseries/lombscargle.html}} based on the algorithms described
in detail in \cite{2014sdmm.book.....I}. For a cross-check, we also used the Generalized Lomb-Scargle Periodogram module from {\tt pyAstronomy},\!\footnote{\url{https://pyastronomy.readthedocs.io/en/latest/pyTimingDoc/pyPeriodDoc/gls.html}} implemented as described by \cite{2009A&A...496..577Z}, and found that these two implementations give very similar results.

The interpretation of the L--S method results should be made
with care. First of all, special attention should be paid to the window function. In most cases it has a shape of a ``comb'' with a minimum interval of one day between the successive observations
and with large gaps when the source is not observable. The convolution of the window function with the true light curve spectral image in the spectral domain can lead to false peak detection.

It was found that for our data the most prominent window function feature was the peak around a period of 365~days in the
$R$-band light curve, caused by the yearly nature of time intervals when the source is unobservable during daytime.

At other frequencies the window function produces
mild comb-like features at the scales of interest (e.g., at long-period scales).

\begin{table}
\centering
\caption{\label{tab:ls} Results of quasi-periodicity analysis}
\begin{tabular}{|c|crcr|}
\hline
band  & all epochs & sign. & low state & sign. \\
\hline
$\gamma$-ray & $6.1\pm0.3$ & 1.4$\sigma$ & $1.4\pm0.1$ & 2$\sigma$ \\
$R$-band       & $6.4\pm0.1$  & 2.4$\sigma$ & $1.7\pm0.1$ & 3$\sigma$\\
230          & $6.0\pm0.3$  & 1$\sigma$ & $1.3\pm0.1$ & 1.8$\sigma$\\
37           & $6.7\pm0.2$  & 1.3$\sigma$ & $2.1\pm0.1$ & 2.8$\sigma$\\
22           & $5.0\pm0.2$  & 3.5$\sigma$ & $2.1\pm0.1$ & 2$\sigma$\\
11           & $6.0\pm0.2$  & 2.6$\sigma$ & $2.3\pm0.1$ & $3\sigma$\\
8            & $6.0\pm0.2$  & 4.2$\sigma$ & $1.4\pm0.1$ & 2.6$\sigma$\\
5            & $5.5\pm0.1$  & 5$\sigma$ & $1.7\pm0.1$ & $3\sigma$\\
2            & $6.9\pm0.2$  & 3.5$\sigma$ & -- & --\\
\hline
\end{tabular}
\end{table}

The L--S periodograms show peaks with significance above $2\sigma$ (with the exception of $\gamma$-rays and the 37 and 230 GHz radio bands) (Table~\ref{tab:ls}, columns~2--3, Fig.~\ref{fig:LS1}) related to periods of \mbox{5.0--6.9}~yrs for all the bands, with a mean value of about 6~yrs. With the exception of three frequencies, 2, 22 and 37~GHz, other light curves give very close periods between 5.5 and 6.4 yrs.

Although the QPO of about 6~yr appears in the periodogram, the time span covers fewer than 4 full cycles for the optical and radio bands and fewer than 3 cycles for the Fermi-LAT data. This presents a challenge for making a robust claim of a periodic signal. Additionally, the influence of red noise has been considered, as it significantly affects the periodogram's power at low frequencies. To address this, we conducted a simulation of red noise (following the procedure described in \cite{2022MNRAS.513.5238R}) and corrected the periodogram by subtracting the mean red noise power. The results indicate that the detected peaks, while present, should be interpreted with caution.

This is consistent with the periods found by \cite{2001A&A...377..396R, 2006A&A...459..731R} (quasi-periodicity with periods of about \mbox{5--6}~yrs in the optical and radio bands) and \cite{2002A&A...381....1F, 2007A&A...462..547F} (about 5.8 yrs in the optical and radio bands), but differs from the newest results with periods between 8 and 9~yrs \citep{2022MNRAS.513.5238R, 2023MNRAS.518.5788O}.

In order to detect the hidden periodicity for AO\,0235$+$164, we performed an analysis of its light curves within the ``quiet'' epoch that was not affected by the influence of powerful outbursts. The L--S periodograms reveal the presence of peaks with significance above $2.5\sigma$ for almost all spectral bands except  $\gamma$-rays and the 22 and 230~GHz radio bands(Fig.~\ref{fig:LS2}). All of  the periods lie in the range from 1.3 to 2.3 yrs (Table~\ref{tab:ls}, columns 4--5), which may indicate the presence of a single quasi-periodic process responsible for the manifestation of brightness variability throughout the studied epoch. The variation of the periods over the electromagnetic spectrum may be explained by a more significant effect of the $1/f$ red noise at lower periodogram frequencies, where the peaks concentrate.

This result is in good agreement with the conclusions of \cite{2021MNRAS.501.5997T}, who analysed the AO\,0235$+$164 radio emission variability over three decades. The authors did not find significant variability at periods of 6--8 yrs, but identified periods with a time-scale from 1.8 to 2.6 yrs.

\section{Variability at shorter time-scales}

\begin{figure}
\begin{minipage}[h]{1.0\linewidth}
\center{\includegraphics[width=1\linewidth]{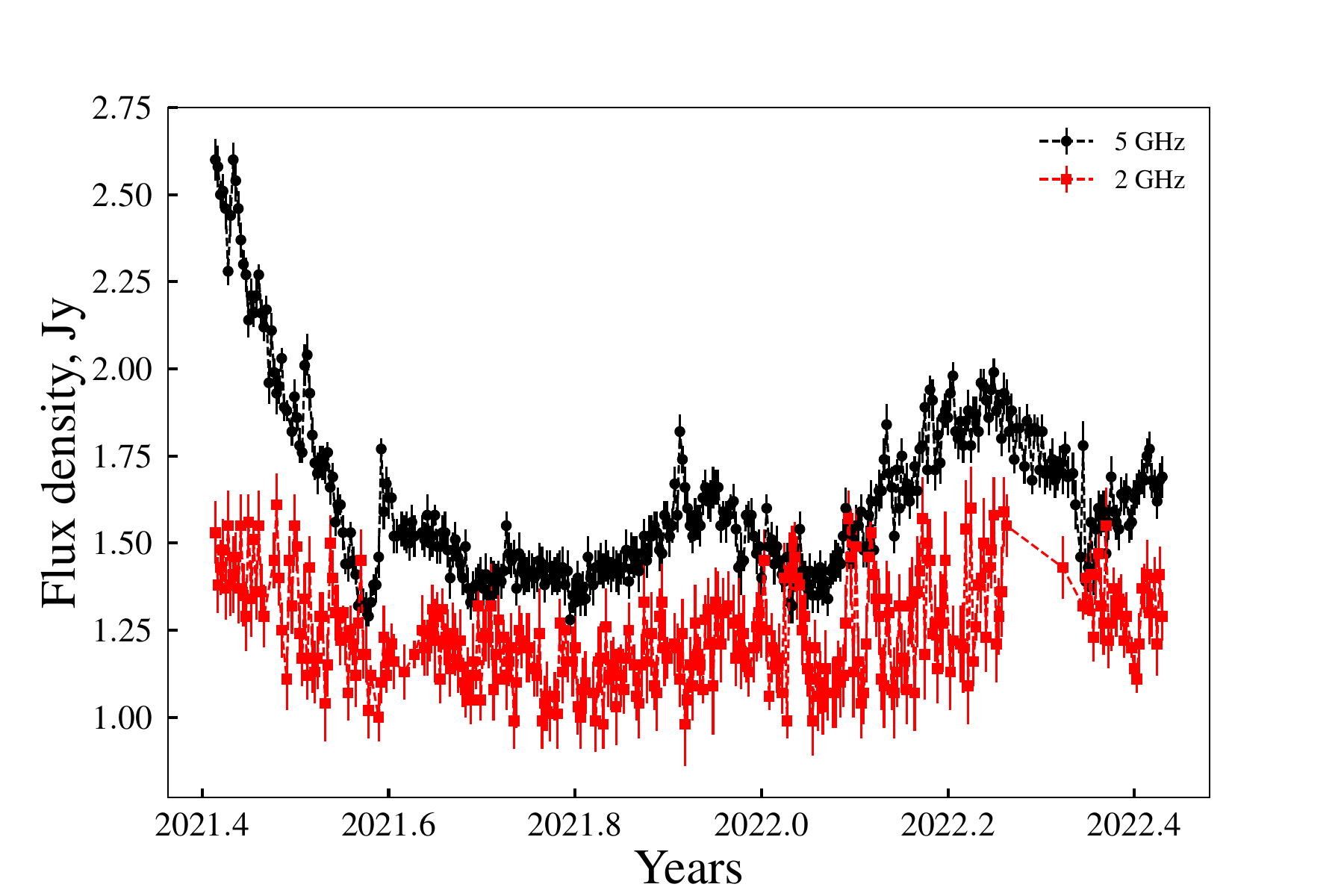}}
\end{minipage}
\hfill
\begin{minipage}[h]{1.0\linewidth}
\center{\includegraphics[width=1\linewidth]{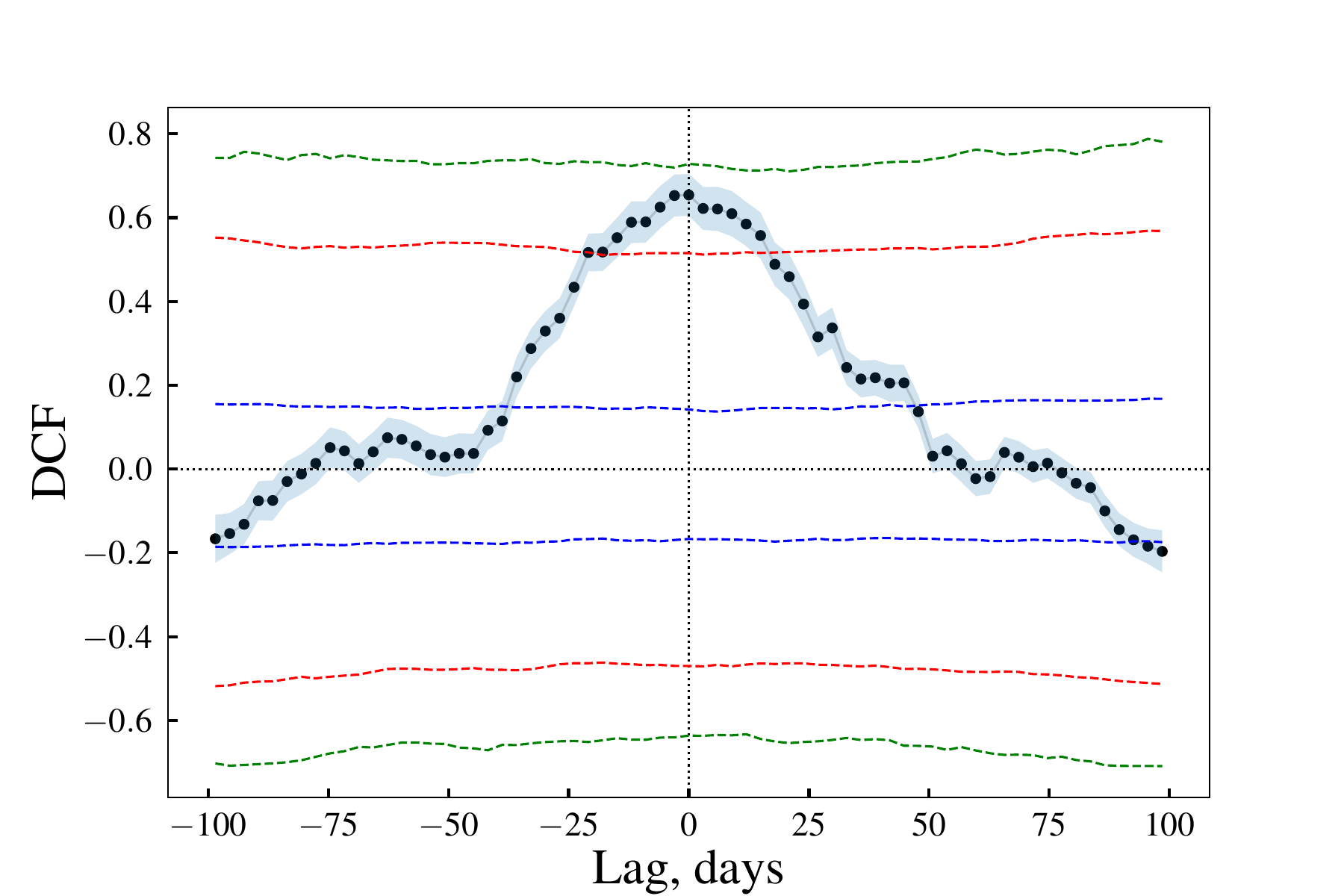}}
\end{minipage}
\caption{Top: the light curves of AO\,0235$+$164 in 2021--2022 at 5 and 2 GHz (top). Bottom: the cross-correlation function between the 5 GHz and 2 GHz light curves with a time lag resolution of 3 days for the daily measurements in 2021--2022. The significance levels of 1, 2, and 3 $\sigma$ are shown with the dashed lines (bottom).}
\label{west_lc}
\end{figure}

\begin{figure}
\begin{minipage}[h]{1.0\linewidth}
\center{\includegraphics[width=1\linewidth]{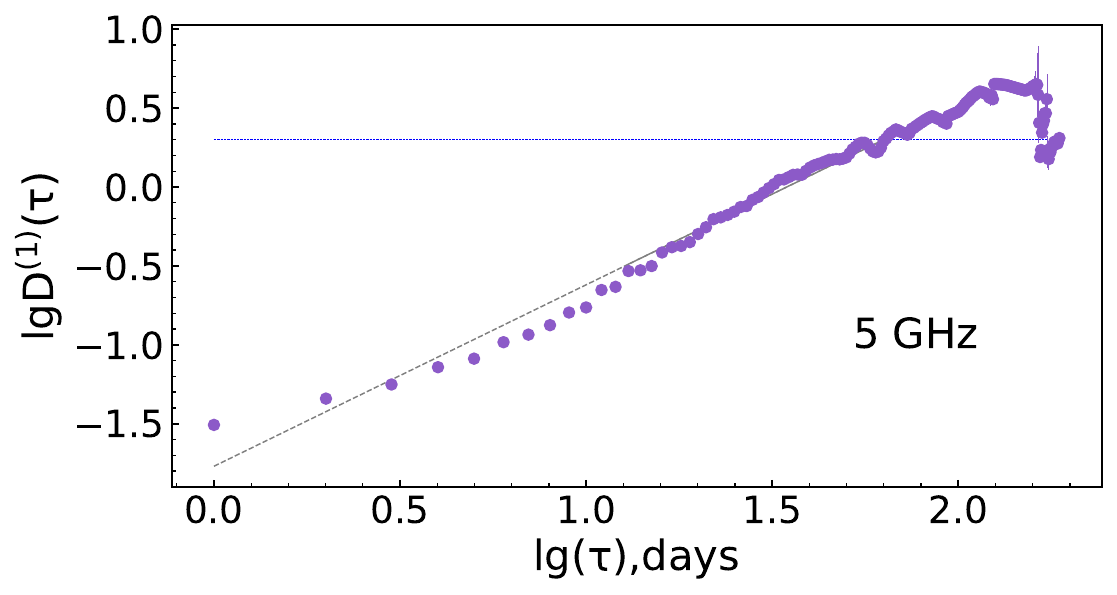}}
\end{minipage}
\hfill
\begin{minipage}[h]{1.0\linewidth}
\center{\includegraphics[width=1\linewidth]{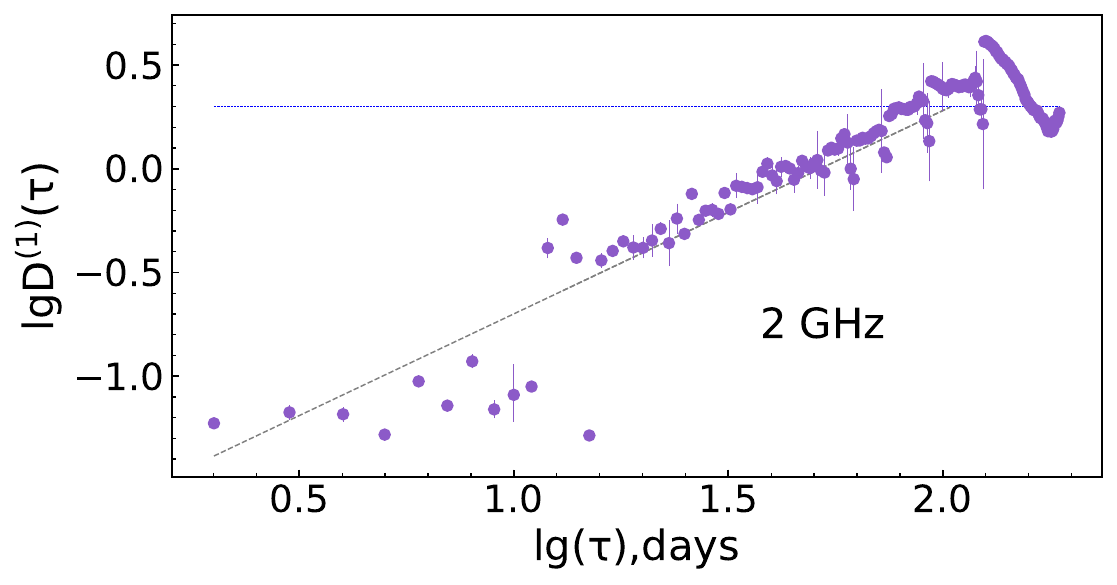}}
\end{minipage}
\caption{The SF for the total flux variations at 5 GHz (top) and 2~GHz (bottom) for daily measurements in 2021--2022. The SFs were constructed with an initial interval of 1 day, the slope between the plateaus is measured as $b=1.2$ at 5 GHz and $b=0.98$ at 2 GHz.}
\label{west-sf}
\end{figure}

Using the 2 and 5 GHz data obtained in daily observations from May 2021 to June 2022 (Fig.~\ref{west_lc}), we searched for the short-term temporal and spectral variation. The observations coincide
with the period of the prolonged activity in epoch~4 (Fig.~\ref{fig:l1}) and cover 376 days. The spectral index $\alpha_{2-5}$ correlates with the 5 GHz flux density and varies from $+0.20$ to $+0.75$, which means that the corresponding emission area is optically thick over the whole period of observations in 2021--2022. We define the spectral index $\alpha$ from the power-law $S_{\nu}\sim\nu^{\alpha}$, where $S_{\nu}$ is the flux density at a frequency $\nu$. The SF reveals a variability time-scale of the order of 100--120 days (Fig.~\ref{west-sf}) at 5 GHz, and similar value at 2 GHz (the low level). This probably indicates the presence of a shorter variability scale than it was determined for the long-term light curves in epochs 2 and 3 at these frequencies.

\begin{figure}
\includegraphics[width=0.99\columnwidth]{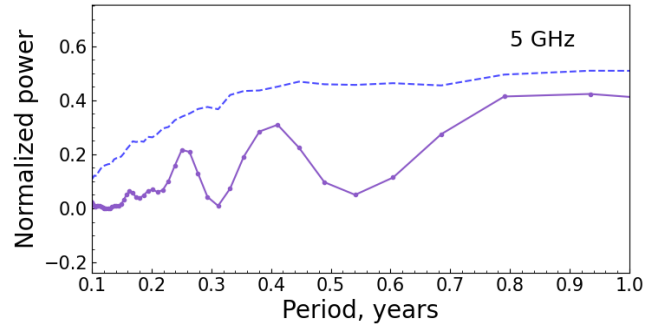}
\caption{The L--S periodogram for the total flux variations at 5~GHz. The dashed line shows the level of ${\rm FAP}=1$ per cent.}
\label{fig:lsw}
\end{figure}

The DCF analysis indicates significant correlation between
the 2 GHz and 5 GHz flux densities with a peak \mbox{${\rm DCF}=0.65\pm0.05$} (at a confidence level $>2\sigma$) and with a time lag closer to zero, ${\rm lag}\approx0\pm1.5$ days (Fig.~\ref{west_lc}). Using light curve interpolation, we obtain a similar result: ${\rm lag}\approx7\pm1$ days (${\rm DCF}=0.86\pm0.06$, $>3\sigma$).
The L--S periodogram at 5~GHz shows a few peaks at significance levels between 1.5 and $2\sigma$, which may hint at QPOs with characteristic times of about 0.25 and 0.4 yrs (Fig.~\ref{fig:lsw}).

\section{Broadband radio spectrum}

In this section we consider the properties of  the quasi-simultaneous and average broadband radio spectra of AO~0235$+$164.

\subsection{Quasi-simultaneous radio spectra}

\begin{figure}
\includegraphics[width=\columnwidth]{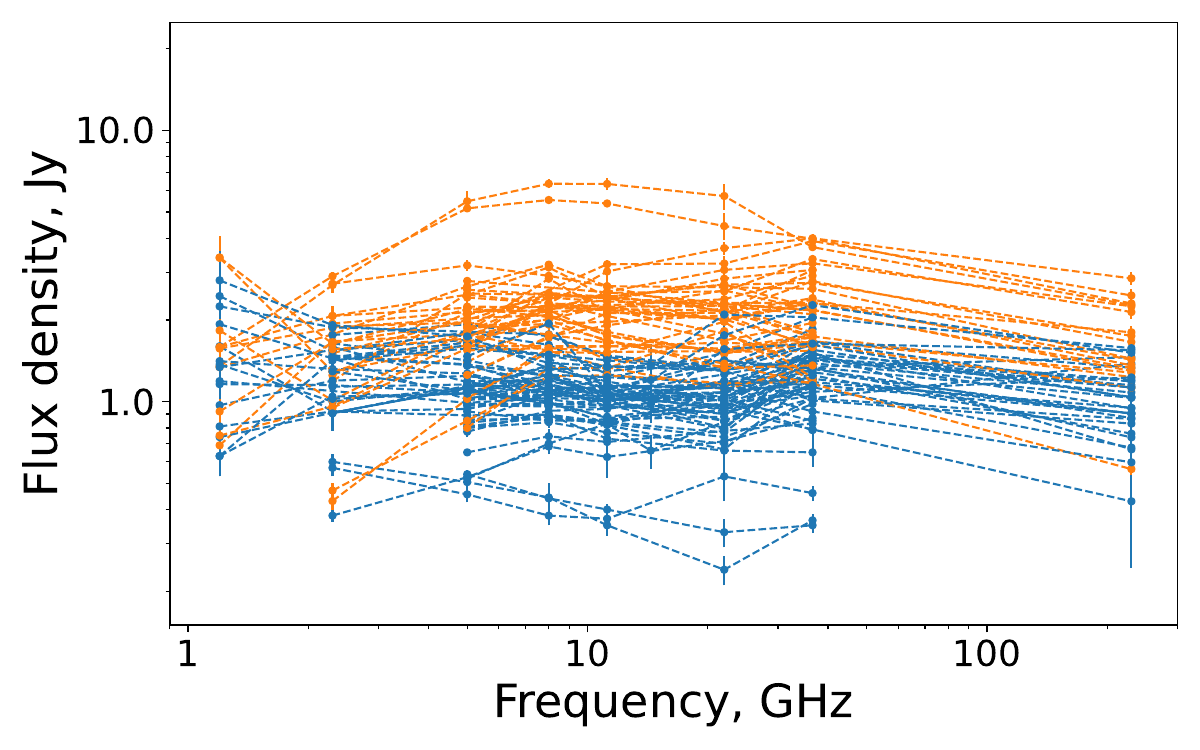}
\caption{Quasi-simultaneous spectra of the high (orange) and low (blue) activity states according to the clustering results.}
\label{fig:sp_2classes}
\end{figure}

To derive quasi-simultaneous broadband radio spectra at \mbox{2--230}~GHz (Fig.~\ref{fig:sp_2classes}), we divided the observing period into equal time intervals of 30~days and averaged flux densities within these intervals. A total of 270 spectra have been obtained in such a way, although flux densities for some frequencies may be missing due to the differing gaps in the multiband light curve (see Fig.~\ref{fig:l1}).

This sample of spectra was investigated using the method of cluster analysis (clustering) with an implementation similar to that described in \cite{Kudryavtsev_2024}. The analysis aims at identifying possible different states of non-thermal radiation properties during selected activity epochs. Cluster analysis allows one to combine investigated objects into groups based on their similarity, which is evaluated based on a number of natural or specially designed characteristics of the objects, named features. The features constitute axes of a multidimensional feature space, and the distance in this space, after proper scaling, is the metric of similarity. The obtained groups can be further investigated as different types/classes of the considered objects. Cluster analysis is a multiparametric technique that makes it possible to study a sample using the entire set of its available properties and their relationships.

To select the necessary features to form the clustering feature space, we first had to trim the quasi-simultaneous radio spectra sample to drop the dates earlier than MJD\,54700 (roughly the year 2008.6) because a number of the observed frequencies had no measurements for these earlier epochs, particularly the $\gamma$-ray measurements that are pretty abundant for later dates. Having then the flux densities at different radio frequencies and in the optical $R$ and $\gamma$-ray bands as well as radio indices $\alpha$ calculated for different radio frequency pairs, we evaluated the multicollinearity and the percentage of missing values for particular characteristics. The rule of thumb is to drop the characteristics with more than 40 per cent of missing values and to choose one of the multicollinear characteristics (correlation greater than 0.7) with the least number of missing values.

As of the multicollinearity, we made an exception for the flux densities at radio frequencies of 5, 8, 11, 22, and 37~GHz. All these characteristics have the percentage of missing values lower than 40 per cent and are strongly correlated, but the gaps in the obtained measurements may be located at different epochs. The emission at radio frequencies is also roughly of the same physical nature. We took all this into consideration and, instead of dropping all but one radio flux densities, combined them into one metafeature, thus preserving a greater number of data points for the clustering. The metafeature, let us call it radio flux, was calculated as the first principal component in the standard principal component analysis (PCA) for the 5, 8, 11, 22, and 37~GHz flux densities. In order to perform the standard PCA, we, nevertheless, had first to impute missing values in the combined features. This was made using probabilistic PCA (PPCA) \citep{DBLP:journals/neco/TippingB99}, particularly its implementation\footnote{\tt https://github.com/el-hult/pyppca} based on \cite{porta:inria-00321476}.

After performing the above mentioned steps of feature selection, we ended up with a total of five features in the clustering dataset: combined radio flux $S_{\rm radio}$, $\gamma$-ray flux $S_{\gamma}$, and radio indices $\alpha_{{11}-{37}}$, $\alpha_{{11}-{22}}$, $\alpha_{{5}-{11}}$. The number of the quasi-simultaneous spectra in the dataset after dropping those with missing feature values amounted to 187.

The following k-means clustering can be performed both in the original 5-dimensional feature space or in a space of primary components after PCA dimensionality reduction. We performed both and obtained the same results. The {\tt scikit-learn} library \citep{2011JMLR...12.2825P} was used. The optimal number of PCA primary components describing the ``true'' dimensionality of the data according to Minka's maximum-likelihood estimation \citep{PCAdim} is two, so the clustering after dimensionality reduction was performed onto a simple 2-dimensional plane.

\begin{figure}
\includegraphics[width=\columnwidth]{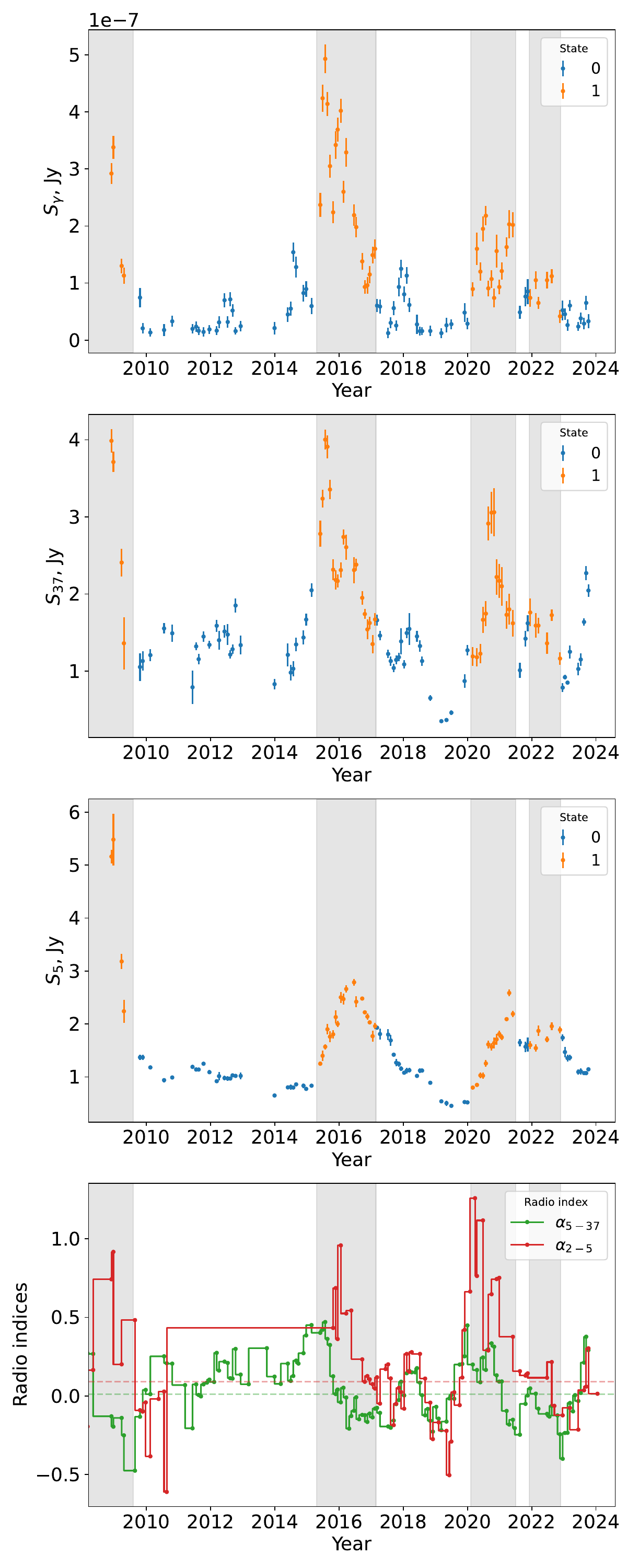}
\caption{Top three panels: light curves in the $\gamma$-ray range and at 37 and 5~GHz, respectively, with the activity states according to the clustering results; bottom: $\alpha_{2-5}$ and $\alpha_{5-37}$ radio indices for the same time intervals.
Grey areas correspond to the high activity states. The horizontal dashed lines show the median values.}
\label{fig:lc_labels}
\end{figure}

As is seen in Fig.~\ref{fig:lc_labels}, the obtained clusters can be attributed to the low and high activity states of the source. Note that this result is obtained based not only on the observed fluxes, the three radio indices mentioned above influenced the clustering in equal measure. The figure shows the light curves in the $\gamma$-ray range and radio frequencies of 37 and 5~GHz. In the bottom panel the variation the $\alpha_{2-5}$ and $\alpha_{5-37}$ radio indices are presented. During the high states, the spectrum becomes optically thick ($\alpha>0$), and remains in this condition during several years.

\begin{figure}
\includegraphics[width=\columnwidth]{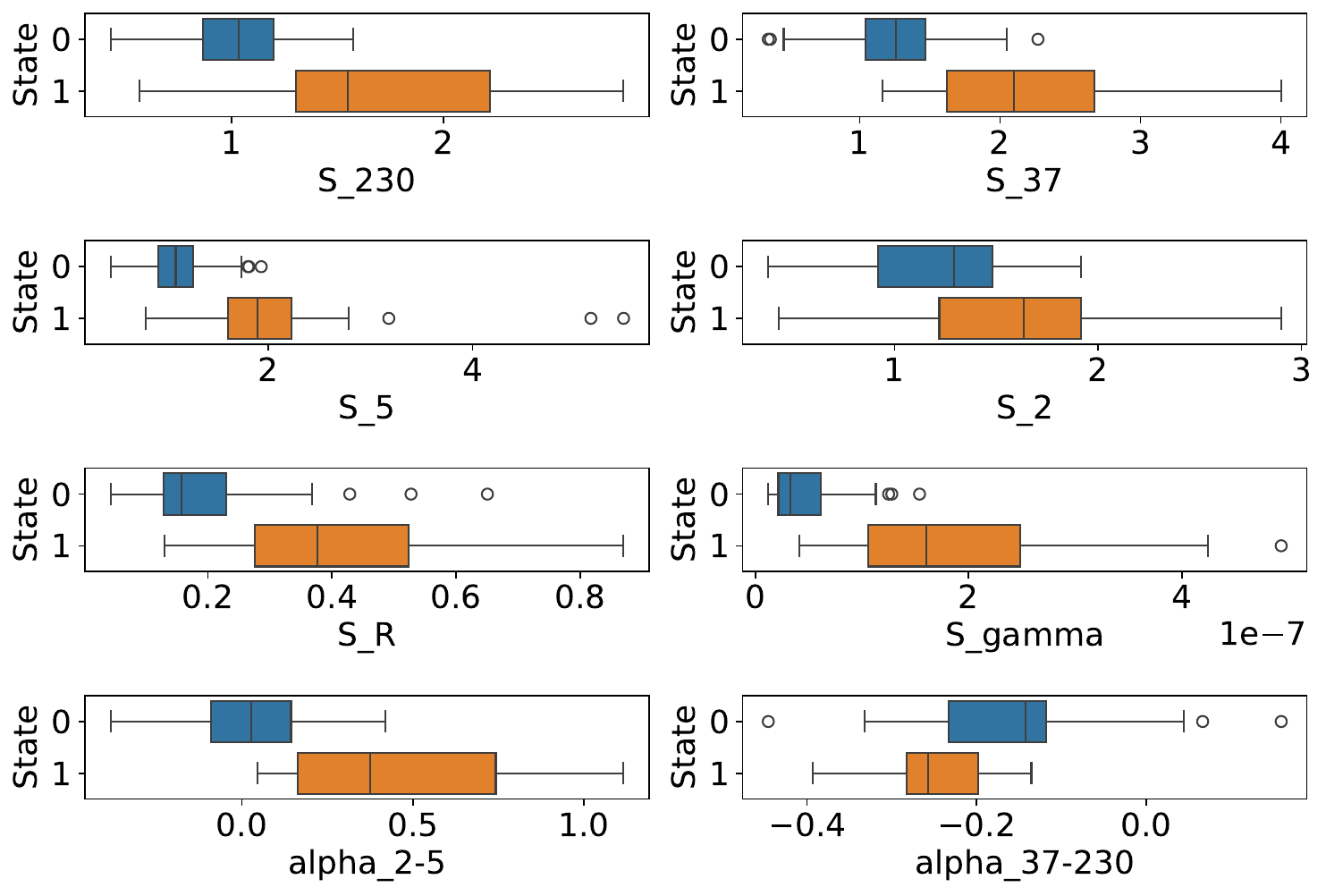}
\caption{Distributions of the parameters of quasi-simultaneous spectra in the low (0) and high (1) activity states.}
\label{fig:distribs}
\end{figure}

The shapes of the quasi-simultaneous spectra of the high and low activity states are clearly different (Fig.~\ref{fig:sp_2classes}). The distributions of flux densities and radio indices are shown in Fig.~\ref{fig:distribs} as box plots. A box is the interquartile range (IQR, 25th to 75th percentile, or Q1 to Q3), the median of a distribution is shown as a vertical line inside the box, and the ``whiskers'' extend to show the rest of the distribution, except for points that are determined as ``outliers,\!'' locating beyond the median $\pm1.5$~IQR range, these outliers are shown by dots.

\subsection{Average spectrum}

\begin{figure}
\includegraphics[width=\columnwidth]{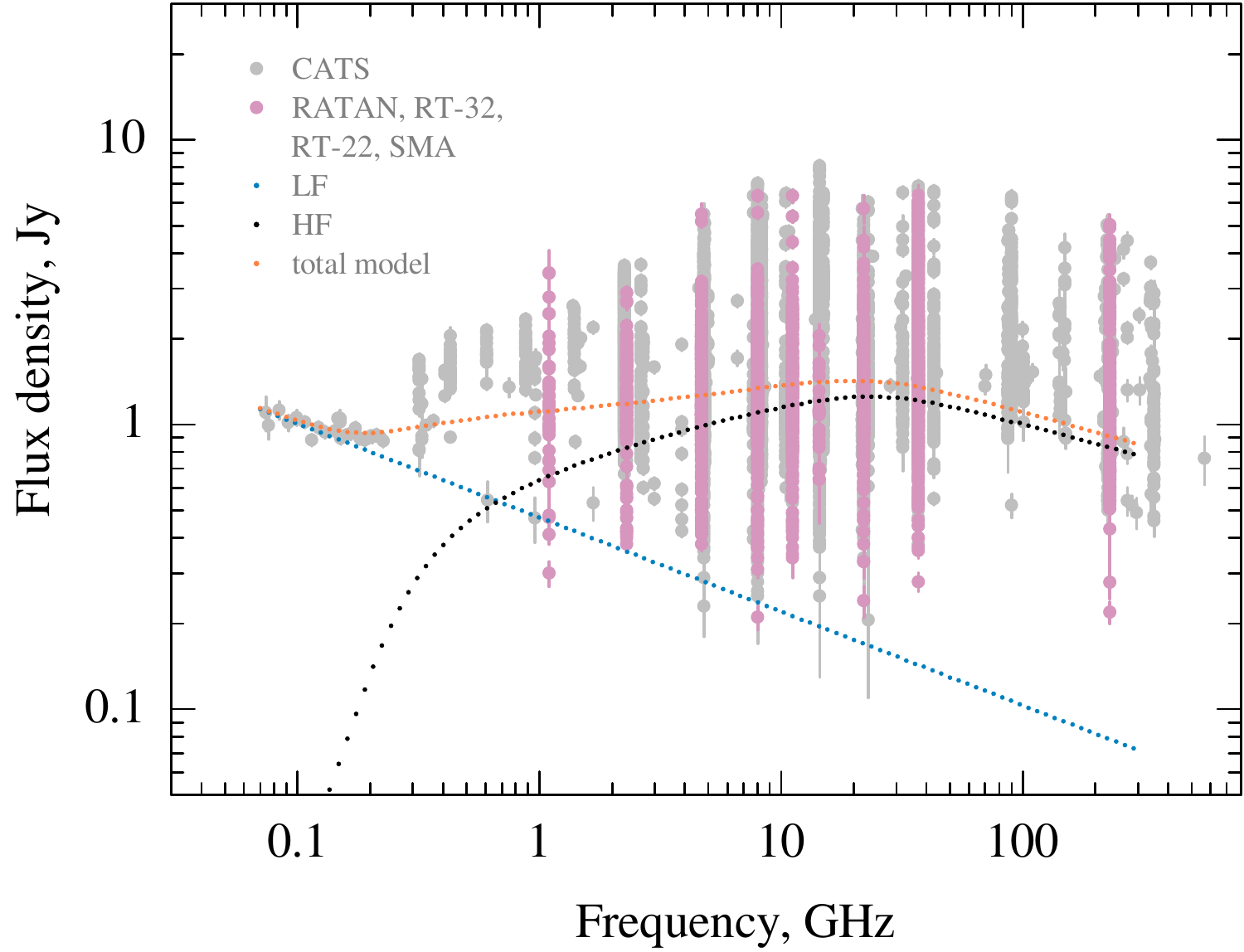}
\caption{The data from RATAN-600, RT-32, RT-22, and SMA obtained in 1997--2023 (pink points) as well as the measurements from the CATS database (grey points). The orange line is the result of fitting the model spectrum to the average total spectrum. This model spectrum is the sum of two components: the variable High Frequency Component (HFC, the black line, from the Hedgehog model) and the non-variable Low Frequency Component (LFC, the blue line, from the optically thin extended structures outer of the jet), similar to \protect\cite{2023AstBu..78..464V}. An additional component to the model may exist at frequencies greater than 80 GHz.}

\label{fig:spectra}
\end{figure}

The AO\,0235$+$164 average spectrum is constructed based on the experimental data described in Section~\ref{sec:radio_studies} and on the literature data taken from the CATS database \citep{1997BaltA...6..275V}. The spectrum has a peak at millimetre waves in the observer's frame of reference. AO\,0235$+$164 is an extremely variable AGN and its broadband radio spectrum (Fig.~\ref{fig:spectra}) show a strong variety of shapes: from steep ($\alpha<-0.5$) to rising ($\alpha>0$). During the 27 yrs of observations, the spectral index $\alpha_{11-22}$ had been less than $-0.5$ only during three epochs: Feb~2010, Oct~2011, and Jan~2023.

According to the Hedgehog jet model \citep{2000PASJ...52.1027K}, suggested by N.~S.~Kardashev in 1969, the strong longitudinal magnetic field $B$ of the radio jet depends on the distance $r$ as $B= B_0\, (r/r_0)^{-2}$. In Fig.~\ref{fig:spectra} the
result of fitting the sum of two components in this model to the averaged observed data is given. Such average model spectra can be observed if the continuous flow $dN(t)/dt$ of emitting particles across the base of the jet is constant for a long time. The variability of $dN(t)/dt$ is converted in the model to the variability of the High Frequency Component (HFC) and the total spectrum. A possible excessive emission at frequencies above 80 GHz is still hardly explained by this model and needs to be analysed in future studies.

Similar to \cite{2023AstBu..78..464V}, we obtained the following fitted physical parameters of the radio-emitting jet (the HFC): the angle to the observer's line of sight $\vartheta \sim 1.0^\circ$, the flux $S_m$ and frequency $\nu_m$ of the jet spectrum maximum $S_m \sim 0.53$ Jy at $\nu_m \sim 60$ GHz, $\gamma \sim 1.5$, and supposed $\gamma_E \sim 300$. Neglecting the redshift, the next parameters can also be estimated \citep{2022evlb.confE...9S}:
\begin{equation}
B_\perp/M_{2e} \sim 0.82\cdot 10^{-6} \cdot \nu_m \cdot \gamma_E^{-2},
\end{equation}
\begin{equation}
T_b/M_{2e} \sim 1.5\cdot10^8\cdot\gamma_E,
\end{equation}

\begin{equation}
\Theta \sim \lambda_m \left(\frac{2S_m}{\pi k_B T_b}\right)^{1/2}.
\end{equation}
Here $\nu_m$ (Hz), $\lambda_m$ (m), and $S_m$ (W m$^{-2}$ Hz$^{-1}$) are the frequency, wavelength, and flux density of the HFC maximum; the gamma-factor $\gamma_E = E/(Mc^2)$ of emitting particles is supposed to have the same value for electrons and protons; $B_\perp$ = $B \sin \vartheta$ (G), $M_{2e} = 1$ for electrons, and $M_{2e} = 1836$ for protons. $\Theta$ (rad), $k_B$, and $T_b$ (K) are the angular diameter of the emitting jet in the picture plane, the Boltzman constant, and the brightness temperature. The estimated parameters are shown in Table~\ref{tab:Estim}.

The results in Table~\ref{tab:Estim} have supported a new idea \citep{2020ApJ...894..101P,2020AdSpR..65..745K,2022evlb.confE...9S,2022muto.confE..27K} that replacing electrons with protons in the synchrotron radio emission of relativistic jets in the strong longitudinal magnetic field of an active nuclei may be an important assumption to describe both the nature of blazars and the generation of high energy neutrinos via the high energy part of the distribution of relativistic protons in blazars.

\begin{table}
\centering
\caption{\label{tab:Estim} Main physical parameters of the jet determined from fitting the HFC to the observed data in Fig.~\ref{fig:spectra}  using (6)--(8) for the two sorts of emitting particles (electrons or protons): jet magnetic field $B$, brightness temperature $T_b$, angular diameter $\Theta$, and the relation of the magnetic energy density $W_H = B^2/8\pi$ to the energy density $W_E$ of emitting particles.}
\begin{tabular}{ccccc}
\hline
Particles  & $B$ & $T_b$  & $\Theta$  & $W_H / W_E$\\
 & Gauss & K & mas     &  \\
\hline
electrons & $30$ & $5\cdot10^{10}$ & $0.07$  &
$W_H \gg W_E$\\
protons & $6\cdot10^4$ & $9\cdot10^{13}$ & $0.002$ & $W_H \gg W_E$ \\
\hline
\end{tabular}
\end{table}

\section{Discussion}
\label{sec:Discussion}

In this study we have enhanced the time coverage for the optical and radio data for the blazar AO\,0235$+$164 up to 2023, achieving a 27-yr-long time period for the MW analysis. This period includes two multiband major flares: in 2007 (marked as epoch 1 in our study) and in 2015--2016 (epoch~3), which matches the $\sim\!8.2$-yr period described by the previous authors mentioned above. The flare in 2007 was followed by another massive outburst in 2009; the flare in 2015 was also followed by an outburst in 2016, but it was less prominent. In the latter case, that second outburst even merges with the major flare at 2.3 and 5 GHz. In addition to that quasi-periodic major flares, we spot other minor flares and flux density variations. At higher frequencies the structure of the flares is more complex and the amplitude variations are more notable compared to the low-frequency radio bands.

The variability time-scales $\tau$ range from 110 to 620 days for individual epochs 1--4. In general time-scales decrease smoothly with an increase of frequency, and at each frequency $\tau$ increases from epoch 1 to epoch 4. This reflects the differences in the physical properties during each flare, such as the size of the new compact components originating from the central region and observed in 2007, 2008, 2014, and 2020. We estimate the upper limit on the physical size of the emitting region ($R$) from $\leq\!0.9$ pc (for the 100-day variability time-scale) to $\leq$5.5 pc (for 600 days), assuming
\begin{equation}
R \leq c \cdot t_{\mathrm{var}} \cdot \delta / (1+z),
\end{equation}
where $c$ is the speed of light, $t_{\mathrm{var}}$ is the variability time-scale, $\delta=21$ \citep{2018MNRAS.475.4994K} is the Doppler factor, and $z=0.94$ is the redshift.

For comparison, we calculated the brightness temperature and associated Doppler factor of each major flare at 37 GHz according to equation~(1) from \cite{1999ApJ...521..493L} using the method of flare decomposition described in Section~3.1. We adopted the Hubble constant $H_0 = 68$~km\,s$^{-1}$\,Mpc$^{-1}$ and the intrinsic brightness temperature $T_{\rm int} = 7\times10^{10}$~K \citep{2018MNRAS.475.4994K}, which is comparable to the equipartition brightness temperature $T_{\rm eq} \backsimeq 5 \times10^{10}$~K suggested by \cite{1994ApJ...426...51R}. As we note from the results (Table~\ref{tab:doppler}), our estimated Doppler factors are quite conservative: for example, \citealt{2018MNRAS.475.4994K} determined $\delta \sim 21$ for the second flare, while we estimated $\delta \sim 6$ for it. The difference arises from the adopted variability time-scale: we estimated the flare rise time-scale ($\tau=189$~days) from the exponential fits \citep{1999ApJS..120...95V}, while \citealt{2018MNRAS.475.4994K} obtained a much shorter time-scale of core variability of $\sim\!33$ days. If we consider our Doppler factors, the physical size of the emitting region at 37 GHz is about \mbox{$R \sim0.4$--$0.5$}~pc.

\begin{table}
    \centering
    \caption{Parameters of the four major flares at 37 GHz: date of the flare, maximum flux density, variability time-scale, brightness temperature, Doppler factor, and the size of the emitting region}
    \label{tab:doppler}
    \begin{tabular}{ccccc}
    \hline
        Flare & 1 & 2 & 3 & 4 \\
    \hline
        MJD & 54233 & 54761 & 57272 & 59151 \\
        $S_{\rm max}$, Jy & 4.97 & 4.66 & 3.14 & 2.78 \\
        $\tau$, days & 97 & 189 & 248 & 286 \\
        $T_{\rm b}$, K & 6$\times$10$^{13}$ & 1.5$\times$10$^{13}$ & 6$\times$10$^{12}$ & 4$\times$10$^{12}$ \\
        $\delta$ & 10 & 6 & 4 & 4 \\
        $R$, pc & 0.42 & 0.49 & 0.42 & 0.49 \\
    \hline
    \end{tabular}
\end{table}

A cross-correlation analysis between MW light curves reveals that flares at longer wavelengths follow the short-wavelength flares, with different time delays during individual episodes. In the period of two major flares in 2006--2007, a 40-day lag is found for many pairs of the emission bands considered. The maximum lags up to 450 days have been measured for the two epochs in 2013--2019 and 2019--2023, and they correspond to the delay between the extreme radio frequencies of 230 and 2 GHz.

Cross-correlation between the optical and $\gamma$-ray flux variations is not so evident: DCF maxima for epochs 2--4 are rather high, lying between 0.66 and 0.86, but their significance does not exceed 2$\sigma$. The time delay at this level is close to 0 days with an uncertainty of about 5 days, but with two major exceptions: the presence of a higher secondary peak (peak value~$= 0.86$) at 50 days for epoch 3 and a DCF peak equal to 0.74 at $-70$~days for epoch 4. This result is caused by a complex structure of both light curves, with ``sterile'' (optical without a $\gamma$-ray counterpart) flares at MJD $\sim\!57050$, 59150, and 60200
and the lack of data in the optical light curve on yearly basis.

The same pairs of light curves (radio--$\gamma$, radio--optical, radio--radio) show different time lags for various epochs, indicating a complex multiwavelength connection between them and, likely, the presence of several sources of electromagnetic radiation in AO\,0235$+$164. Based on our calculated time lags, we can evaluate the distance between the emitting regions using the following equation \citep{2010ApJ...722L...7P}:
\begin{equation}
D = \beta \cdot c \cdot t_{\mathrm{lag,rest}} / \mathrm{sin}~\theta,
\end{equation}
where $c$ is the speed of light, $t_{\mathrm{lag,rest}}$ is the time lag in
the source's frame of reference, $\beta\sim10$ is the apparent angular jet speed, and $\theta=1.7^{\circ}$ is the viewing angle \citep{2018MNRAS.475.4994K}. We obtain for a time lag of 10 days a distance $D\sim 1.4$ pc, and for the greatest time lag of 450 days a distance $D\sim 65$~pc.

The assumption for these estimates is that the synchrotron opacity in the nuclear region causes the lags, and different production zones in a jet for the radio, optical, and $\gamma$-ray emissions become observable at different distances. This theory is consistent with the findings that the flares in higher energy bands leads the lower energy flares (e.g., \citealt{2010ApJ...722L...7P,2014MNRAS.445..428M,2022MNRAS.510..469K}), hence suggesting that the former emission originates upstream with respect to the latter emission. The cases with the shortest time lags (10 days) indicate almost co-spatial (within several pc) origin for some $\gamma$-ray and high-frequency radio flares \citep{2011A&A...532A.146L}.

Epochs 3 and 4 show the typical behaviour of time lag for the $\gamma$-ray and radio band emission in quasars: the high-energy flare followed by the 230 GHz flare in about 10 days, which in turn is followed by the 37 and 22 GHz flares after 1~month, then the flare subsequently appearing at 11, 8, and 5 GHz in 6--7~months, and, finally, emerging after 10 months at 2~GHz. For example, \cite{2022MNRAS.510..469K} found the delays for the $\gamma$-ray and 15 GHz core emission of 331 AGNs to lie between 1 and 8 months with the most significant delays lying in the range of 3--5 months.

\begin{figure}
\includegraphics[width=\columnwidth]{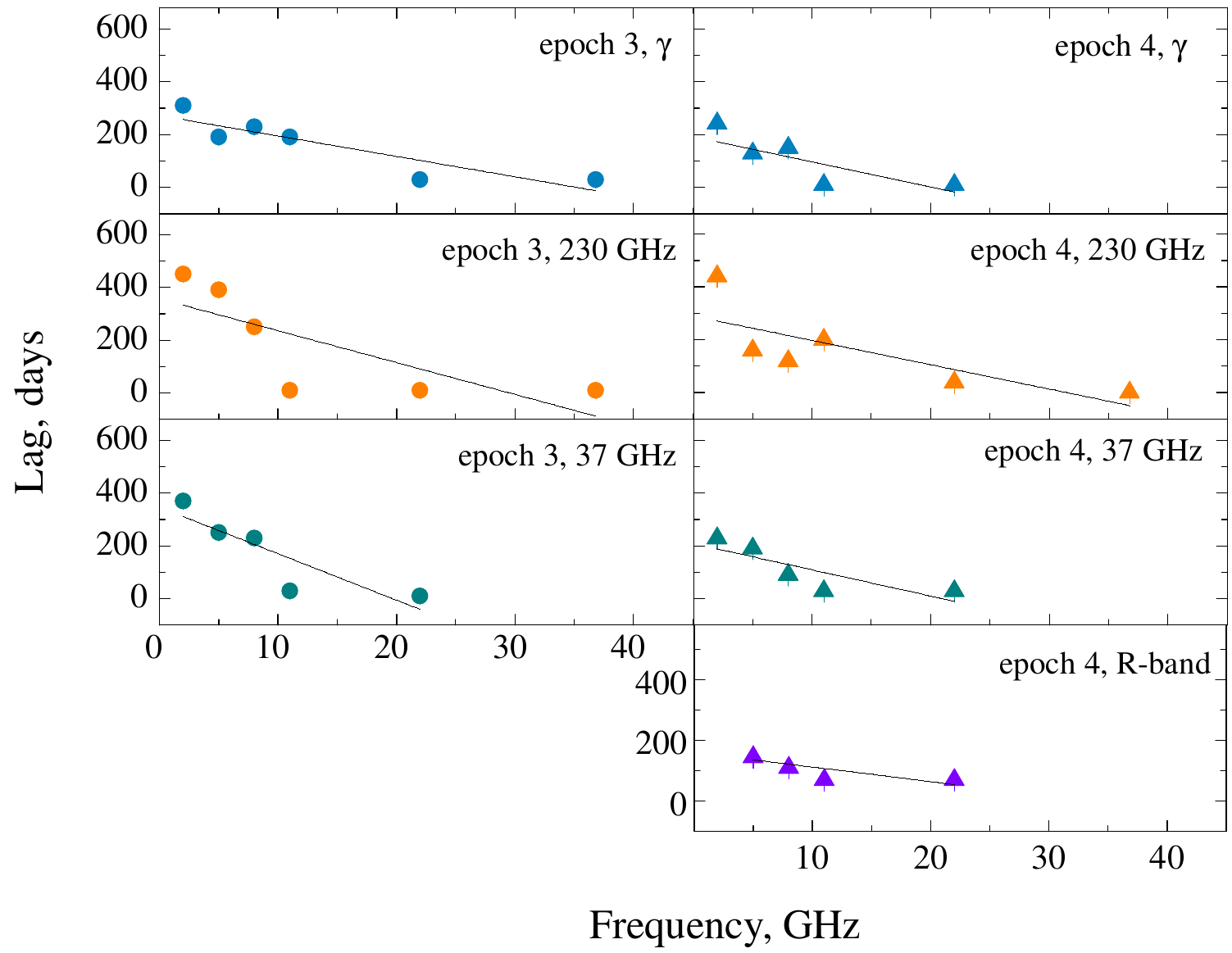}
\caption{Lag-versus-frequency relation for 37, 230 GHz, \mbox{$R$-band}, and  $\gamma$-rays, taken from the DCF analysis for epochs 3 and 4. The solid black line represents the straight-line fit to the data.}
\label{fig:lag-band}
\end{figure}

The relationships between time lag and frequency (Fig.~\ref{fig:lag-band}) for 37, 230 GHz, $R$-band, and  $\gamma$-rays are well fit for epoch 3 by the following straight lines with a negative slope:
\begin{equation}
{\rm lag}_{\rm 37GH}  =  350 (\pm65) - 20 (\pm5) \times \nu
\end{equation}
\begin{equation}
{\rm lag}_{\rm 230GH} = 360 (\pm90) - 12 (\pm5) \times \nu
\end{equation}
\begin{equation}
{\rm lag}_{\gamma} = 270 (\pm35) - 8 (\pm2) \times \nu
\end{equation}

For epoch 4:
\begin{equation}
{\rm lag}_{\rm 37GH} = 210 (\pm45) - 10 (\pm5) \times \nu
\end{equation}
\begin{equation}
{\rm lag}_{\rm 230GH} = 290 (\pm70) - 9 (\pm4) \times \nu
\end{equation}
\begin{equation}
{\rm lag}_{\gamma} = 190 (\pm50) - 10 (\pm4) \times \nu
\end{equation}
\begin{equation}
{\rm lag}_{R} = 160 (\pm20) - 5 (\pm2) \times \nu
\end{equation}

The negative slope reveals that the lower-frequency emission lags that at higher frequencies. The quantitative estimate of these slopes is about $-10$ d GHz$^{-1}$ (5 of 7 better-defined slopes) and corresponds to the rate at which the lag decreases with increasing frequency. A possible reason of this trend is that synchrotron self-absorption is greater at low frequencies than at the higher ones due to different optical opacities. For the blazar 3C\,279, \cite{2024MNRAS.527.6970K} found a slope of about $-30$ d GHz$^{-1}$ at the radio frequencies. The differences can be explained by many reasons, mainly the physical conditions in a compact relativistic jet. The redshift of the object should also be taken into account for calculation of the ``lag vs frequency'' relation in the observer's frame of reference. Thus, the measurements of time lags in MW light curves give important information about the properties of AGN relativistic jets.

We have carried out an analysis of the multifrequency \mbox{1--230}~GHz spectra obtained with the radio telescopes RATAN-600 at 1--22 GHz in 1997--2023, RT-32 at 5 and 8~GHz, and RT-22 at 37 GHz and the data taken from the CATS database. As a result, we have estimated the physical parameters for AO\,0235$+$164 based on the Hedgehog model of variable synchrotron radio emission in the strong magnetic field of the jet from an active nucleus of the blazar. A high-frequency component with a flux about 0.3--0.5 Jy may emit at frequencies above 100 GHz near the jet base in addition to the jet model used. The results have supported a new idea that replacement of electrons with protons in the synchrotron radio emission model may be important both for the investigation of the nature of AGNs and for the generation of high energy neutrinos by relativistic protons in blazars.

\subsection{Low activity state in 2009--2014}

During the extended period of low activity in November 2009~-- March 2014, a clear variability in all wavelength bands was observed with the highest flux variations of 35 and 45 per cent occurring in the optical range and $\gamma$-rays, respectively. The fractional variability for all spectral ranges is factor 2--3 less compared to epochs 1--4 with the exception of $\gamma$-rays at epoch~4. We estimate almost same variability time-scales, $\tau=150$--$180$ days, at all wavelengths (Table~\ref{tab:sf}). For several wavelength pairs, strong positive correlations are found ($\geq\!2\sigma$, Table~\ref{tab:ccf-lags}, Fig.~\ref{fig:dcf2}) with time lags of 100--200 days for $\gamma$--22 GHz and R--230 GHz, and with zero lags for the 230--37, 22--11, 11--8, and 8--5 GHz light curves. This can mean that the mechanisms that dominate the radio, optical, and $\gamma$-ray variations during the low state are not substantially different from those that are responsible for the emission during flaring states. Some indications of flux variation quasi-periodicity is detected in most of the wavelength bands in 2009--2014 (Fig.~\ref{fig:LS2}, Table~\ref{tab:ls}) with main periods at 1.4--2.1 yrs, which is much shorter than for the whole period of 1997--2023.

A similar result was obtained for the blazar Mrk\,501 during a period of its extremely low broadband activity \citep{2023ApJS..266...37A}. For this epoch the authors found significant flux variations in all wavebands with the highest values in X-rays and very-high-energy ($E>0.1$ TeV) $\gamma$-rays. A strong correlation between $\gamma$-rays and radio with a time delay of 100 days was discovered. These properties are considered as the Mrk\,501 basic emission properties.

\subsection{Periodicity search: two periods may exist}

AO\,0235$+$164 is known for its MW emission variations on various time-scales, from hours to years. Continuous systematic observations of the source since 1975 in the radio \citep{1985ApJS...59..513A} and optical \citep{1975ApJ...201..275S} bands allowed us to reveal long-term QPO patterns in its optical and radio light curves.

Based on the estimates of the $\sim\!6$~yr period of multiwave variability for AO\,0235$+$164, \cite{2001A&A...377..396R} predicted subsequent outburst for it, but it did not occur (e.g., \cite{2006A&A...459..731R, 2008A&A...480..339R}). The persistent attempts to detect the quasi-periodicity of its brightness are explained by the idea that this object is a good candidate for the binary black hole scenario \citep{2003ChJAA...3..513R, 2004A&A...419..913O}.
The binary system would allow one to explain the $\sim\!6$ yr variability time-scale by the precessing jet model, as it is shown in \cite{2024arXiv240510141E}. The absence of the predicted flare as well as the fact that the quasi-period for AO\,0235$+$164 varies significantly for different spectral ranges or analysed data temporal coverage (see, \cite{2001A&A...377..396R, 2008A&A...480..339R, 2021MNRAS.501.5997T, 2022MNRAS.513.5238R}) need to be explained.

The Lomb--Scargle method used in the search for periodicity in light curves primarily takes the fluxes at the moments of increased brightness due to their greater statistical significance. Moreover, the first $\sim\!6$-yr period estimate was done by averaging the mutual distances between the peaks of five large-amplitude outbursts in the 8 GHz radio band throughout the analysed time period (\cite{2001A&A...377..396R}).

Thus, powerful flares in all ranges of the electromagnetic spectrum, which are largely stochastic in nature (\cite{2017ApJ...849..138K}), can mask the manifestations of
the real periodicity of the brightness due to the regular properties of the active nucleus, like its binary nature or any features of nuclear structures (e.g., a disk or a torus). Obviously, the stochastic behaviour of the active nucleus itself may provoke regular processes in  surrounding media, thereby regularizing the observed phenomenon. Moreover, the results of the DCF analysis may be affected by the lack of data for long time intervals, mainly in the optical bands because of
the solar conjunctions or inappropriate weather conditions.
Another feature of optical data is the presence of fast flux variations, which may affect the results of frequency analysis (\cite{2018ApJS..236...16V}).

The reliability of our result is given by the fact that the analysis of the total LCs gives similar periods of variability from the data obtained in different ranges of the spectrum and with various instruments: 6.1 and 6.4 yrs for $\gamma$-rays and the optical $R$-band and from 5.0 to 6.9 yrs for all radio ranges, respectively. Thus, we have been able to get rid of systematic errors that affect the final result. As it is shown in Fig.~\ref{fig:LS1}, the significance of the L--S periodograms has decreased after applying the red-noise model. As a result, we can assign $6.0\pm0.5$ yrs as the most robust but debatable quasi-period estimate. This value may be related to the quasi-periodicity of the outburst events in AO\,0235$+$164.

At the same time, the 2-yr period we have suspected throughout the ``low state'' epoch can more reliably describe the intrinsic properties of the active nucleus and its environment. The detected periods, $1.9\pm0.3$ yrs, are also mutually consistent, which strengthens our argument in favor of this result. As it is noted above, taking into consideration the red-noise influence has decreased the significance of the L--S peaks (Fig.~\ref{fig:LS2}).

Further note that a systematic analysis of the $\gamma$-ray AGN data measured by Fermi-LAT resulted in hints of periodic oscillations of $\sim\!2$~yr in 7 BL Lacs and 4 FSRQs \citep{2020ApJ...896..134P}. Super-massive binary black hole systems have been suggested as a plausible origin of periodicity \citep{2017MNRAS.471.3036P}. Indeed, \citet{2007Ap&SS.309..271R} suggested AO\,0235$+$164 as a super-massive binary black hole candidate.

In conclusion, we express the idea that the general picture of the brightness variability in AO\,0235$+$164 can be represented by a composition of the well-pronounced variations with an average repetition time of about 6 yrs, probably due to the stochastic nature of the active nucleus radiation, and the latent variability with a 2-yr periodicity, which may reflect the real features of the nucleus internal structure and can be revealed reliably within its low-state periods.

\section{Summary}

We report a comprehensive analysis of the MW behaviour of AO\,0235$+$164 during its several activity epochs in 1997--2023, including four major flares and relatively low state in 2009--2014. We implemented the long-term broadband radio measurements with RATAN-600, RT-32, RT-22, and optical observations with Zeiss-1000 and AS-500/2. Additionally we used the literature data at 230 GHz (SMA) and in the $\gamma$-rays (Fermi-LAT). From our study we sum up the following conclusions.

\begin{enumerate}
\item The inspection of the light curves by cross-correlation analysis has shown
that time lag decreases with frequency in the range from 0 to 450 days. The relation between lag and frequency is described by a linear fit with a negative slope of $-10$ day GHz$^{-1}$ for the pairs of most radio frequencies versus 230 GHz, 37 GHz, R-band, and $\gamma$-rays. This behaviour is common for many blazars, and we can assume that the higher-energy emission originates closer to the central object upstream of the jet, whereas lower-energy synchrotron emission is self-absorbed and becomes visible at farther distances downstream the jet. Additionally, the time lag differs from epoch to epoch for the same pairs of bands. This indicates a complicated relationship for multi-band flares and a possible overlap of several emission sources for different flaring processes.

\item A smooth decrease of variability time-scale from radio waves to $\gamma$-rays is observed for all the epochs. Typical values of $\tau$ range from 0.3 to 1.7 yrs. During the low state, the variability time-scale $\tau=0.4$--$0.5$ yrs practically does not vary with frequency, suggesting a similar size of the non-thermal emission region for all the bands. The variability time-scale in 2021--2022, obtained using the data at 2 and 5 GHz with daily time resolution, is about 100--120 days, which is also seen for the epochs with longer time periods and caused by the appearance of a new compact component responsible for the flare in epoch 4 (2019--2024).

\item For four years of low activity in 2009--2014, we have measured clear variability throughout the electromagnetic spectrum. The highest flux density variations occurred in the $R$-band and $\gamma$-rays and reached 30 and 50 per cent, respectively. This variability might be caused by interaction between the remnants of the shocks followed from major outbursts. A correlation is revealed between the radio, optical, and $\gamma$-ray light curves for this period, which means that the mechanisms dominating the radio--$\gamma$-ray variations in the low state are not different from those during the
active phase.

\item The Lomb--Scargle periodogram reveals disputable peaks with close values, near 6 yrs, for almost all of the wavelengths. However, because our data span less than 4 full estimated periods, we cannot assert confidently the presence of the 6 yr quasi-period. For the low state in 2009--2014, a shorter period of $\sim\!2$ yrs has also been suspected. The disputable quasi-period of 6 yrs may reflects the time between the most prominent outbursts in the light curves. These flares have probably a stochastic nature, and the detected quasi-periodicity does not have tight connection with the characteristics of the active nucleus and relativistic jet. On the other side, we suppose that the periodicity found for the low state reflects general AGN properties.
\end{enumerate}

\section*{Acknowledgements}

The reported study was funded by the Ministry of Science and Higher Education of the Russian Federation under contract 075-15-2022-1227. The observations were carried out with the RATAN-600 scientific facility, Zeiss-1000 and AS-500/2 optical reflectors of SAO RAS, and RT-22 of CrAO~RAS. The observations at 5.05 and 8.63 GHz were performed with the Badary, Svetloe, and Zelenchukskaya RT-32 radiotelescopes operated by the Shared Research Facility Center for the Quasar VLBI Network of IAA RAS (\url{https://iaaras.ru/cu-center/}).

YYK was supported by the M2FINDERS project which has received funding from the European Research Council (ERC) under the European Union’s Horizon2020 Research and Innovation Programme (grant agreement No.~101018682). VAE and VLN are grateful to the staff of the Radio Astronomy Department of CrAO RAS for their participation in the observations.

SR acknowledges the support from the National Research Foundation, South Africa, through the BRICS Joint Science and Research Collaboration program with grant No.~150504.

This research has made use of the NASA/IPAC Extragalactic Database (NED), which is operated by the Jet Propulsion Laboratory, California Institute of Technology, under contract with the National Aeronautics and Space Administration; the CATS database, available on the Special Astrophysical Observatory website; the SIMBAD database, operated at CDS, Strasbourg, France. This research has made use of the VizieR catalogue access tool, CDS, Strasbourg, France.
The development of the Fermi-LAT Light Curve Repository has been funded in part through the Fermi Guest Investigator Program (NASA Research Announcements NNH19ZDA001N and NNH20ZDA001N).
The Submillimeter Array is a joint project between the Smithsonian Astrophysical Observatory and the Academia Sinica Institute of Astronomy and Astrophysics and is funded by the Smithsonian Institution and the Academia Sinica. We recognize that Maunakea is a culturally important site for the indigenous Hawaiian people; we are privileged to study the cosmos from its summit.

\section*{Data Availability}

The data underlying this article are available in the article and in its online supplementary material. The measured flux densities are distributed in the VizieR Information System.

The SMA data are available at \url{http:// sma1.sma.hawaii.edu/callist/callist.html}, the Fermi-LAT data are presented in public Light Curve Repository at \url{https://fermi.gsfc.nasa.gov/ssc/data/access/lat/LightCurveRepository/about.html}, the RATAN-600 data partly available in the BLcat on-line catalogue at \url{https://www.sao.ru/blcat/}.

\medskip\noindent
\textit{Facilities:} RATAN-600, Zeiss-1000, AS-500/2, RT-22, RT-32, \textit{Fermi} LAT, SMA.

\bibliographystyle{mnras}
\bibliography{vlasyuk}

\begin{thebibliography}{}
\makeatletter
\relax
\def\mn@urlcharsother{\let\do\@makeother \do\$\do\&\do\#\do\^\do\_\do\%\do\~}
\def\mn@doi{\begingroup\mn@urlcharsother \@ifnextchar [ {\mn@doi@} {\mn@doi@[]}}
\def\mn@doi@[#1]#2{\def\@tempa{#1}\ifx\@tempa\@empty \href {http://dx.doi.org/#2} {doi:#2}\else \href {http://dx.doi.org/#2} {#1}\fi \endgroup}
\def\mn@eprint#1#2{\mn@eprint@#1:#2::\@nil}
\def\mn@eprint@arXiv#1{\href {http://arxiv.org/abs/#1} {{\tt arXiv:#1}}}
\def\mn@eprint@dblp#1{\href {http://dblp.uni-trier.de/rec/bibtex/#1.xml} {dblp:#1}}
\def\mn@eprint@#1:#2:#3:#4\@nil{\def\@tempa {#1}\def\@tempb {#2}\def\@tempc {#3}\ifx \@tempc \@empty \let \@tempc \@tempb \let \@tempb \@tempa \fi \ifx \@tempb \@empty \def\@tempb {arXiv}\fi \@ifundefined {mn@eprint@\@tempb}{\@tempb:\@tempc}{\expandafter \expandafter \csname mn@eprint@\@tempb\endcsname \expandafter{\@tempc}}}

\bibitem[\protect\citeauthoryear{{Abdo} et~al.,}{{Abdo} et~al.}{2009}]{2009ApJS..183...46A}
{Abdo} A.~A.,  et~al., 2009, \mn@doi [\apjs] {10.1088/0067-0049/183/1/46}, \href {https://ui.adsabs.harvard.edu/abs/2009ApJS..183...46A} {183, 46}

\bibitem[\protect\citeauthoryear{{Abdo}, {Ackermann}, {Agudo}, {Ajello}, {Aller}, {Aller}, {Angelakis}  \& {Arkharov}}{{Abdo} et~al.}{2010}]{2010ApJ...716...30A}
{Abdo} A.~A.,  {Ackermann} M.,  {Agudo} I.,  {Ajello} M.,  {Aller} H.~D.,  {Aller} M.~F.,  {Angelakis} E.,   {Arkharov} A.~A.,  2010, \mn@doi [\apj] {10.1088/0004-637X/716/1/30}, \href {http://adsabs.harvard.edu/abs/2010ApJ...716...30A} {716, 30}

\bibitem[\protect\citeauthoryear{{Abdollahi} et~al.,}{{Abdollahi} et~al.}{2023}]{2023ApJS..265...31A}
{Abdollahi} S.,  et~al., 2023, \mn@doi [\apjs] {10.3847/1538-4365/acbb6a}, \href {https://ui.adsabs.harvard.edu/abs/2023ApJS..265...31A} {265, 31}

\bibitem[\protect\citeauthoryear{{Abe} et~al.,}{{Abe} et~al.}{2023}]{2023ApJS..266...37A}
{Abe} H.,  et~al., 2023, \mn@doi [\apjs] {10.3847/1538-4365/acc181}, \href {https://ui.adsabs.harvard.edu/abs/2023ApJS..266...37A} {266, 37}

\bibitem[\protect\citeauthoryear{{Ackermann} et~al.,}{{Ackermann} et~al.}{2012}]{2012ApJ...751..159A}
{Ackermann} M.,  et~al., 2012, \mn@doi [\apj] {10.1088/0004-637X/751/2/159}, \href {https://ui.adsabs.harvard.edu/abs/2012ApJ...751..159A} {751, 159}

\bibitem[\protect\citeauthoryear{{Aller}, {Aller}, {Latimer}  \& {Hodge}}{{Aller} et~al.}{1985}]{1985ApJS...59..513A}
{Aller} H.~D.,  {Aller} M.~F.,  {Latimer} G.~E.,   {Hodge} P.~E.,  1985, \mn@doi [\apjs] {10.1086/191083}, \href {http://adsabs.harvard.edu/abs/1985ApJS...59..513A} {59, 513}

\bibitem[\protect\citeauthoryear{{Ballet}, {Burnett}, {Digel}  \& {Lott}}{{Ballet} et~al.}{2020}]{2020arXiv200511208B}
{Ballet} J.,  {Burnett} T.~H.,  {Digel} S.~W.,   {Lott} B.,  2020, arXiv e-prints, \href {https://ui.adsabs.harvard.edu/abs/2020arXiv200511208B} {p. arXiv:2005.11208}

\bibitem[\protect\citeauthoryear{{Cheong} et~al.,}{{Cheong} et~al.}{2024}]{2024MNRAS.527..882C}
{Cheong} W.~Y.,  et~al., 2024, \mn@doi [\mnras] {10.1093/mnras/stad3250}, \href {https://ui.adsabs.harvard.edu/abs/2024MNRAS.527..882C} {527, 882}

\bibitem[\protect\citeauthoryear{{Cohen}, {Smith}, {Junkkarinen}  \& {Burbidge}}{{Cohen} et~al.}{1987}]{1987ApJ...318..577C}
{Cohen} R.~D.,  {Smith} H.~E.,  {Junkkarinen} V.~T.,   {Burbidge} E.~M.,  1987, \mn@doi [\apj] {10.1086/165393}, \href {https://ui.adsabs.harvard.edu/abs/1987ApJ...318..577C} {318, 577}

\bibitem[\protect\citeauthoryear{{D'Ammando} et~al.,}{{D'Ammando} et~al.}{2019}]{2019MNRAS.490.5300D}
{D'Ammando} F.,  et~al., 2019, \mn@doi [\mnras] {10.1093/mnras/stz2792}, \href {https://ui.adsabs.harvard.edu/abs/2019MNRAS.490.5300D} {490, 5300}

\bibitem[\protect\citeauthoryear{{Edelson} \& {Krolik}}{{Edelson} \& {Krolik}}{1988}]{1988ApJ...333..646E}
{Edelson} R.~A.,  {Krolik} J.~H.,  1988, \mn@doi [\apj] {10.1086/166773}, \href {https://ui.adsabs.harvard.edu/abs/1988ApJ...333..646E/abstract} {333, 646}

\bibitem[\protect\citeauthoryear{{Emmanoulopoulos}, {McHardy}  \& {Uttley}}{{Emmanoulopoulos} et~al.}{2010}]{2010MNRAS.404..931E}
{Emmanoulopoulos} D.,  {McHardy} I.~M.,   {Uttley} P.,  2010, \mn@doi [\mnras] {10.1111/j.1365-2966.2010.16328.x}, \href {https://ui.adsabs.harvard.edu/abs/2010MNRAS.404..931E} {404, 931}

\bibitem[\protect\citeauthoryear{{Emmanoulopoulos}, {McHardy}  \& {Papadakis}}{{Emmanoulopoulos} et~al.}{2013}]{2013MNRAS.433..907E}
{Emmanoulopoulos} D.,  {McHardy} I.~M.,   {Papadakis} I.~E.,  2013, \mn@doi [\mnras] {10.1093/mnras/stt764}, \href {https://ui.adsabs.harvard.edu/abs/2013MNRAS.433..907E} {433, 907}

\bibitem[\protect\citeauthoryear{{Escudero Pedrosa} et~al.,}{{Escudero Pedrosa} et~al.}{2024}]{2024arXiv240510141E}
{Escudero Pedrosa} J.,  et~al., 2024, \mn@doi [arXiv e-prints] {10.48550/arXiv.2405.10141}, \href {https://ui.adsabs.harvard.edu/abs/2024arXiv240510141E} {p. arXiv:2405.10141}

\bibitem[\protect\citeauthoryear{{Fan}, {Lin}, {Xie}, {Zhang}, {Mei}, {Su}  \& {Peng}}{{Fan} et~al.}{2002}]{2002A&A...381....1F}
{Fan} J.~H.,  {Lin} R.~G.,  {Xie} G.~Z.,  {Zhang} L.,  {Mei} D.~C.,  {Su} C.~Y.,   {Peng} Z.~M.,  2002, \mn@doi [\aap] {10.1051/0004-6361:20011356}, \href {https://ui.adsabs.harvard.edu/abs/2002A&A...381....1F} {381, 1}

\bibitem[\protect\citeauthoryear{{Fan} et~al.,}{{Fan} et~al.}{2007}]{2007A&A...462..547F}
{Fan} J.~H.,  et~al., 2007, \mn@doi [\aap] {10.1051/0004-6361:20054775}, \href {https://ui.adsabs.harvard.edu/abs/2007A&A...462..547F} {462, 547}

\bibitem[\protect\citeauthoryear{{Fan} et~al.,}{{Fan} et~al.}{2017}]{2017ApJ...837...45F}
{Fan} J.~H.,  et~al., 2017, \mn@doi [\apj] {10.3847/1538-4357/aa5def}, \href {https://ui.adsabs.harvard.edu/abs/2017ApJ...837...45F} {837, 45}

\bibitem[\protect\citeauthoryear{{Gaur} et~al.,}{{Gaur} et~al.}{2015}]{2015A&A...582A.103G}
{Gaur} H.,  et~al., 2015, \mn@doi [\aap] {10.1051/0004-6361/201526536}, \href {https://ui.adsabs.harvard.edu/abs/2015A&A...582A.103G} {582, A103}

\bibitem[\protect\citeauthoryear{{Gonz{\'a}lez-P{\'e}rez}, {Kidger}  \& {Mart{\'\i}n-Luis}}{{Gonz{\'a}lez-P{\'e}rez} et~al.}{2001}]{2001AJ....122.2055G}
{Gonz{\'a}lez-P{\'e}rez} J.~N.,  {Kidger} M.~R.,   {Mart{\'\i}n-Luis} F.,  2001, \mn@doi [\aj] {10.1086/322129}, \href {https://ui.adsabs.harvard.edu/abs/2001AJ....122.2055G} {122, 2055}

\bibitem[\protect\citeauthoryear{{Gopal-Krishna}, {Sagar}  \& {Wiita}}{{Gopal-Krishna} et~al.}{1993}]{1993MNRAS.262..963G}
{Gopal-Krishna} {Sagar} R.,   {Wiita} P.~J.,  1993, \mn@doi [\mnras] {10.1093/mnras/262.4.963}, \href {https://ui.adsabs.harvard.edu/abs/1993MNRAS.262..963G} {262, 963}

\bibitem[\protect\citeauthoryear{{Gupta}, {Banerjee}, {Ashok}  \& {Joshi}}{{Gupta} et~al.}{2004}]{2004A&A...422..505G}
{Gupta} A.~C.,  {Banerjee} D.~P.~K.,  {Ashok} N.~M.,   {Joshi} U.~C.,  2004, \mn@doi [\aap] {10.1051/0004-6361:20040306}, \href {https://ui.adsabs.harvard.edu/abs/2004A&A...422..505G} {422, 505}

\bibitem[\protect\citeauthoryear{{Gupta} et~al.,}{{Gupta} et~al.}{2017}]{2017MNRAS.472..788G}
{Gupta} A.~C.,  et~al., 2017, \mn@doi [\mnras] {10.1093/mnras/stx2072}, \href {https://ui.adsabs.harvard.edu/abs/2017MNRAS.472..788G} {472, 788}

\bibitem[\protect\citeauthoryear{{Gurwell}, {Peck}, {Hostler}, {Darrah}  \& {Katz}}{{Gurwell} et~al.}{2007}]{2007ASPC..375..234G}
{Gurwell} M.~A.,  {Peck} A.~B.,  {Hostler} S.~R.,  {Darrah} M.~R.,   {Katz} C.~A.,  2007, in {Baker} A.~J.,  {Glenn} J.,  {Harris} A.~I.,  {Mangum} J.~G.,   {Yun} M.~S.,  eds,  Astronomical Society of the Pacific Conference Series Vol. 375, From Z-Machines to ALMA: (Sub)Millimeter Spectroscopy of Galaxies. p.~234

\bibitem[\protect\citeauthoryear{{Hagen-Thorn}, {Larionov}, {Jorstad}, {Arkharov}, {Hagen-Thorn}, {Efimova}, {Larionova}  \& {Marscher}}{{Hagen-Thorn} et~al.}{2008}]{2008ApJ...672...40H}
{Hagen-Thorn} V.~A.,  {Larionov} V.~M.,  {Jorstad} S.~G.,  {Arkharov} A.~A.,  {Hagen-Thorn} E.~I.,  {Efimova} N.~V.,  {Larionova} L.~V.,   {Marscher} A.~P.,  2008, \mn@doi [\apj] {10.1086/523841}, \href {https://ui.adsabs.harvard.edu/abs/2008ApJ...672...40H} {672, 40}

\bibitem[\protect\citeauthoryear{{Hagen-Thorn}, {Larionov}, {Morozova}, {Arkharov}, {Hagen-Thorn}, {Shablovinskaya}, {Prokop'eva}  \& {Yakovleva}}{{Hagen-Thorn} et~al.}{2018}]{2018ARep...62..103H}
{Hagen-Thorn} V.~A.,  {Larionov} V.~M.,  {Morozova} D.~A.,  {Arkharov} A.~A.,  {Hagen-Thorn} E.~I.,  {Shablovinskaya} E.~S.,  {Prokop'eva} M.~S.,   {Yakovleva} V.~A.,  2018, \mn@doi [Astronomy Reports] {10.1134/S106377291802004X}, \href {https://ui.adsabs.harvard.edu/abs/2018ARep...62..103H} {62, 103}

\bibitem[\protect\citeauthoryear{{Heidt} \& {Wagner}}{{Heidt} \& {Wagner}}{1996}]{1996A&A...305...42H}
{Heidt} J.,  {Wagner} S.~J.,  1996, \mn@doi [\aap] {10.48550/arXiv.astro-ph/9506032}, \href {https://ui.adsabs.harvard.edu/abs/1996A&A...305...42H} {305, 42}

\bibitem[\protect\citeauthoryear{{Hovatta}, {Valtaoja}, {Tornikoski}  \& {L{\"a}hteenm{\"a}ki}}{{Hovatta} et~al.}{2009}]{2009A&A...494..527H}
{Hovatta} T.,  {Valtaoja} E.,  {Tornikoski} M.,   {L{\"a}hteenm{\"a}ki} A.,  2009, \mn@doi [\aap] {10.1051/0004-6361:200811150}, \href {https://ui.adsabs.harvard.edu/abs/2009A&A...494..527H} {494, 527}

\bibitem[\protect\citeauthoryear{{Hufnagel} \& {Bregman}}{{Hufnagel} \& {Bregman}}{1992}]{1992ApJ...386..473H}
{Hufnagel} B.~R.,  {Bregman} J.~N.,  1992, \mn@doi [\apj] {10.1086/171033}, \href {https://ui.adsabs.harvard.edu/abs/1992ApJ...386..473H} {386, 473}

\bibitem[\protect\citeauthoryear{{Hughes}, {Aller}  \& {Aller}}{{Hughes} et~al.}{1992}]{1992ApJ...396..469H}
{Hughes} P.~A.,  {Aller} H.~D.,   {Aller} M.~F.,  1992, \mn@doi [\apj] {10.1086/171734}, \href {https://ui.adsabs.harvard.edu/abs/1992ApJ...396..469H} {396, 469}

\bibitem[\protect\citeauthoryear{{Ivezi{\'c}}, {Connolly}, {VanderPlas}  \& {Gray}}{{Ivezi{\'c}} et~al.}{2014}]{2014sdmm.book.....I}
{Ivezi{\'c}} {\v{Z}}.,  {Connolly} A.~J.,  {VanderPlas} J.~T.,   {Gray} A.,  2014, {Statistics, Data Mining, and Machine Learning in Astronomy: A Practical Python Guide for the Analysis of Survey Data}, \mn@doi{10.1515/9781400848911.
}

\bibitem[\protect\citeauthoryear{{Jorstad}, {Marscher}, {Mattox}, {Wehrle}, {Bloom}  \& {Yurchenko}}{{Jorstad} et~al.}{2001}]{2001ApJS..134..181J}
{Jorstad} S.~G.,  {Marscher} A.~P.,  {Mattox} J.~R.,  {Wehrle} A.~E.,  {Bloom} S.~D.,   {Yurchenko} A.~V.,  2001, \mn@doi [\apjs] {10.1086/320858}, \href {https://ui.adsabs.harvard.edu/abs/2001ApJS..134..181J} {134, 181}

\bibitem[\protect\citeauthoryear{{Jorstad} et~al.,}{{Jorstad} et~al.}{2017}]{2017ApJ...846...98J}
{Jorstad} S.~G.,  et~al., 2017, \mn@doi [\apj] {10.3847/1538-4357/aa8407}, \href {https://ui.adsabs.harvard.edu/abs/2017ApJ...846...98J} {846, 98}

\bibitem[\protect\citeauthoryear{{Kharinov}, {Konnikova}, {Ipatov}, {Ipatova}  \& {Erkenov}}{{Kharinov} et~al.}{2020}]{2020ARep...64..350K}
{Kharinov} M.~A.,  {Konnikova} V.~K.,  {Ipatov} A.~V.,  {Ipatova} I.~A.,   {Erkenov} A.~K.,  2020, \mn@doi [Astronomy Reports] {10.1134/S1063772920050029}, \href {https://ui.adsabs.harvard.edu/abs/2020ARep...64..350K} {64, 350}

\bibitem[\protect\citeauthoryear{{Kovalev}, {Nizhelsky}, {Kovalev}, {Berlin}, {Zhekanis}, {Mingaliev}  \& {Bogdantsov}}{{Kovalev} et~al.}{1999}]{1999A&AS..139..545K}
{Kovalev} Y.~Y.,  {Nizhelsky} N.~A.,  {Kovalev} Y.~A.,  {Berlin} A.~B.,  {Zhekanis} G.~V.,  {Mingaliev} M.~G.,   {Bogdantsov} A.~V.,  1999, \mn@doi [\aaps] {10.1051/aas:1999406}, \href {https://ui.adsabs.harvard.edu/abs/1999A&AS..139..545K} {139, 545}

\bibitem[\protect\citeauthoryear{{Kovalev}, {Kovalev}  \& {Nizhelsky}}{{Kovalev} et~al.}{2000}]{2000PASJ...52.1027K}
{Kovalev} Y.~A.,  {Kovalev} Y.~Y.,   {Nizhelsky} N.~A.,  2000, \mn@doi [\pasj] {10.1093/pasj/52.6.1027}, \href {https://ui.adsabs.harvard.edu/abs/2000PASJ...52.1027K} {52, 1027}

\bibitem[\protect\citeauthoryear{{Kovalev} et~al.,}{{Kovalev} et~al.}{2020}]{2020AdSpR..65..745K}
{Kovalev} Y.~A.,  et~al., 2020, \mn@doi [Advances in Space Research] {10.1016/j.asr.2019.04.034}, \href {https://ui.adsabs.harvard.edu/abs/2020AdSpR..65..745K} {65, 745}

\bibitem[\protect\citeauthoryear{{Kovalev} et~al.,}{{Kovalev} et~al.}{2022}]{2022muto.confE..27K}
{Kovalev} Y.~A.,  et~al., 2022, in The Multifaceted Universe: Theory and Observations - 2000. p.~27

\bibitem[\protect\citeauthoryear{{Kramarenko}, {Pushkarev}, {Kovalev}, {Lister}, {Hovatta}  \& {Savolainen}}{{Kramarenko} et~al.}{2022}]{2022MNRAS.510..469K}
{Kramarenko} I.~G.,  {Pushkarev} A.~B.,  {Kovalev} Y.~Y.,  {Lister} M.~L.,  {Hovatta} T.,   {Savolainen} T.,  2022, \mn@doi [\mnras] {10.1093/mnras/stab3358}, \href {https://ui.adsabs.harvard.edu/abs/2022MNRAS.510..469K} {510, 469}

\bibitem[\protect\citeauthoryear{{Krishna Mohana} et~al.,}{{Krishna Mohana} et~al.}{2024}]{2024MNRAS.527.6970K}
{Krishna Mohana} A.,  et~al., 2024, \mn@doi [\mnras] {10.1093/mnras/stad3583}, \href {https://ui.adsabs.harvard.edu/abs/2024MNRAS.527.6970K} {527, 6970}

\bibitem[\protect\citeauthoryear{{Kudryashova}, {Bursov}  \& {Trushkin}}{{Kudryashova} et~al.}{2024}]{2024AstBu..79...36K}
{Kudryashova} A.~A.,  {Bursov} N.~N.,   {Trushkin} S.~A.,  2024, \mn@doi [Astrophysical Bulletin] {10.1134/S1990341324700263}, \href {https://ui.adsabs.harvard.edu/abs/2024AstBu..79...36K} {79, 36}

\bibitem[\protect\citeauthoryear{Kudryavtsev, Sotnikova, Stolyarov, Mufakharov, Vlasyuk, Khabibullina, Mikhailov  \& Cherepkova}{Kudryavtsev et~al.}{2024}]{Kudryavtsev_2024}
Kudryavtsev D.~O.,  Sotnikova Y.~V.,  Stolyarov V.~A.,  Mufakharov T.~V.,  Vlasyuk V.~V.,  Khabibullina M.~L.,  Mikhailov A.~G.,   Cherepkova Y.~V.,  2024, \mn@doi [Research in Astronomy and Astrophysics] {10.1088/1674-4527/ad3d14}, 24, 055011

\bibitem[\protect\citeauthoryear{{Kushwaha}, {Sinha}, {Misra}, {Singh}  \& {de Gouveia Dal Pino}}{{Kushwaha} et~al.}{2017}]{2017ApJ...849..138K}
{Kushwaha} P.,  {Sinha} A.,  {Misra} R.,  {Singh} K.~P.,   {de Gouveia Dal Pino} E.~M.,  2017, \mn@doi [\apj] {10.3847/1538-4357/aa8ef5}, \href {https://ui.adsabs.harvard.edu/abs/2017ApJ...849..138K} {849, 138}

\bibitem[\protect\citeauthoryear{{Kutkin} et~al.,}{{Kutkin} et~al.}{2018}]{2018MNRAS.475.4994K}
{Kutkin} A.~M.,  et~al., 2018, \mn@doi [\mnras] {10.1093/mnras/sty144}, \href {https://ui.adsabs.harvard.edu/abs/2018MNRAS.475.4994K} {475, 4994}

\bibitem[\protect\citeauthoryear{{L{\"a}hteenm{\"a}ki} \& {Valtaoja}}{{L{\"a}hteenm{\"a}ki} \& {Valtaoja}}{1999}]{1999ApJ...521..493L}
{L{\"a}hteenm{\"a}ki} A.,  {Valtaoja} E.,  1999, \mn@doi [\apj] {10.1086/307587}, \href {https://ui.adsabs.harvard.edu/abs/1999ApJ...521..493L} {521, 493}

\bibitem[\protect\citeauthoryear{{Larionov} et~al.,}{{Larionov} et~al.}{2020}]{2020MNRAS.492.3829L}
{Larionov} V.~M.,  et~al., 2020, \mn@doi [\mnras] {10.1093/mnras/staa082}, \href {https://ui.adsabs.harvard.edu/abs/2020MNRAS.492.3829L} {492, 3829}

\bibitem[\protect\citeauthoryear{{Le{\'o}n-Tavares}, {Valtaoja}, {Tornikoski}, {L{\"a}hteenm{\"a}ki}  \& {Nieppola}}{{Le{\'o}n-Tavares} et~al.}{2011}]{2011A&A...532A.146L}
{Le{\'o}n-Tavares} J.,  {Valtaoja} E.,  {Tornikoski} M.,  {L{\"a}hteenm{\"a}ki} A.,   {Nieppola} E.,  2011, \mn@doi [\aap] {10.1051/0004-6361/201116664}, \href {https://ui.adsabs.harvard.edu/abs/2011A&A...532A.146L} {532, A146}

\bibitem[\protect\citeauthoryear{{Lomb}}{{Lomb}}{1976}]{1976Ap&SS..39..447L}
{Lomb} N.~R.,  1976, \mn@doi [\apss] {10.1007/BF00648343}, \href {https://ui.adsabs.harvard.edu/abs/1976Ap&SS..39..447L} {39, 447}

\bibitem[\protect\citeauthoryear{{Majorova}, {Bursov}  \& {Trushkin}}{{Majorova} et~al.}{2023}]{2023AstBu..78..429M}
{Majorova} E.~K.,  {Bursov} N.~N.,   {Trushkin} S.~A.,  2023, \mn@doi [Astrophysical Bulletin] {10.1134/S1990341323700141}, \href {https://ui.adsabs.harvard.edu/abs/2023AstBu..78..429M} {78, 429}

\bibitem[\protect\citeauthoryear{{Marcha}, {Browne}, {Impey}  \& {Smith}}{{Marcha} et~al.}{1996}]{1996MNRAS.281..425M}
{Marcha} M.~J.~M.,  {Browne} I.~W.~A.,  {Impey} C.~D.,   {Smith} P.~S.,  1996, \mn@doi [\mnras] {10.1093/mnras/281.2.425}, \href {http://adsabs.harvard.edu/abs/1996MNRAS.281..425M} {281, 425}

\bibitem[\protect\citeauthoryear{{Max-Moerbeck} et~al.,}{{Max-Moerbeck} et~al.}{2014}]{2014MNRAS.445..428M}
{Max-Moerbeck} W.,  et~al., 2014, \mn@doi [\mnras] {10.1093/mnras/stu1749}, \href {https://ui.adsabs.harvard.edu/abs/2014MNRAS.445..428M} {445, 428}

\bibitem[\protect\citeauthoryear{{Mead}, {Ballard}, {Brand}, {Hough}, {Brindle}  \& {Bailey}}{{Mead} et~al.}{1990}]{1990A&AS...83..183M}
{Mead} A.~R.~G.,  {Ballard} K.~R.,  {Brand} P.~W.~J.~L.,  {Hough} J.~H.,  {Brindle} C.,   {Bailey} J.~A.,  1990, \aaps, \href {https://ui.adsabs.harvard.edu/abs/1990A&AS...83..183M} {83, 183}

\bibitem[\protect\citeauthoryear{{Miller}, {Carini}  \& {Goodrich}}{{Miller} et~al.}{1989}]{1989Natur.337..627M}
{Miller} H.~R.,  {Carini} M.~T.,   {Goodrich} B.~D.,  1989, \mn@doi [\nat] {10.1038/337627a0}, \href {https://ui.adsabs.harvard.edu/abs/1989Natur.337..627M} {337, 627}

\bibitem[\protect\citeauthoryear{Minka}{Minka}{2000}]{PCAdim}
Minka T.~P.,  2000, M.I.T Media Laboratory Perceptual Computing Section Technical Report No. 514

\bibitem[\protect\citeauthoryear{{Ostorero}, {Villata}  \& {Raiteri}}{{Ostorero} et~al.}{2004}]{2004A&A...419..913O}
{Ostorero} L.,  {Villata} M.,   {Raiteri} C.~M.,  2004, \mn@doi [\aap] {10.1051/0004-6361:20035813}, \href {https://ui.adsabs.harvard.edu/abs/2004A&A...419..913O} {419, 913}

\bibitem[\protect\citeauthoryear{{Otero-Santos}, {Pe{\~n}il}, {Acosta-Pulido}, {Becerra Gonz{\'a}lez}, {Raiteri}, {Carnerero}  \& {Villata}}{{Otero-Santos} et~al.}{2023}]{2023MNRAS.518.5788O}
{Otero-Santos} J.,  {Pe{\~n}il} P.,  {Acosta-Pulido} J.~A.,  {Becerra Gonz{\'a}lez} J.,  {Raiteri} C.~M.,  {Carnerero} M.~I.,   {Villata} M.,  2023, \mn@doi [\mnras] {10.1093/mnras/stac3142}, \href {https://ui.adsabs.harvard.edu/abs/2023MNRAS.518.5788O} {518, 5788}

\bibitem[\protect\citeauthoryear{{Parijskij}}{{Parijskij}}{1993}]{1993IAPM...35....7P}
{Parijskij} Y.~N.,  1993, \mn@doi [IEEE Antennas and Propagation Magazine] {10.1109/74.229840}, \href {http://adsabs.harvard.edu/abs/1993IAPM...35....7P} {35, 7}

\bibitem[\protect\citeauthoryear{{Pe{\~n}il} et~al.,}{{Pe{\~n}il} et~al.}{2020}]{2020ApJ...896..134P}
{Pe{\~n}il} P.,  et~al., 2020, \mn@doi [\apj] {10.3847/1538-4357/ab910d}, \href {https://ui.adsabs.harvard.edu/abs/2020ApJ...896..134P} {896, 134}

\bibitem[\protect\citeauthoryear{{Pedregosa} et~al.,}{{Pedregosa} et~al.}{2011}]{2011JMLR...12.2825P}
{Pedregosa} F.,  et~al., 2011, \mn@doi [Journal of Machine Learning Research] {10.48550/arXiv.1201.0490}, \href {https://ui.adsabs.harvard.edu/abs/2011JMLR...12.2825P} {12, 2825}

\bibitem[\protect\citeauthoryear{{Peng} \& {de Bruyn}}{{Peng} \& {de Bruyn}}{2004}]{2004ApJ...610..151P}
{Peng} B.,  {de Bruyn} A.~G.,  2004, \mn@doi [\apj] {10.1086/421485}, \href {https://ui.adsabs.harvard.edu/abs/2004ApJ...610..151P} {610, 151}

\bibitem[\protect\citeauthoryear{{Perley} \& {Butler}}{{Perley} \& {Butler}}{2017}]{2017ApJS..230....7P}
{Perley} R.~A.,  {Butler} B.~J.,  2017, \mn@doi [\apjs] {10.3847/1538-4365/aa6df9}, \href {https://ui.adsabs.harvard.edu/abs/2017ApJS..230....7P} {230, 7}

\bibitem[\protect\citeauthoryear{{Piner}, {Bhattarai}, {Edwards}  \& {Jones}}{{Piner} et~al.}{2006}]{2006ApJ...640..196P}
{Piner} B.~G.,  {Bhattarai} D.,  {Edwards} P.~G.,   {Jones} D.~L.,  2006, \mn@doi [\apj] {10.1086/500006}, \href {https://ui.adsabs.harvard.edu/abs/2006ApJ...640..196P} {640, 196}

\bibitem[\protect\citeauthoryear{{Plavin}, {Kovalev}, {Kovalev}  \& {Troitsky}}{{Plavin} et~al.}{2020}]{2020ApJ...894..101P}
{Plavin} A.,  {Kovalev} Y.~Y.,  {Kovalev} Y.~A.,   {Troitsky} S.,  2020, \mn@doi [\apj] {10.3847/1538-4357/ab86bd}, \href {https://ui.adsabs.harvard.edu/abs/2020ApJ...894..101P} {894, 101}

\bibitem[\protect\citeauthoryear{Porta, Verbeek  \& Krose}{Porta et~al.}{2005}]{porta:inria-00321476}
Porta J.,  Verbeek J.,   Krose B.,  2005, \mn@doi [{Autonomous Robots}] {10.1023/B:AURO.0000047287.00119.b6}, 18, 59

\bibitem[\protect\citeauthoryear{{Prokhorov} \& {Moraghan}}{{Prokhorov} \& {Moraghan}}{2017}]{2017MNRAS.471.3036P}
{Prokhorov} D.~A.,  {Moraghan} A.,  2017, \mn@doi [\mnras] {10.1093/mnras/stx1742}, \href {https://ui.adsabs.harvard.edu/abs/2017MNRAS.471.3036P} {471, 3036}

\bibitem[\protect\citeauthoryear{{Pushkarev}, {Kovalev}  \& {Lister}}{{Pushkarev} et~al.}{2010}]{2010ApJ...722L...7P}
{Pushkarev} A.~B.,  {Kovalev} Y.~Y.,   {Lister} M.~L.,  2010, \mn@doi [\apjl] {10.1088/2041-8205/722/1/L7}, \href {https://ui.adsabs.harvard.edu/abs/2010ApJ...722L...7P} {722, L7}

\bibitem[\protect\citeauthoryear{{Raiteri} et~al.,}{{Raiteri} et~al.}{2001}]{2001A&A...377..396R}
{Raiteri} C.~M.,  et~al., 2001, \mn@doi [\aap] {10.1051/0004-6361:20011112}, \href {http://adsabs.harvard.edu/abs/2001A%26A...377..396R} {377, 396}

\bibitem[\protect\citeauthoryear{{Raiteri} et~al.,}{{Raiteri} et~al.}{2005}]{2005A&A...438...39R}
{Raiteri} C.~M.,  et~al., 2005, \mn@doi [\aap] {10.1051/0004-6361:20042567}, \href {https://ui.adsabs.harvard.edu/abs/2005A&A...438...39R} {438, 39}

\bibitem[\protect\citeauthoryear{{Raiteri} et~al.,}{{Raiteri} et~al.}{2006}]{2006A&A...459..731R}
{Raiteri} C.~M.,  et~al., 2006, \mn@doi [\aap] {10.1051/0004-6361:20065744}, \href {https://ui.adsabs.harvard.edu/abs/2006A&A...459..731R} {459, 731}

\bibitem[\protect\citeauthoryear{{Raiteri} et~al.,}{{Raiteri} et~al.}{2008}]{2008A&A...480..339R}
{Raiteri} C.~M.,  et~al., 2008, \mn@doi [\aap] {10.1051/0004-6361:20079044}, \href {https://ui.adsabs.harvard.edu/abs/2008A&A...480..339R} {480, 339}

\bibitem[\protect\citeauthoryear{{Rajput}, {Shah}, {Stalin}, {Sahayanathan}  \& {Rakshit}}{{Rajput} et~al.}{2021}]{2021MNRAS.504.1772R}
{Rajput} B.,  {Shah} Z.,  {Stalin} C.~S.,  {Sahayanathan} S.,   {Rakshit} S.,  2021, \mn@doi [\mnras] {10.1093/mnras/stab970}, \href {https://ui.adsabs.harvard.edu/abs/2021MNRAS.504.1772R} {504, 1772}

\bibitem[\protect\citeauthoryear{{Readhead}}{{Readhead}}{1994}]{1994ApJ...426...51R}
{Readhead} A. C.~S.,  1994, \mn@doi [\apj] {10.1086/174038}, \href {https://ui.adsabs.harvard.edu/abs/1994ApJ...426...51R} {426, 51}

\bibitem[\protect\citeauthoryear{{Rieger}}{{Rieger}}{2007}]{2007Ap&SS.309..271R}
{Rieger} F.~M.,  2007, \mn@doi [\apss] {10.1007/s10509-007-9467-y}, \href {https://ui.adsabs.harvard.edu/abs/2007Ap&SS.309..271R} {309, 271}

\bibitem[\protect\citeauthoryear{{Robertson}, {Gallo}, {Zoghbi}  \& {Fabian}}{{Robertson} et~al.}{2015}]{2015MNRAS.453.3455R}
{Robertson} D.~R.~S.,  {Gallo} L.~C.,  {Zoghbi} A.,   {Fabian} A.~C.,  2015, \mn@doi [\mnras] {10.1093/mnras/stv1575}, \href {https://ui.adsabs.harvard.edu/abs/2015MNRAS.453.3455R} {453, 3455}

\bibitem[\protect\citeauthoryear{{Romero}, {Combi}, {Benaglia}, {Azcarate}, {Cersosimo}  \& {Wilkes}}{{Romero} et~al.}{1997}]{1997A&A...326...77R}
{Romero} G.~E.,  {Combi} J.~A.,  {Benaglia} P.,  {Azcarate} I.~N.,  {Cersosimo} J.~C.,   {Wilkes} L.~M.,  1997, \aap, \href {https://ui.adsabs.harvard.edu/abs/1997A&A...326...77R} {326, 77}

\bibitem[\protect\citeauthoryear{{Romero}, {Cellone}  \& {Combi}}{{Romero} et~al.}{2000}]{2000A&A...360L..47R}
{Romero} G.~E.,  {Cellone} S.~A.,   {Combi} J.~A.,  2000, \mn@doi [\aap] {10.48550/arXiv.astro-ph/0007407}, \href {https://ui.adsabs.harvard.edu/abs/2000A&A...360L..47R} {360, L47}

\bibitem[\protect\citeauthoryear{{Romero}, {Fan}  \& {Nuza}}{{Romero} et~al.}{2003}]{2003ChJAA...3..513R}
{Romero} G.~E.,  {Fan} J.-H.,   {Nuza} S.~E.,  2003, \mn@doi [\cjaa] {10.1088/1009-9271/3/6/513}, \href {https://ui.adsabs.harvard.edu/abs/2003ChJAA...3..513R} {3, 513}

\bibitem[\protect\citeauthoryear{{Roy} et~al.,}{{Roy} et~al.}{2022}]{2022MNRAS.513.5238R}
{Roy} A.,  et~al., 2022, \mn@doi [\mnras] {10.1093/mnras/stac1287}, \href {https://ui.adsabs.harvard.edu/abs/2022MNRAS.513.5238R} {513, 5238}

\bibitem[\protect\citeauthoryear{{Scargle}}{{Scargle}}{1982}]{1982ApJ...263..835S}
{Scargle} J.~D.,  1982, \mn@doi [\apj] {10.1086/160554}, \href {https://ui.adsabs.harvard.edu/abs/1982ApJ...263..835S} {263, 835}

\bibitem[\protect\citeauthoryear{{Scargle}, {Norris}, {Jackson}  \& {Chiang}}{{Scargle} et~al.}{2013}]{2013ApJ...764..167S}
{Scargle} J.~D.,  {Norris} J.~P.,  {Jackson} B.,   {Chiang} J.,  2013, \mn@doi [\apj] {10.1088/0004-637X/764/2/167}, \href {https://ui.adsabs.harvard.edu/abs/2013ApJ...764..167S} {764, 167}

\bibitem[\protect\citeauthoryear{Shuygina et~al.,}{Shuygina et~al.}{2019}]{2019..VLBI..Quasar}
Shuygina N.,  et~al., 2019, \mn@doi [Geodesy and Geodynamics] {10.1016/j.geog.2018.09.008}, 10, 150

\bibitem[\protect\citeauthoryear{{Simonetti}, {Cordes}  \& {Heeschen}}{{Simonetti} et~al.}{1985}]{1985ApJ...296...46S}
{Simonetti} J.~H.,  {Cordes} J.~M.,   {Heeschen} D.~S.,  1985, \mn@doi [\apj] {10.1086/163418}, \href {https://ui.adsabs.harvard.edu/abs/1985ApJ...296...46S} {296, 46}

\bibitem[\protect\citeauthoryear{{Sotnikova}}{{Sotnikova}}{2020}]{2020gbar.conf...32S}
{Sotnikova} Y.~V.,  2020, in {Romanyuk} I.~I.,  {Yakunin} I.~A.,  {Valeev} A.~F.,   {Kudryavtsev} D.~O.,  eds, Ground-Based Astronomy in Russia. 21st Century. pp 32--40, \mn@doi{10.26119/978-5-6045062-0-2_2020_32}

\bibitem[\protect\citeauthoryear{{Sotnikova} et~al.,}{{Sotnikova} et~al.}{2022a}]{2022AstBu..77..246S}
{Sotnikova} Y.~V.,  et~al., 2022a, \mn@doi [Astrophysical Bulletin] {10.1134/S1990341322030117}, \href {https://ui.adsabs.harvard.edu/abs/2022AstBu..77..246S} {77, 246}

\bibitem[\protect\citeauthoryear{{Sotnikova}, {Kovalev}, {Kovalev}, {Erkenov}  \& {Plavin}}{{Sotnikova} et~al.}{2022b}]{2022evlb.confE...9S}
{Sotnikova} Y.,  {Kovalev} Y.~A.,  {Kovalev} Y.~Y.,  {Erkenov} A.~K.,   {Plavin} A.~V.,  2022b, in European VLBI Network Mini-Symposium and Users' Meeting 2021. p.~9, \mn@doi{10.22323/1.399.0009}

\bibitem[\protect\citeauthoryear{{Spinrad} \& {Smith}}{{Spinrad} \& {Smith}}{1975}]{1975ApJ...201..275S}
{Spinrad} H.,  {Smith} H.~E.,  1975, \mn@doi [\apj] {10.1086/153883}, \href {https://ui.adsabs.harvard.edu/abs/1975ApJ...201..275S} {201, 275}

\bibitem[\protect\citeauthoryear{{Stickel}, {Padovani}, {Urry}, {Fried}  \& {Kuehr}}{{Stickel} et~al.}{1991}]{1991ApJ...374..431S}
{Stickel} M.,  {Padovani} P.,  {Urry} C.~M.,  {Fried} J.~W.,   {Kuehr} H.,  1991, \mn@doi [\apj] {10.1086/170133}, \href {http://adsabs.harvard.edu/abs/1991ApJ...374..431S} {374, 431}

\bibitem[\protect\citeauthoryear{Tipping \& Bishop}{Tipping \& Bishop}{1999}]{DBLP:journals/neco/TippingB99}
Tipping M.~E.,  Bishop C.~M.,  1999, \mn@doi [Neural Comput.] {10.1162/089976699300016728}, 11, 443

\bibitem[\protect\citeauthoryear{{Tripathi}, {Gupta}, {Aller}, {Wiita}, {Bambi}, {Aller}  \& {Gu}}{{Tripathi} et~al.}{2021}]{2021MNRAS.501.5997T}
{Tripathi} A.,  {Gupta} A.~C.,  {Aller} M.~F.,  {Wiita} P.~J.,  {Bambi} C.,  {Aller} H.,   {Gu} M.,  2021, \mn@doi [\mnras] {10.1093/mnras/stab058}, \href {https://ui.adsabs.harvard.edu/abs/2021MNRAS.501.5997T} {501, 5997}

\bibitem[\protect\citeauthoryear{{Tsybulev}}{{Tsybulev}}{2011}]{2011AstBu..66..109T}
{Tsybulev} P.~G.,  2011, \mn@doi [Astrophysical Bulletin] {10.1134/S199034131101010X}, \href {http://adsabs.harvard.edu/abs/2011AstBu..66..109T} {66, 109}

\bibitem[\protect\citeauthoryear{{Tsybulev}, {Nizhelskii}, {Dugin}, {Borisov}, {Kratov}  \& {Udovitskii}}{{Tsybulev} et~al.}{2018}]{2018AstBu..73..494T}
{Tsybulev} P.~G.,  {Nizhelskii} N.~A.,  {Dugin} M.~V.,  {Borisov} A.~N.,  {Kratov} D.~V.,   {Udovitskii} R.~Y.,  2018, \mn@doi [Astrophysical Bulletin] {10.1134/S1990341318040132}, \href {https://ui.adsabs.harvard.edu/abs/2018AstBu..73..494T} {73, 494}

\bibitem[\protect\citeauthoryear{{Udovitskiy}, {Sotnikova}, {Mingaliev}, {Tsybulev}, {Zhekanis}  \& {Nizhelskij}}{{Udovitskiy} et~al.}{2016}]{2016AstBu..71..496U}
{Udovitskiy} R.~Y.,  {Sotnikova} Y.~V.,  {Mingaliev} M.~G.,  {Tsybulev} P.~G.,  {Zhekanis} G.~V.,   {Nizhelskij} N.~A.,  2016, \mn@doi [Astrophysical Bulletin] {10.1134/S1990341316040131}, \href {https://ui.adsabs.harvard.edu/abs/2016AstBu..71..496U} {71, 496}

\bibitem[\protect\citeauthoryear{{Urry} \& {Padovani}}{{Urry} \& {Padovani}}{1995}]{1995PASP..107..803U}
{Urry} C.~M.,  {Padovani} P.,  1995, \mn@doi [\pasp] {10.1086/133630}, \href {http://adsabs.harvard.edu/abs/1995PASP..107..803U} {107, 803}

\bibitem[\protect\citeauthoryear{{Valtaoja}, {Terasranta}, {Urpo}, {Nesterov}, {Lainela}  \& {Valtonen}}{{Valtaoja} et~al.}{1992}]{1992A&A...254...71V}
{Valtaoja} E.,  {Terasranta} H.,  {Urpo} S.,  {Nesterov} N.~S.,  {Lainela} M.,   {Valtonen} M.,  1992, \aap, \href {https://ui.adsabs.harvard.edu/abs/1992A&A...254...71V} {254, 71}

\bibitem[\protect\citeauthoryear{{Valtaoja}, {L{\"a}hteenm{\"a}ki}, {Ter{\"a}sranta}  \& {Lainela}}{{Valtaoja} et~al.}{1999}]{1999ApJS..120...95V}
{Valtaoja} E.,  {L{\"a}hteenm{\"a}ki} A.,  {Ter{\"a}sranta} H.,   {Lainela} M.,  1999, \mn@doi [\apjs] {10.1086/313170}, \href {https://ui.adsabs.harvard.edu/abs/1999ApJS..120...95V} {120, 95}

\bibitem[\protect\citeauthoryear{{Valyavin} et~al.,}{{Valyavin} et~al.}{2022a}]{2022Photo...9..950V}
{Valyavin} G.,  et~al., 2022a, \mn@doi [Photonics] {10.3390/photonics9120950}, \href {https://ui.adsabs.harvard.edu/abs/2022Photo...9..950V} {9, 950}

\bibitem[\protect\citeauthoryear{{Valyavin} et~al.,}{{Valyavin} et~al.}{2022b}]{2022AstBu..77..495V}
{Valyavin} G.,  et~al., 2022b, \mn@doi [Astrophysical Bulletin] {10.1134/S1990341322040186}, \href {https://ui.adsabs.harvard.edu/abs/2022AstBu..77..495V} {77, 551}

\bibitem[\protect\citeauthoryear{{VanderPlas}}{{VanderPlas}}{2018}]{2018ApJS..236...16V}
{VanderPlas} J.~T.,  2018, \mn@doi [\apjs] {10.3847/1538-4365/aab766}, \href {https://ui.adsabs.harvard.edu/abs/2018ApJS..236...16V} {236, 16}

\bibitem[\protect\citeauthoryear{{Vaughan}, {Edelson}, {Warwick}  \& {Uttley}}{{Vaughan} et~al.}{2003}]{2003MNRAS.345.1271V}
{Vaughan} S.,  {Edelson} R.,  {Warwick} R.~S.,   {Uttley} P.,  2003, \mn@doi [\mnras] {10.1046/j.1365-2966.2003.07042.x}, \href {https://ui.adsabs.harvard.edu/abs/2003MNRAS.345.1271V} {345, 1271}

\bibitem[\protect\citeauthoryear{{Verhodanov}, {Andernach}  \& {Chernenkov}}{{Verhodanov} et~al.}{2005}]{2005AstBu..58..118V}
{Verhodanov} O.~V.and~{Trushkin} S.~A.,  {Andernach} H.,   {Chernenkov} V.~N.,  2005, Astrophysical Bulletin, 58, 118

\bibitem[\protect\citeauthoryear{{Verkhodanov}}{{Verkhodanov}}{1997}]{1997ASPC..125...46V}
{Verkhodanov} O.~V.,  1997, Astronomical Data Analysis Software and Systems VI, A.S.P. Conference Series, \href {http://adsabs.harvard.edu/abs/1997ASPC..125...46V} {125, 46}

\bibitem[\protect\citeauthoryear{{Verkhodanov}, {Trushkin}  \& {Chernenkov}}{{Verkhodanov} et~al.}{1997}]{1997BaltA...6..275V}
{Verkhodanov} O.~V.,  {Trushkin} S.~A.,   {Chernenkov} V.~N.,  1997, Baltic Astronomy, \href {http://adsabs.harvard.edu/abs/1997BaltA...6..275V} {6, 275}

\bibitem[\protect\citeauthoryear{{Vlasyuk}}{{Vlasyuk}}{1993}]{1993BSAO...36..107V}
{Vlasyuk} V.~V.,  1993, Bulletin of Special Astrophysical Observatory, \href {https://ui.adsabs.harvard.edu/abs/1993BSAO...36..107V} {36, 107}

\bibitem[\protect\citeauthoryear{{Vlasyuk} et~al.,}{{Vlasyuk} et~al.}{2023}]{2023AstBu..78..464V}
{Vlasyuk} V.~V.,  et~al., 2023, \mn@doi [Astrophysical Bulletin] {10.1134/S1990341323600229}, \href {https://ui.adsabs.harvard.edu/abs/2023AstBu..78..464V} {78, 464}

\bibitem[\protect\citeauthoryear{{Volvach}, {Larionov}, {Volvach}, {Lahteenmaki}, {Tornikoski}, {Aller}, {Aller}  \& {Sasada}}{{Volvach} et~al.}{2015}]{2015ARep...59..145V}
{Volvach} A.~E.,  {Larionov} M.~G.,  {Volvach} L.~N.,  {Lahteenmaki} A.,  {Tornikoski} M.,  {Aller} M.~F.,  {Aller} H.~D.,   {Sasada} M.,  2015, \mn@doi [Astronomy Reports] {10.1134/S1063772914120117}, \href {https://ui.adsabs.harvard.edu/abs/2015ARep...59..145V} {59, 145}

\bibitem[\protect\citeauthoryear{{Volvach}, {Volvach}  \& {Larionov}}{{Volvach} et~al.}{2023}]{2023Galax..11...96V}
{Volvach} A.,  {Volvach} L.,   {Larionov} M.,  2023, \mn@doi [Galaxies] {10.3390/galaxies11050096}, \href {https://ui.adsabs.harvard.edu/abs/2023Galax..11...96V} {11, 96}

\bibitem[\protect\citeauthoryear{{Wagner} \& {Witzel}}{{Wagner} \& {Witzel}}{1995}]{1995ARA&A..33..163W}
{Wagner} S.~J.,  {Witzel} A.,  1995, \mn@doi [\araa] {10.1146/annurev.aa.33.090195.001115}, \href {https://ui.adsabs.harvard.edu/abs/1995ARA&A..33..163W} {33, 163}

\bibitem[\protect\citeauthoryear{{Wagner} et~al.,}{{Wagner} et~al.}{2022}]{2022icrc.confE.868W}
{Wagner} S.~M.,  et~al., 2022, in 37th International Cosmic Ray Conference. p.~868 (\mn@eprint {arXiv} {2110.14797}), \mn@doi{10.22323/1.395.0868}

\bibitem[\protect\citeauthoryear{{Wang}}{{Wang}}{2014}]{2014Ap&SS.351..281W}
{Wang} H.,  2014, \mn@doi [\apss] {10.1007/s10509-014-1840-z}, \href {https://ui.adsabs.harvard.edu/abs/2014Ap&SS.351..281W} {351, 281}

\bibitem[\protect\citeauthoryear{{Wang} \& {Jiang}}{{Wang} \& {Jiang}}{2020}]{2020ApJ...902...41W}
{Wang} Y.-F.,  {Jiang} Y.-G.,  2020, \mn@doi [\apj] {10.3847/1538-4357/abb36c}, \href {https://ui.adsabs.harvard.edu/abs/2020ApJ...902...41W} {902, 41}

\bibitem[\protect\citeauthoryear{{Webb}, {Howard}, {Ben{\'\i}tez}, {Balonek}, {McGrath}, {Shrader}, {Robson}  \& {Jenkins}}{{Webb} et~al.}{2000}]{2000AJ....120...41W}
{Webb} J.~R.,  {Howard} E.,  {Ben{\'\i}tez} E.,  {Balonek} T.,  {McGrath} E.,  {Shrader} C.,  {Robson} I.,   {Jenkins} P.,  2000, \mn@doi [\aj] {10.1086/301432}, \href {https://ui.adsabs.harvard.edu/abs/2000AJ....120...41W} {120, 41}

\bibitem[\protect\citeauthoryear{{Zechmeister} \& {K{\"u}rster}}{{Zechmeister} \& {K{\"u}rster}}{2009}]{2009A&A...496..577Z}
{Zechmeister} M.,  {K{\"u}rster} M.,  2009, \mn@doi [\aap] {10.1051/0004-6361:200811296}, \href {https://ui.adsabs.harvard.edu/abs/2009A&A...496..577Z} {496, 577}

\makeatother
\end{thebibliography}

\appendix
\onecolumn

\section{Plots}
\clearpage

\begin{figure*}
\centerline{\includegraphics[height=3.2cm]{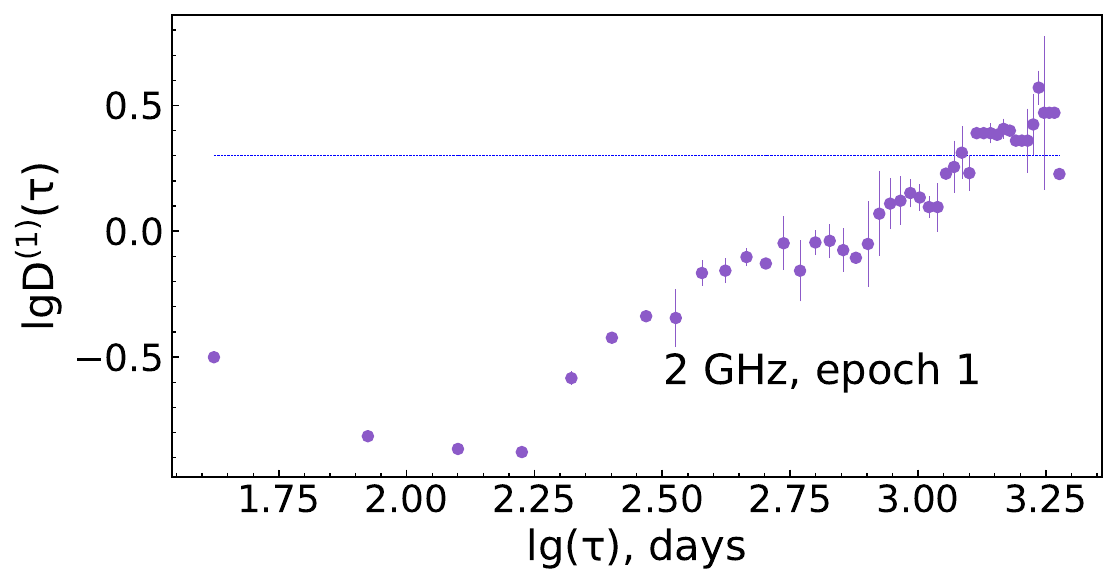}\includegraphics[height=3.2cm]{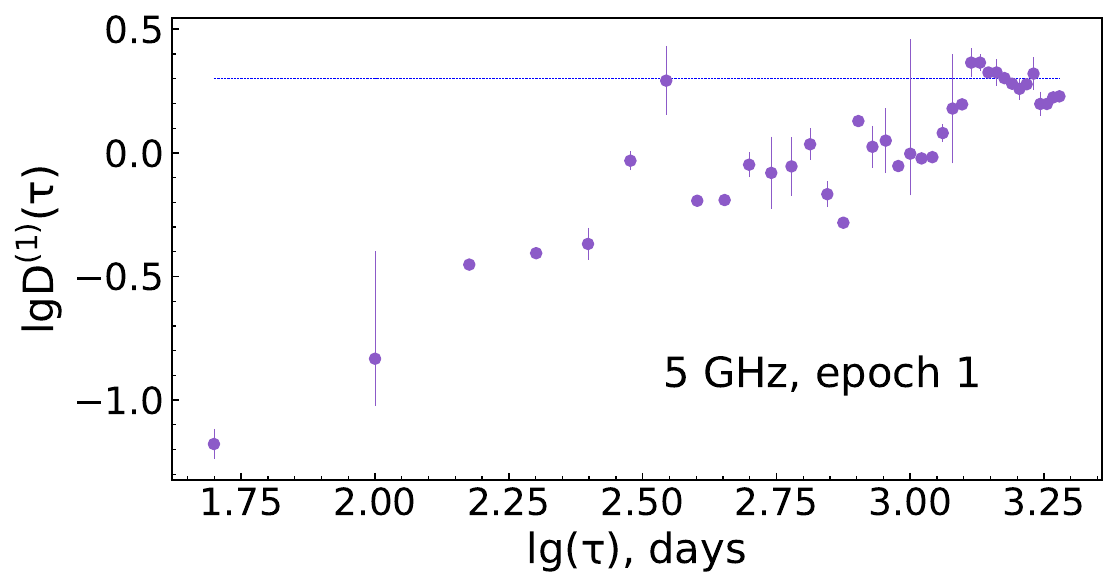}\includegraphics[height=3.2cm]{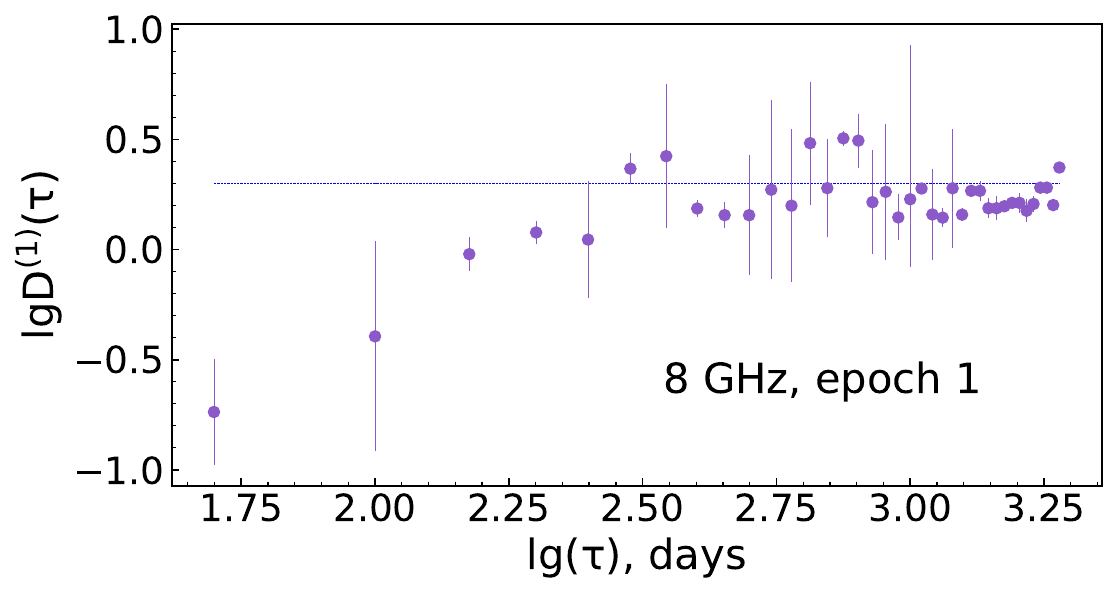}}\centerline{\includegraphics[height=3.2cm]{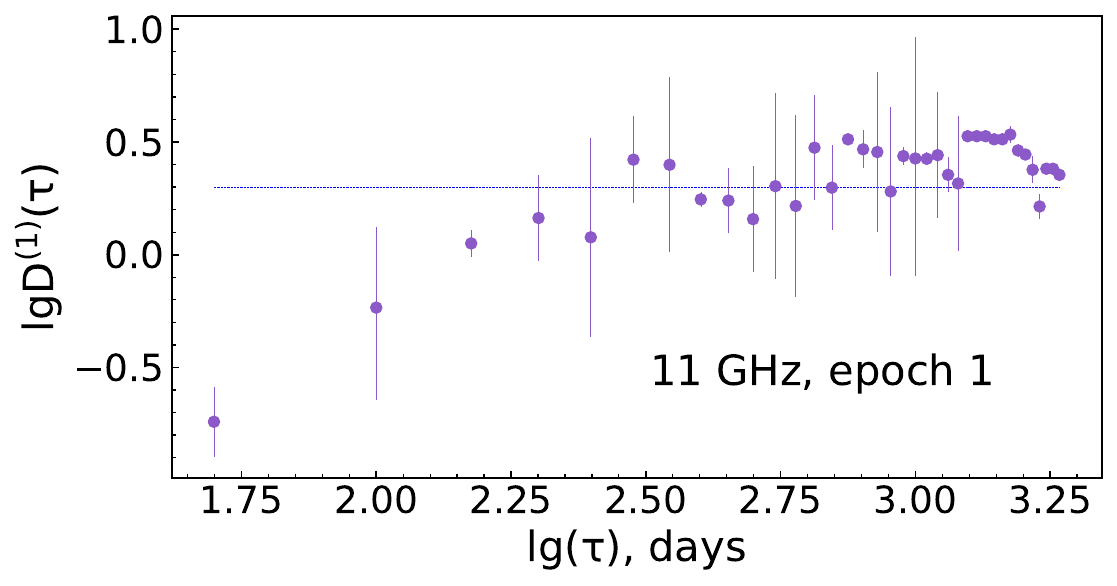}\includegraphics[height=3.2cm]{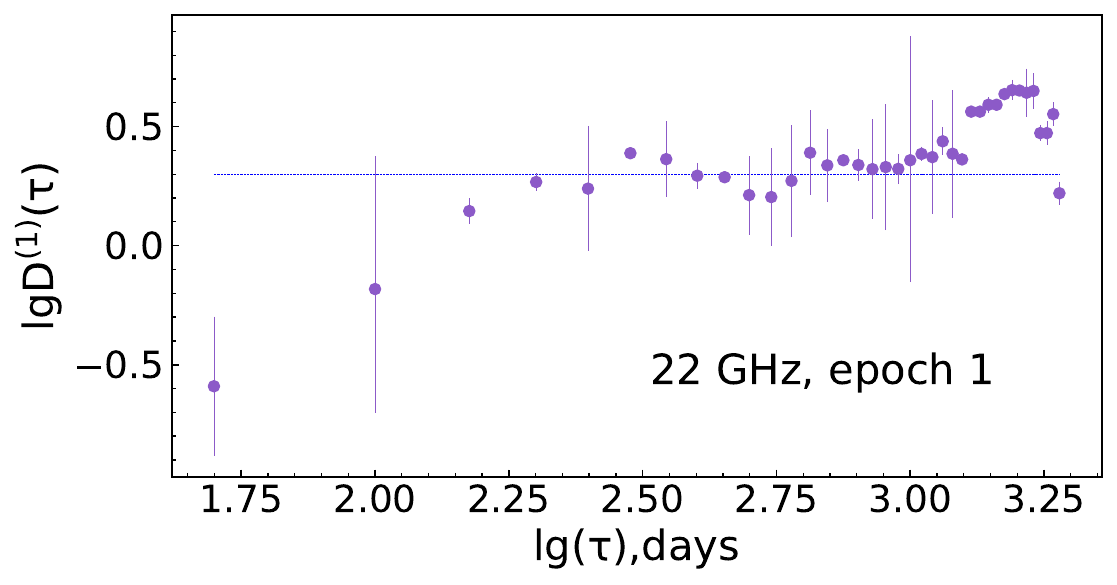}\includegraphics[height=3.2cm]{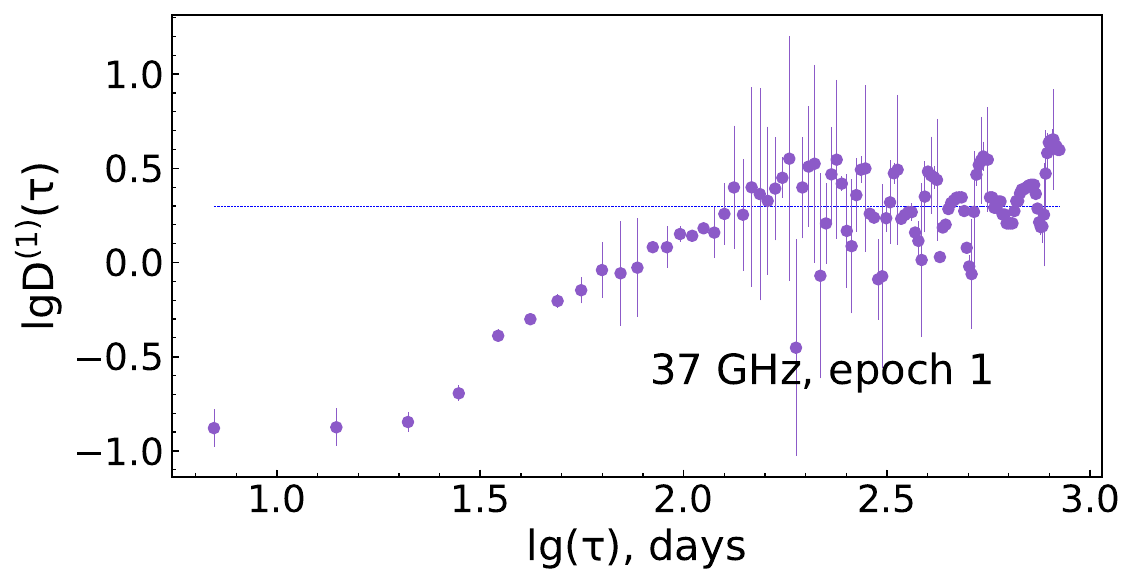}}
\centerline{\includegraphics[height=3.2cm]{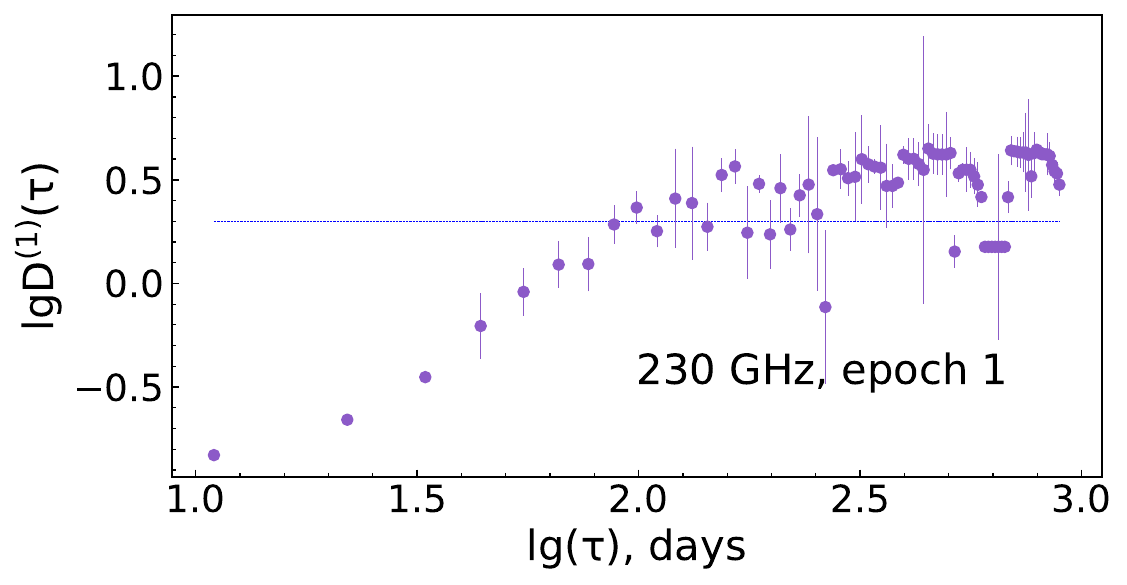}\includegraphics[height=3.2cm]{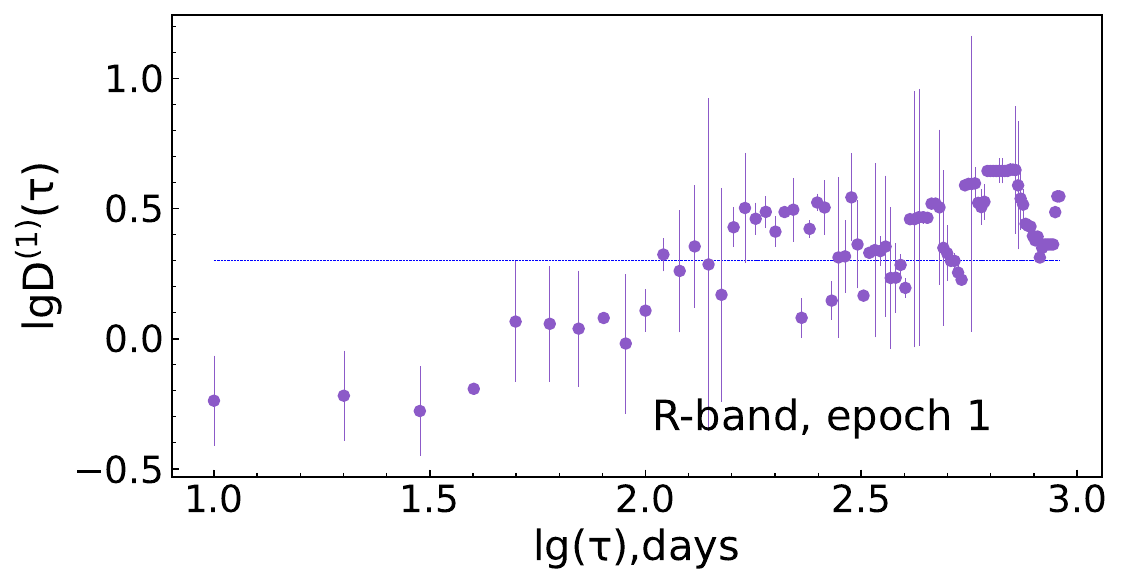}}
\caption{The SFs for the total flux variations at 2, 5, 8, 11.2, 22.2, 37, 230 GHz, and in the optical $R$-band for epoch 1.
}
\label{fig:SF1}
\end{figure*}

\begin{figure*}
\centerline{\includegraphics[height=3.2cm]{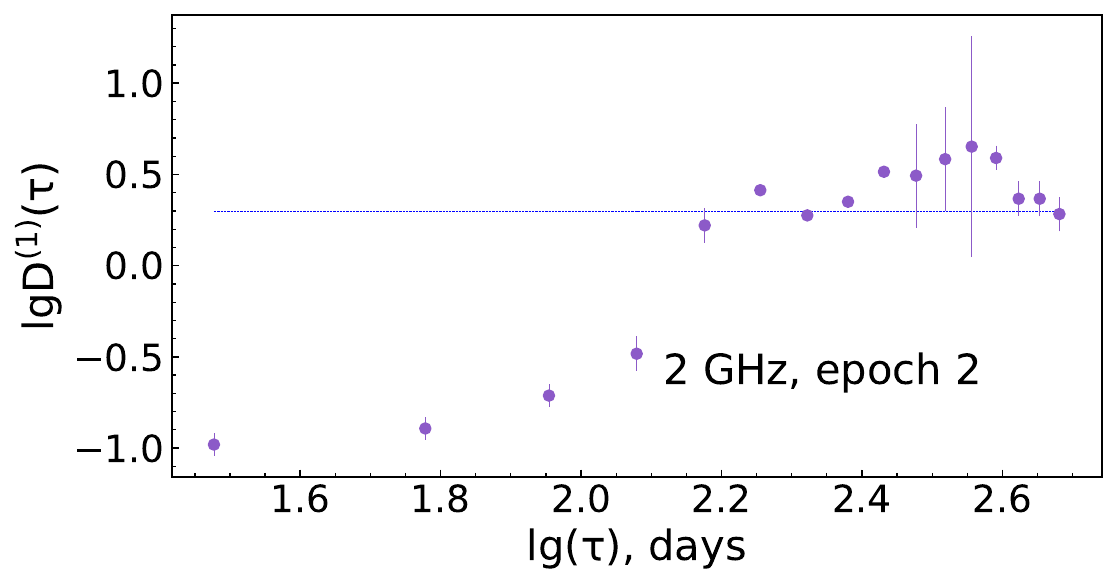}\includegraphics[height=3.2cm]{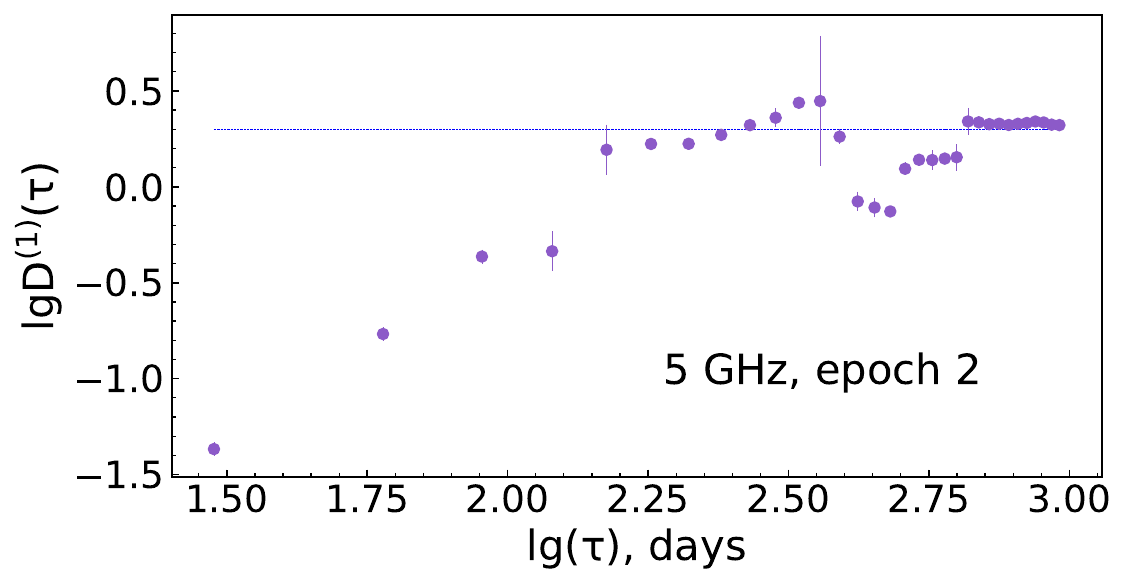}\includegraphics[height=3.2cm]{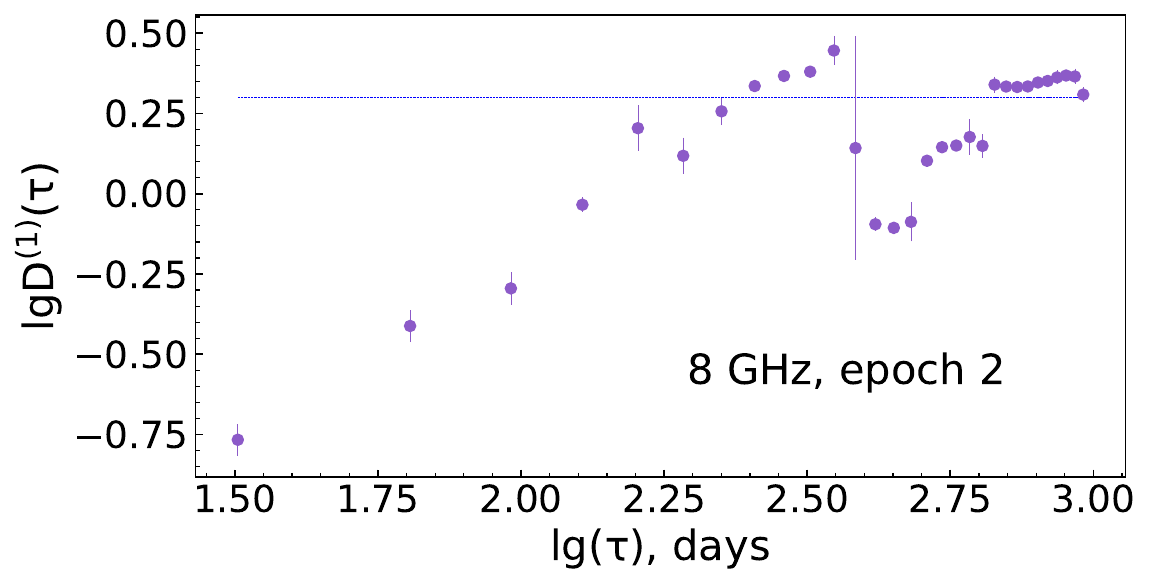}}\centerline{\includegraphics[height=3.2cm]{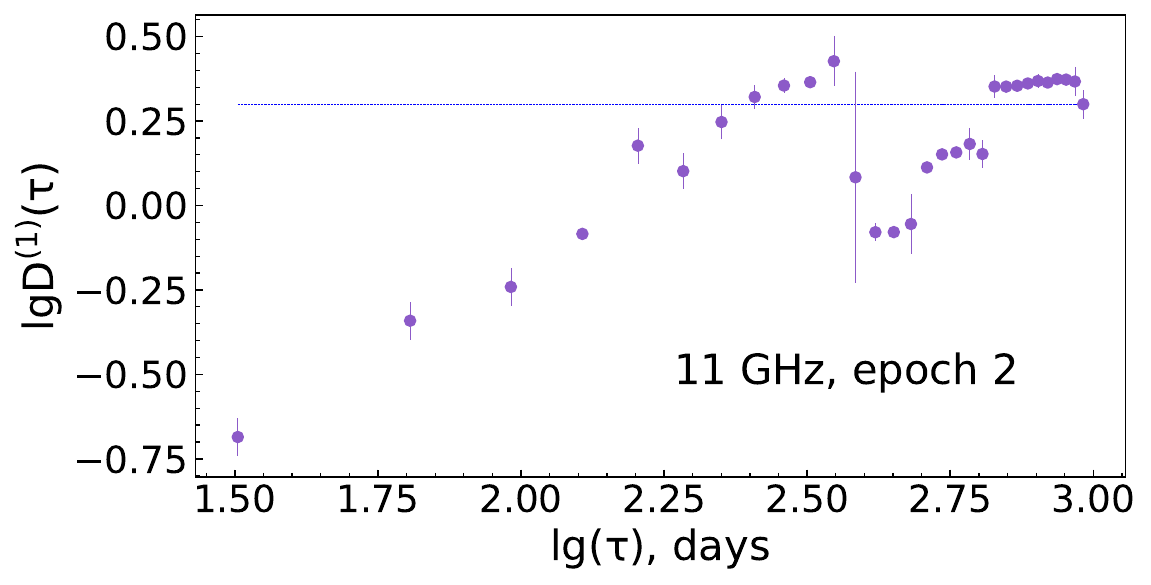}\includegraphics[height=3.2cm]{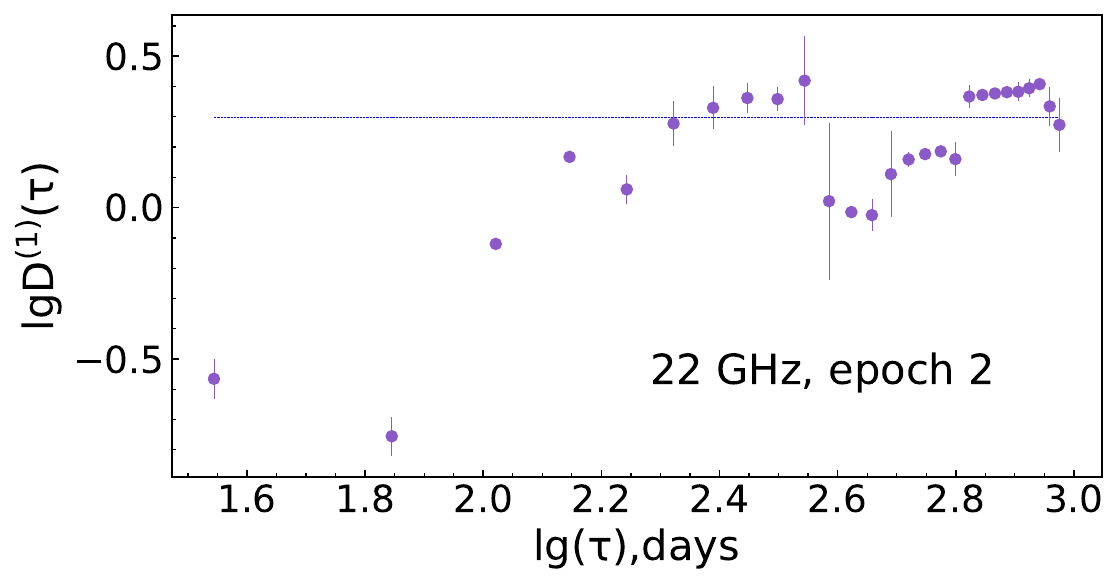}\includegraphics[height=3.2cm]{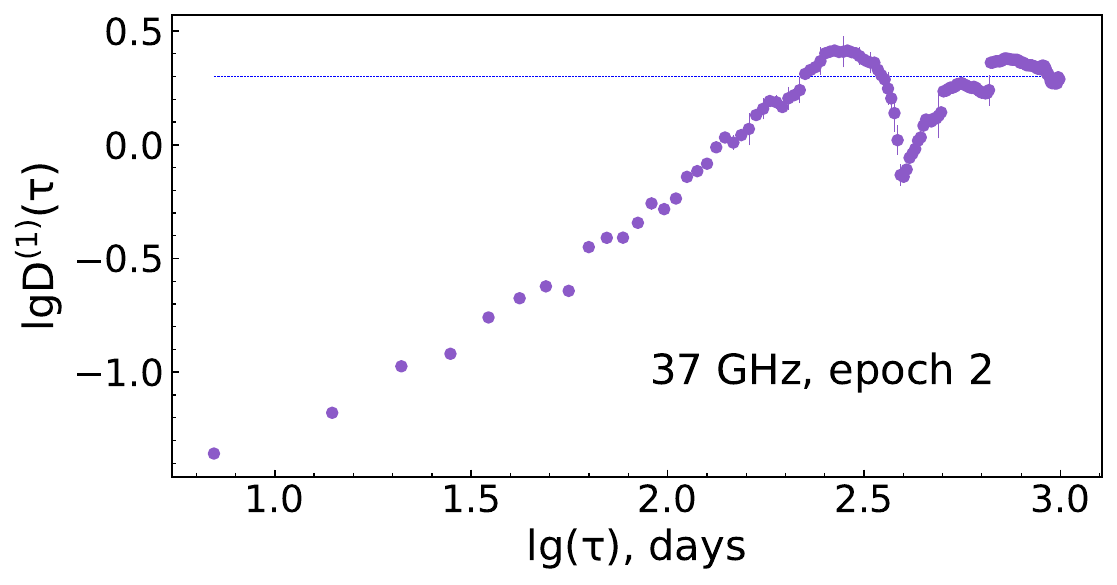}}
\centerline{\includegraphics[height=3.2cm]{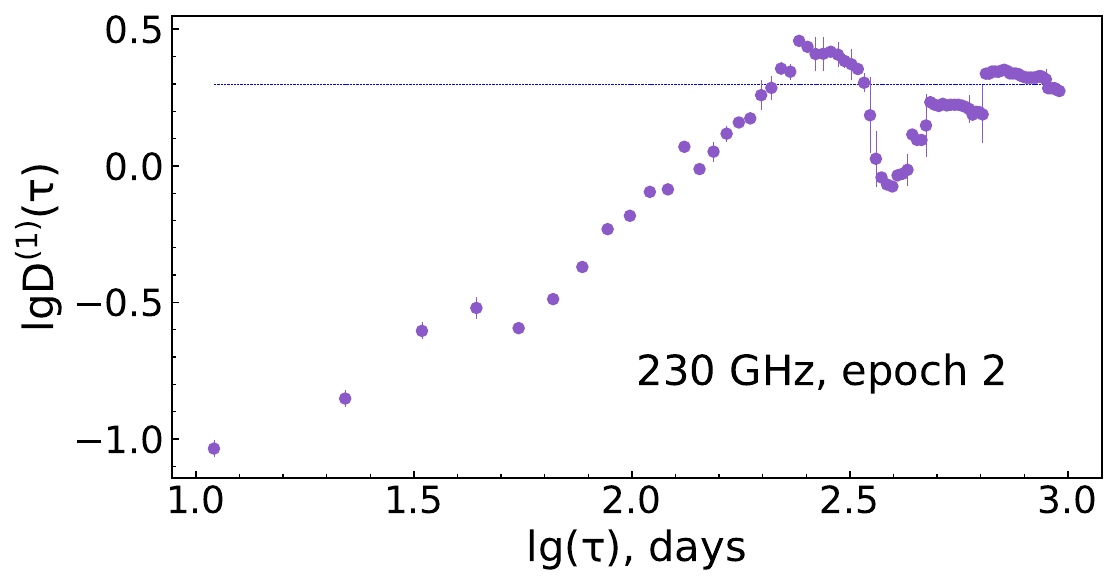}\includegraphics[height=3.2cm]{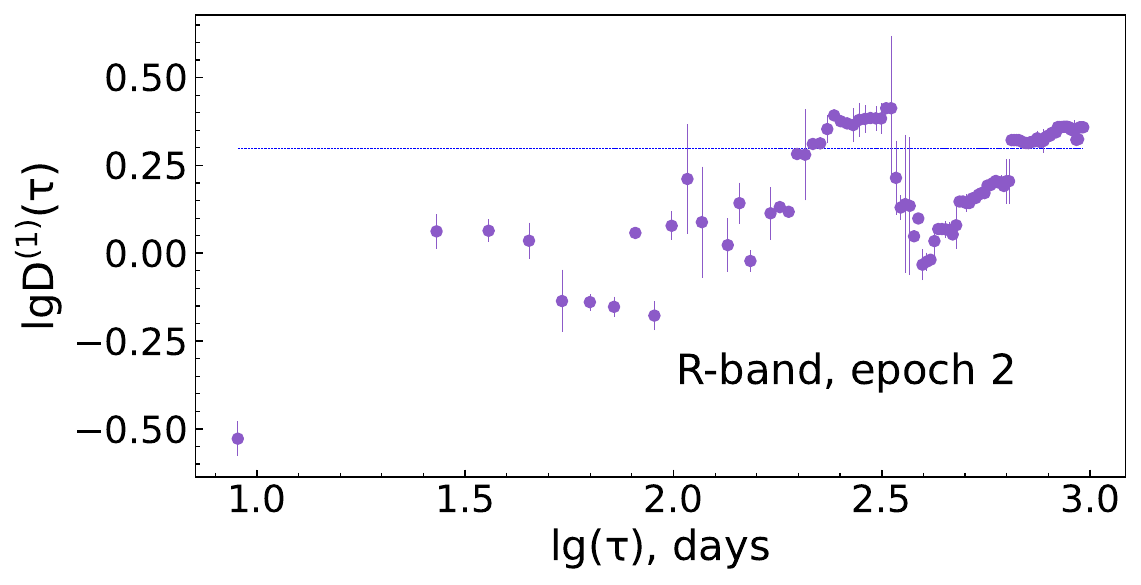}\includegraphics[height=3.2cm]{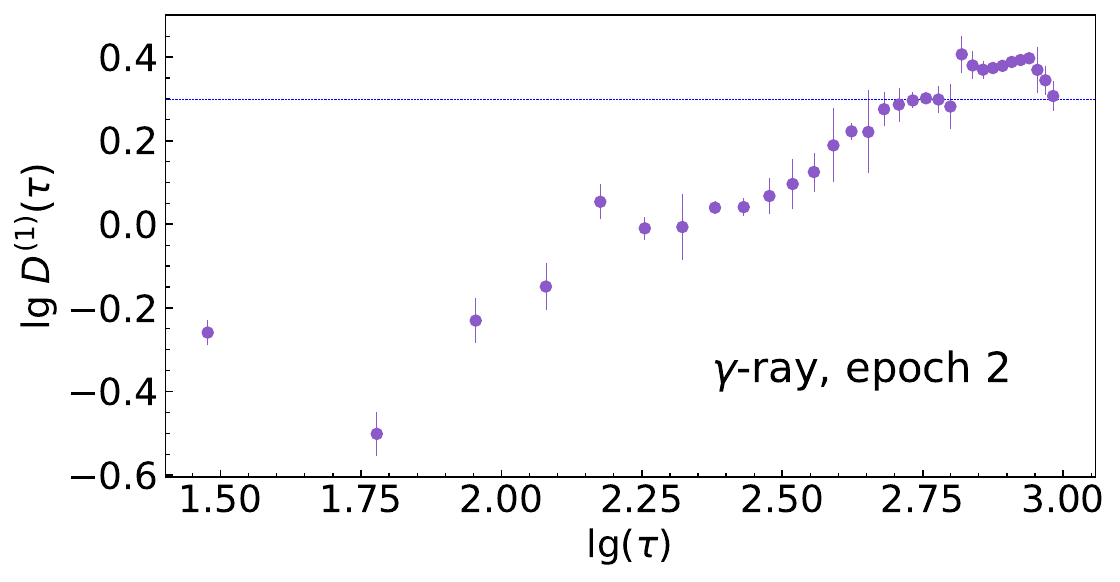}}
\caption{The SFs for the total flux variations at 2, 5, 8, 11.2, 22.2, 37, 230 GHz, in the optical $R$-band, and in $\gamma$-rays for epoch 2.}
\label{fig:SF2}
\end{figure*}


\begin{figure*}
\centerline{\includegraphics[height=3.2cm]{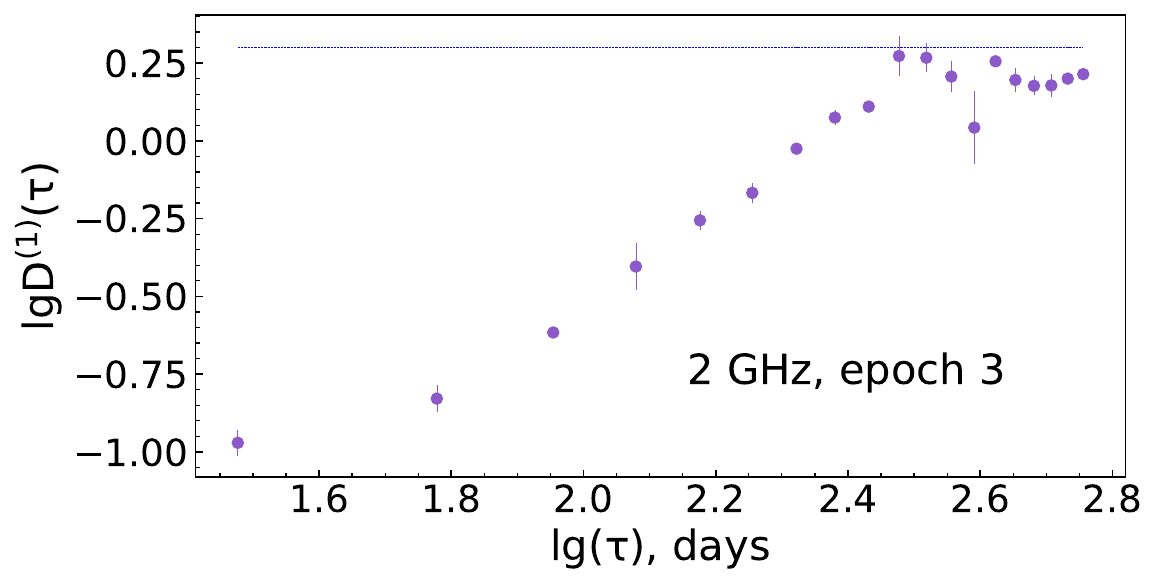}\includegraphics[height=3.2cm]{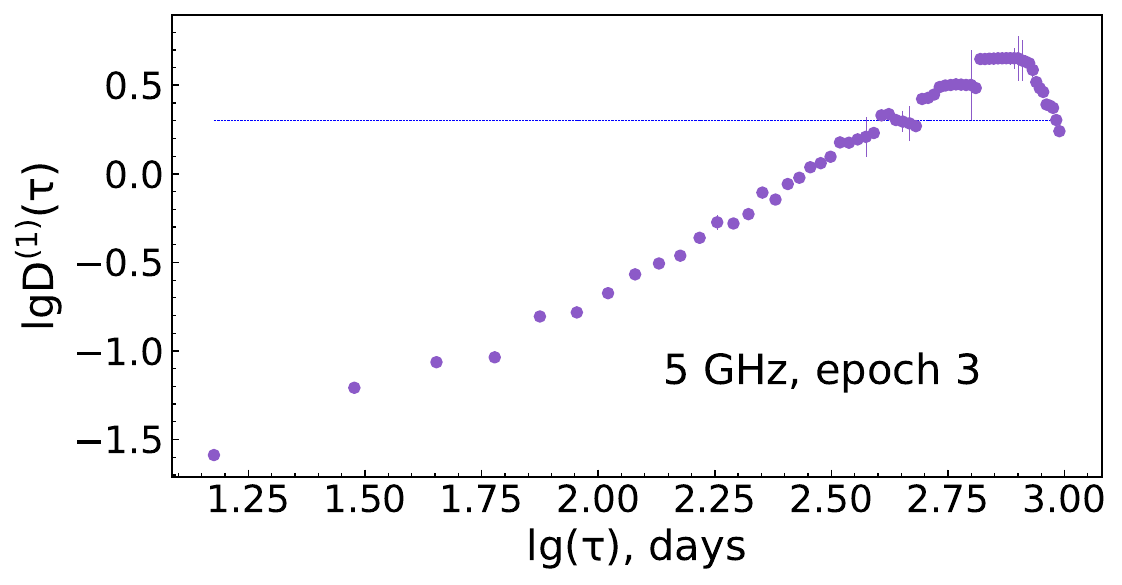}\includegraphics[height=3.2cm]{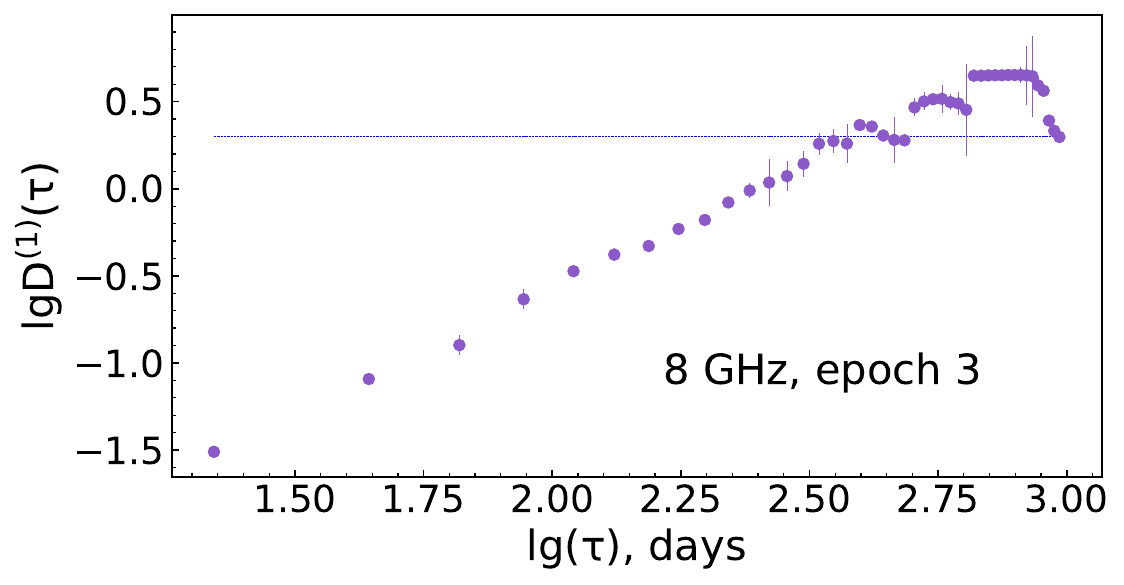}}\centerline{\includegraphics[height=3.2cm]{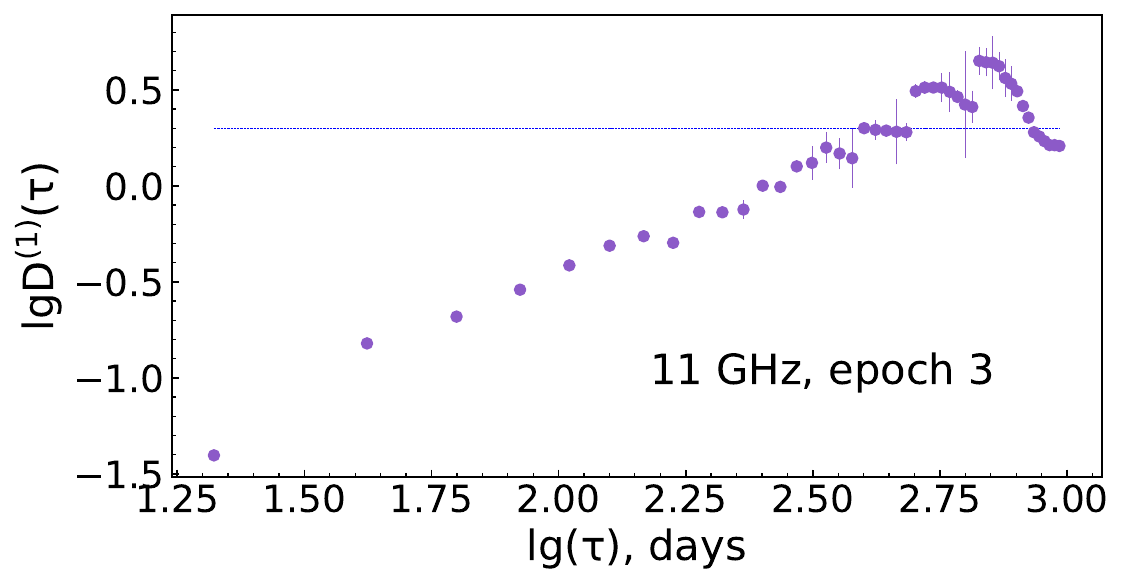}\includegraphics[height=3.2cm]{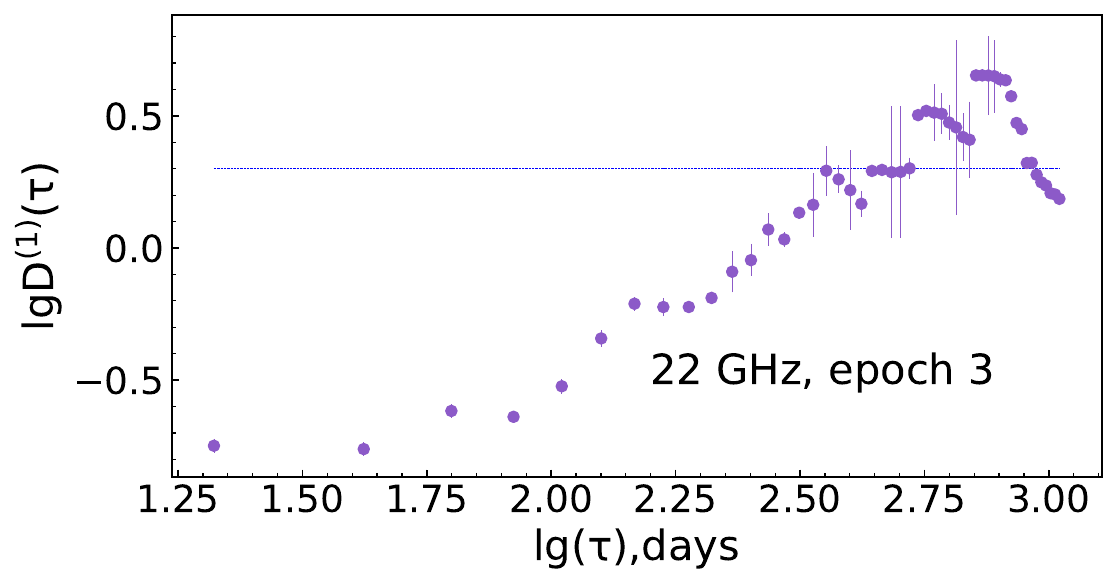}\includegraphics[height=3.2cm]{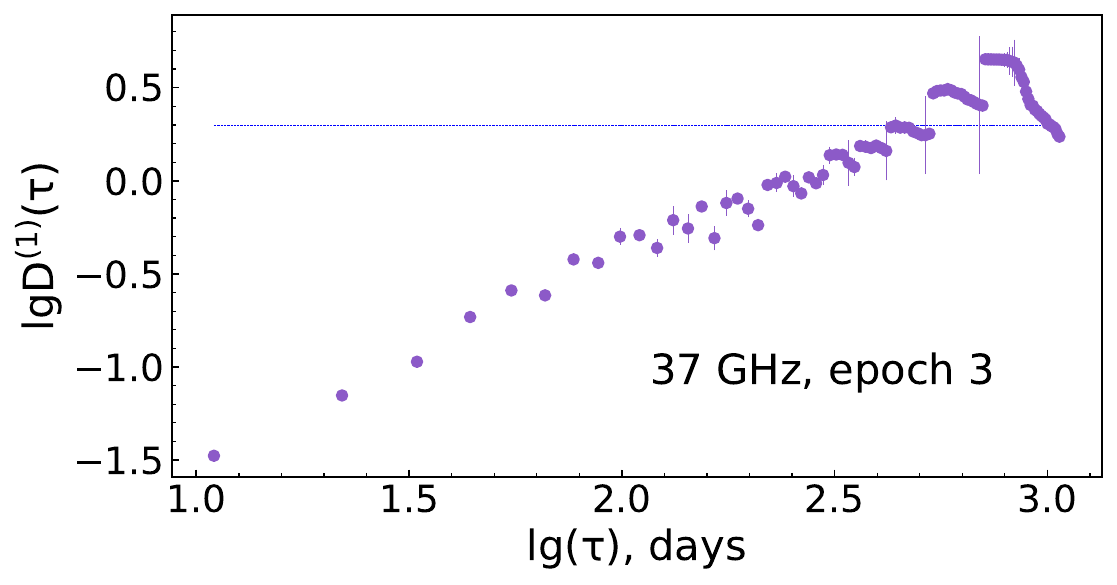}}
\centerline{\includegraphics[height=3.2cm]{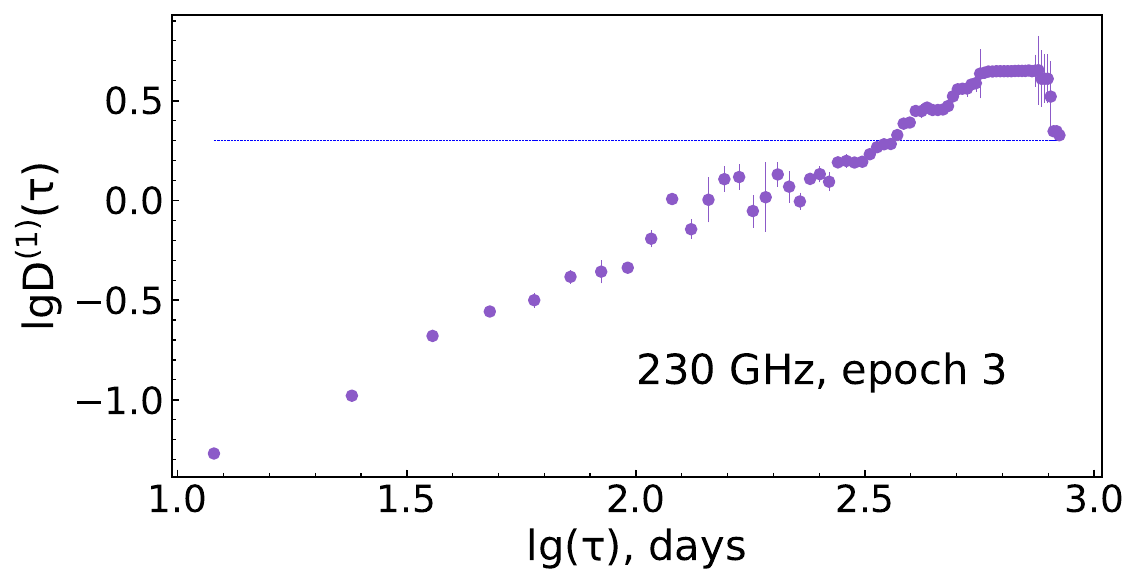}\includegraphics[height=3.2cm]{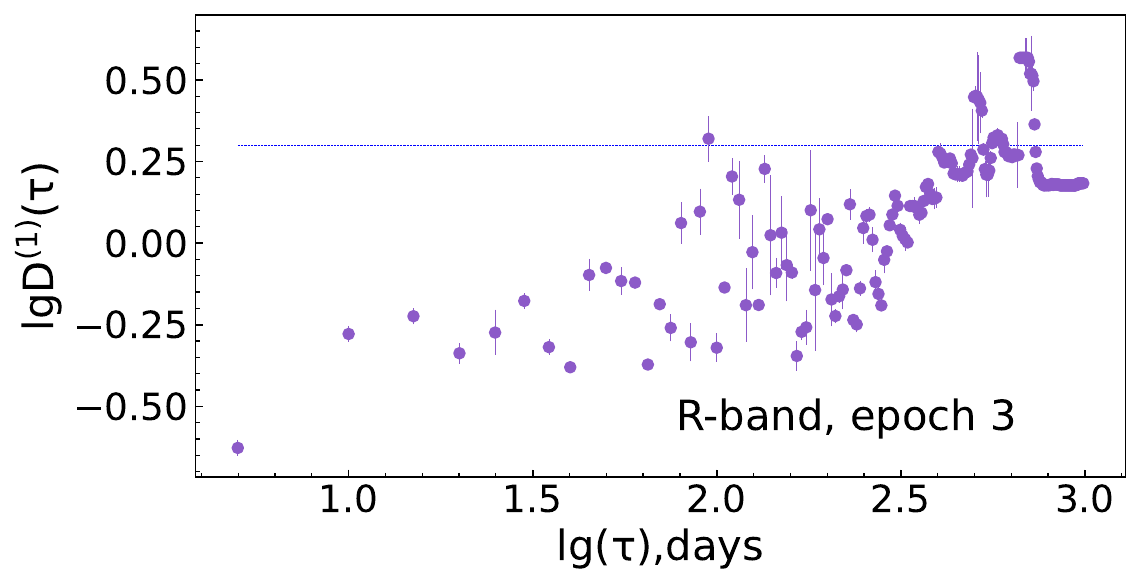}\includegraphics[height=3.2cm]{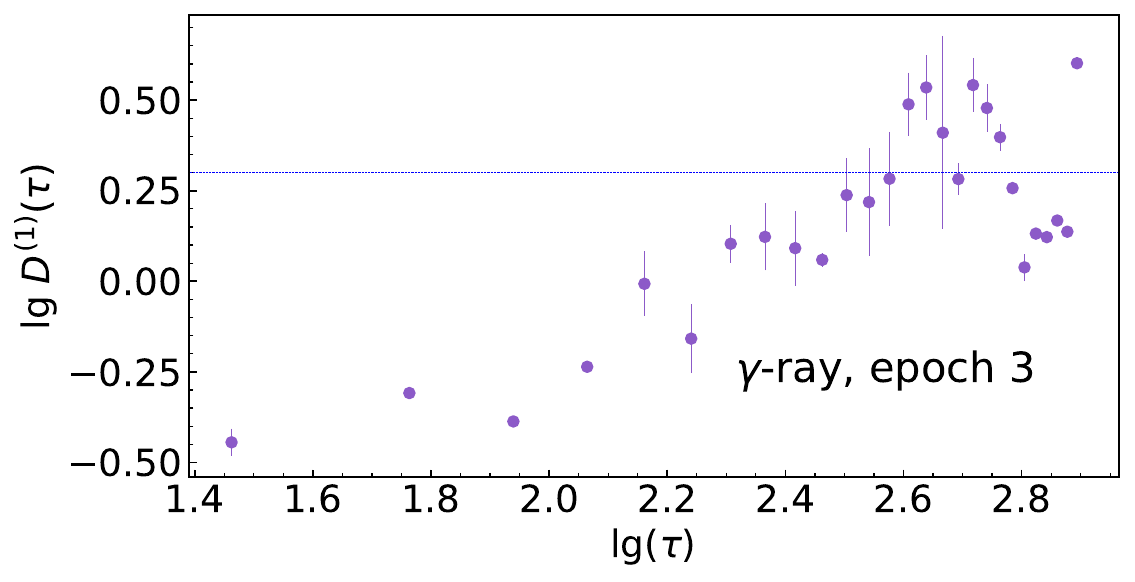}}
\caption{The SFs for the total flux variations at 2, 5, 8, 11.2, 22.2, 37, 230 GHz, in the optical $R$-band, and in $\gamma$-rays for epoch 3.}
\label{fig:SF3}
\end{figure*}

\begin{figure*}
\centerline{\includegraphics[height=3.2cm]{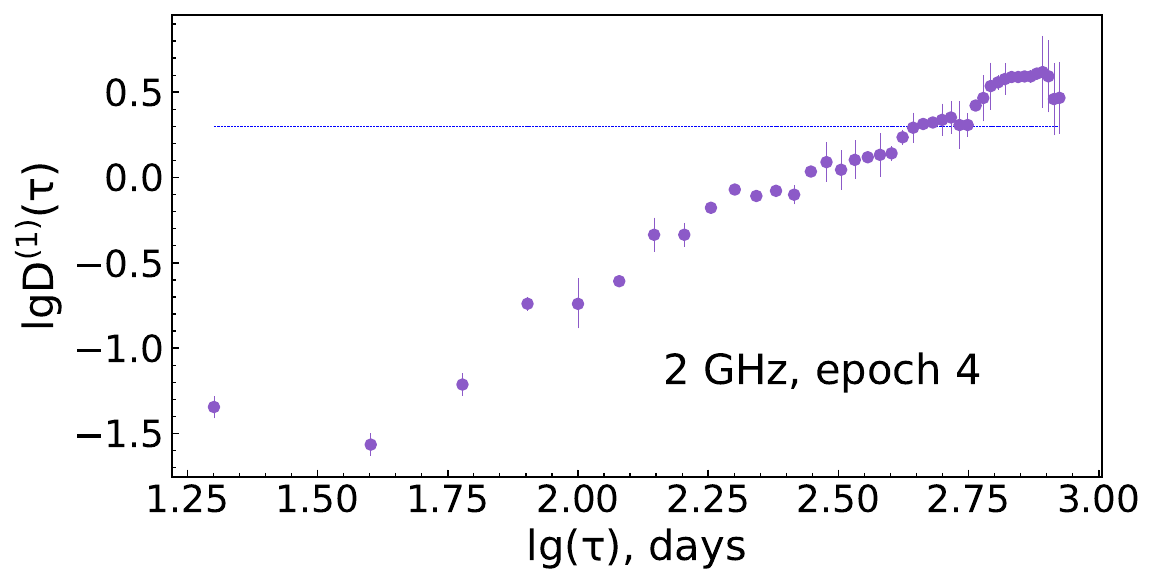}\includegraphics[height=3.2cm]{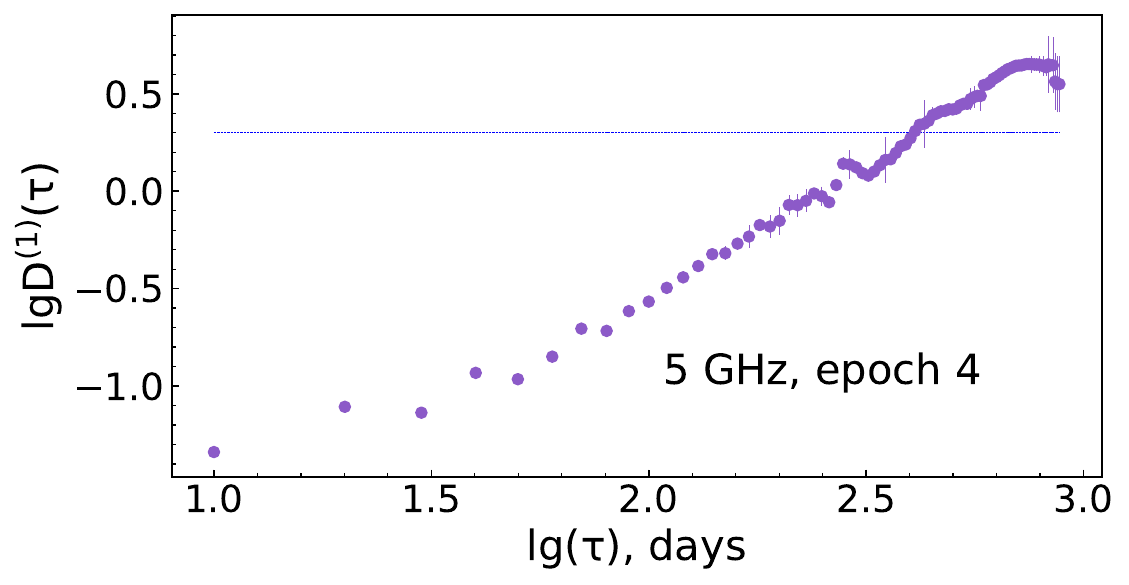}\includegraphics[height=3.2cm]{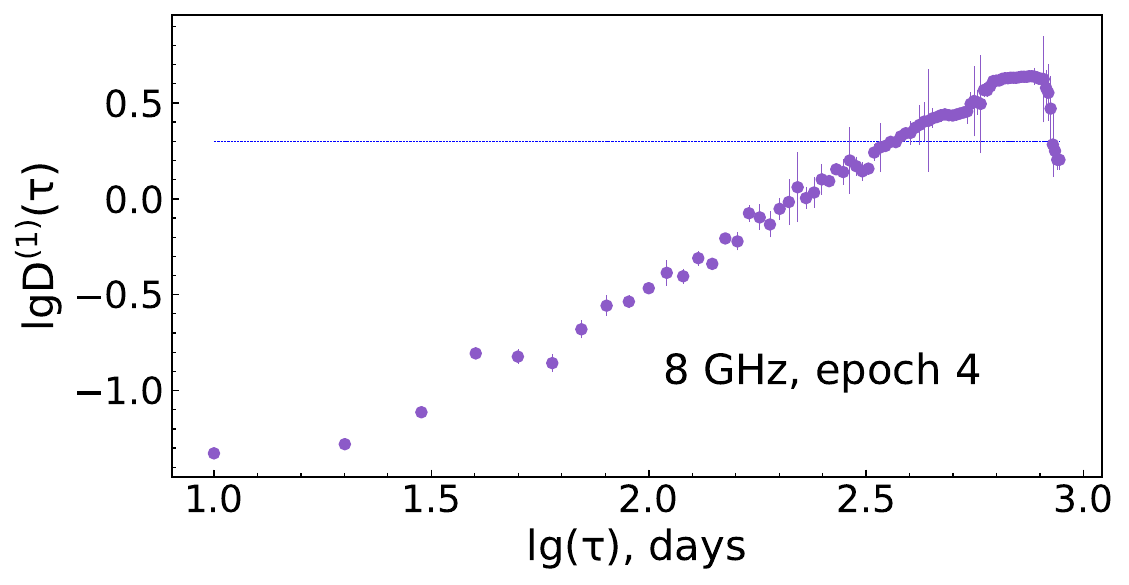}}\centerline{\includegraphics[height=3.2cm]{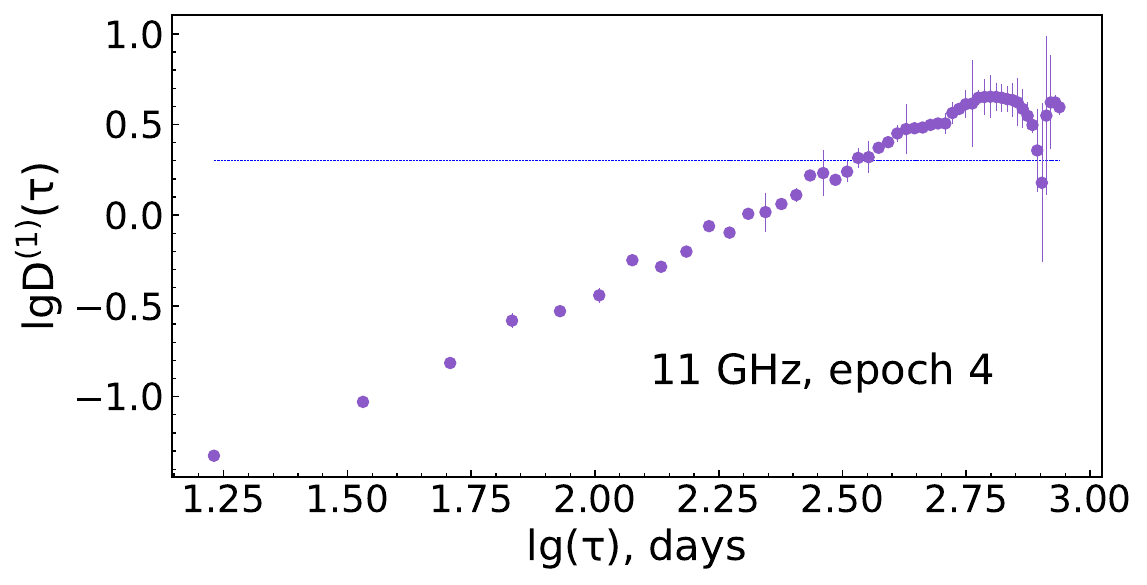}\includegraphics[height=3.2cm]{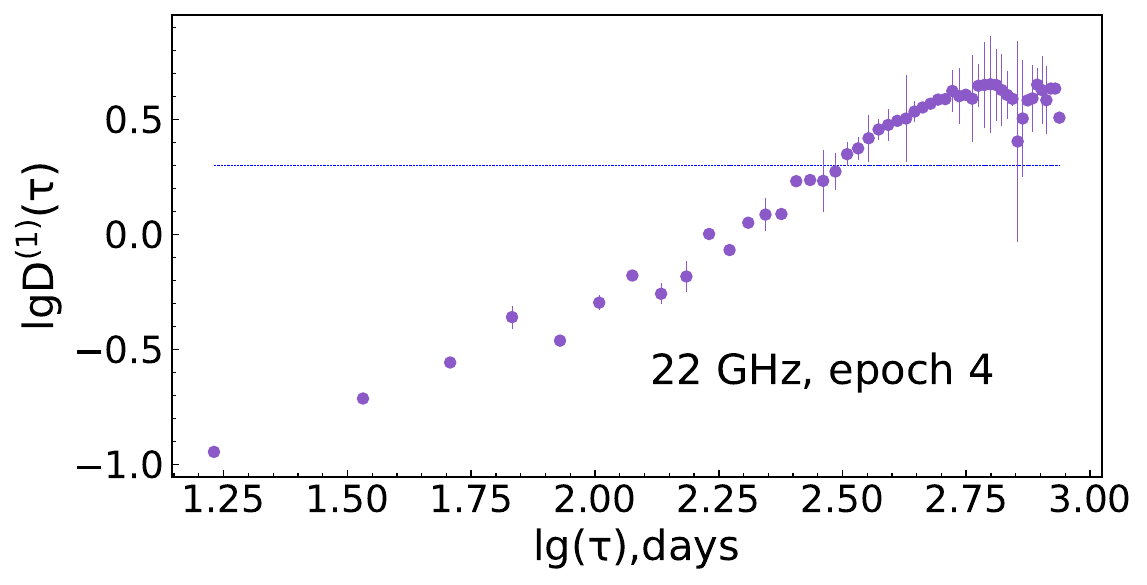}\includegraphics[height=3.2cm]{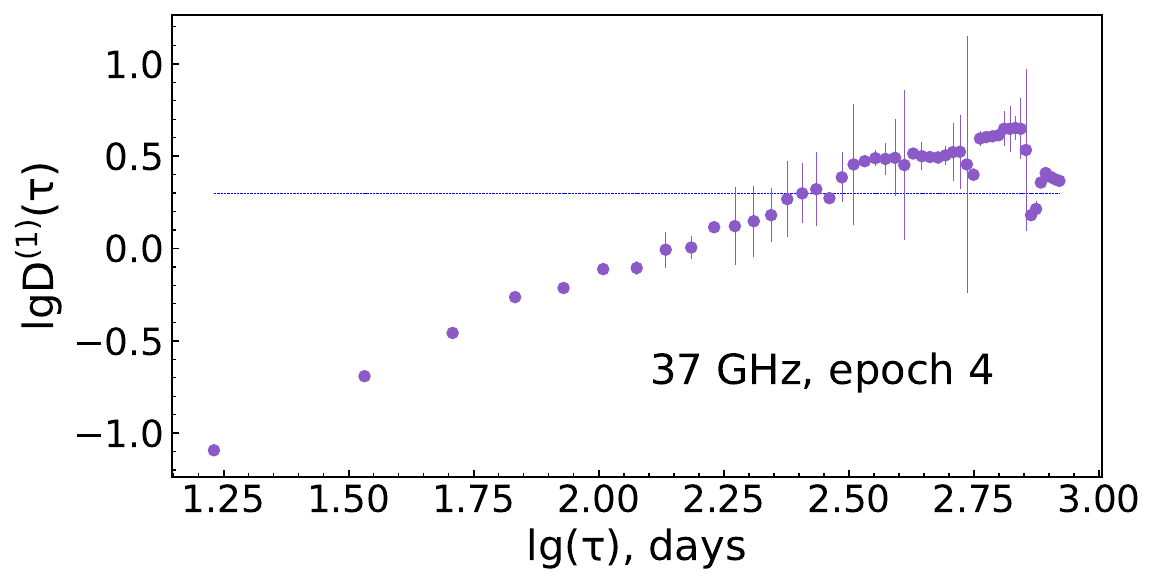}}
\centerline{\includegraphics[height=3.2cm]{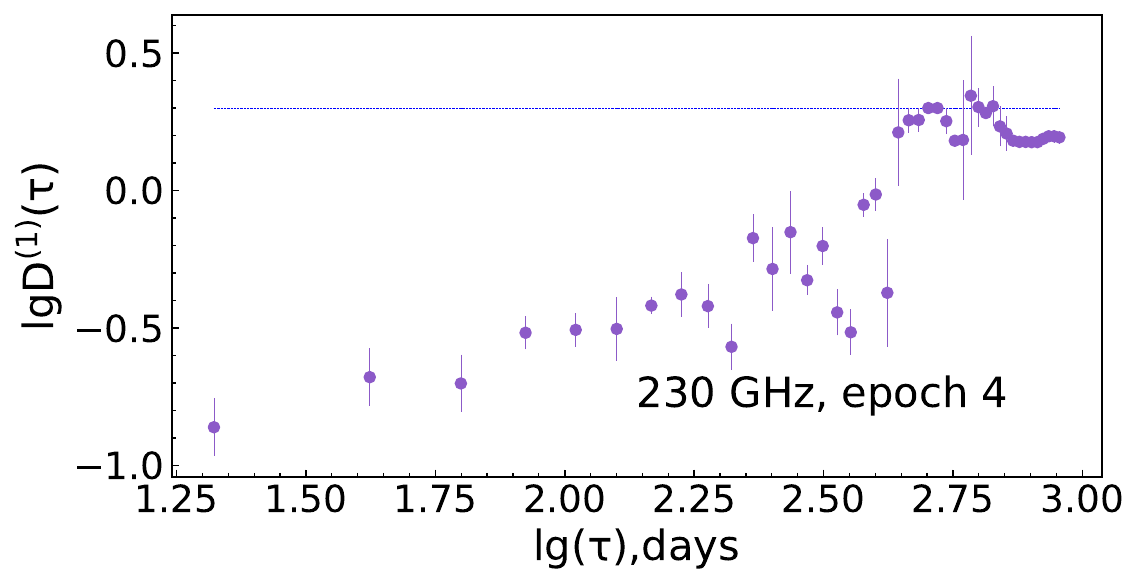}\includegraphics[height=3.2cm]{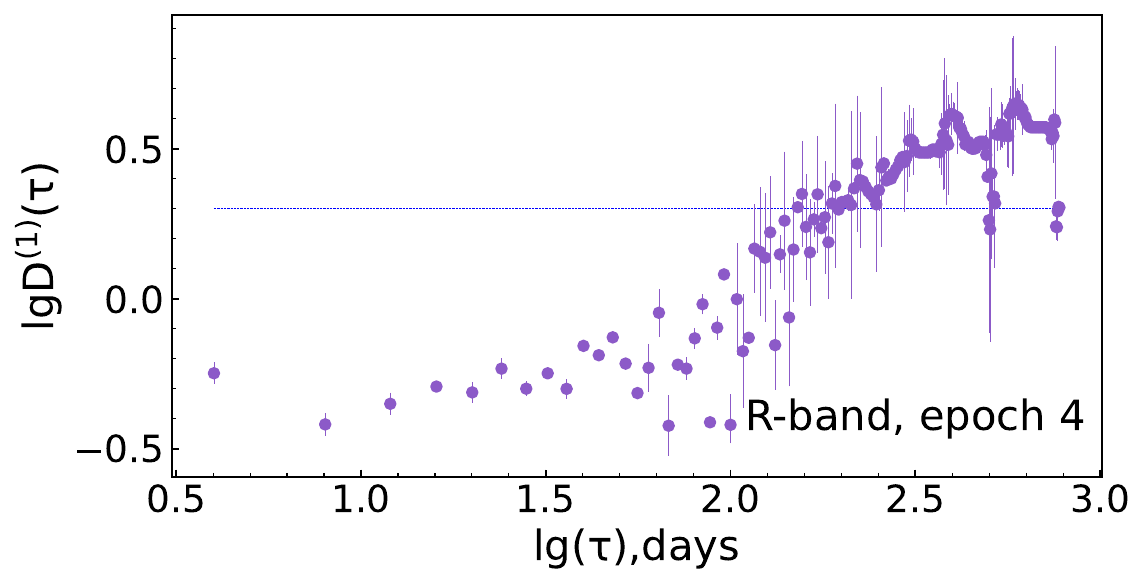}\includegraphics[height=3.2cm]{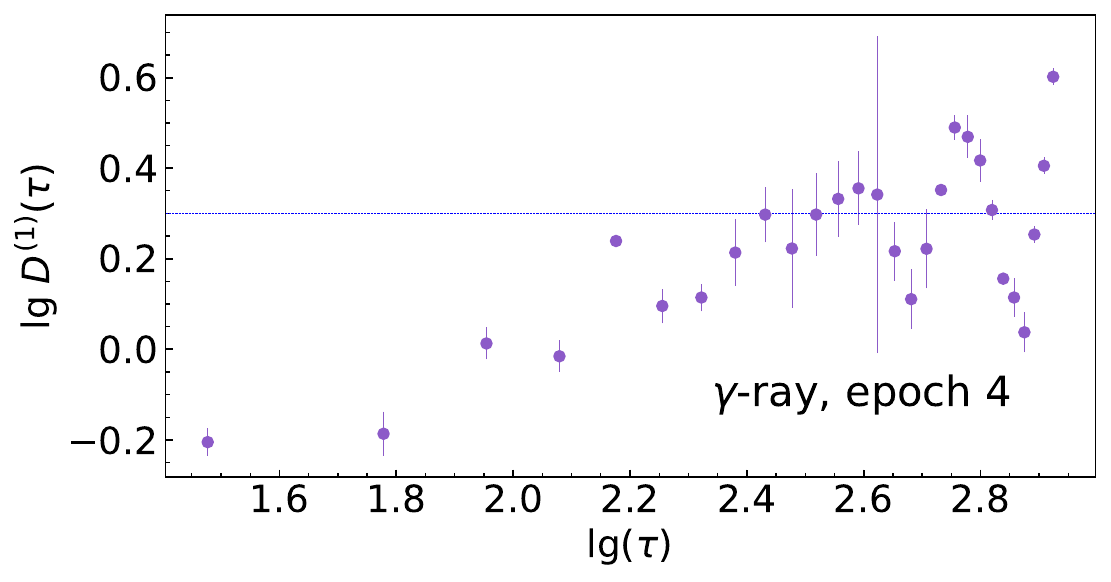}}
\caption{The SFs for the total flux variations at 2, 5, 8, 11.2, 22.2, 37, 230 GHz, in the optical $R$-band, and in $\gamma$-rays for epoch 4.}
\label{fig:SF4}
\end{figure*}

\begin{figure*}
\centerline{\includegraphics[height=3.2cm]{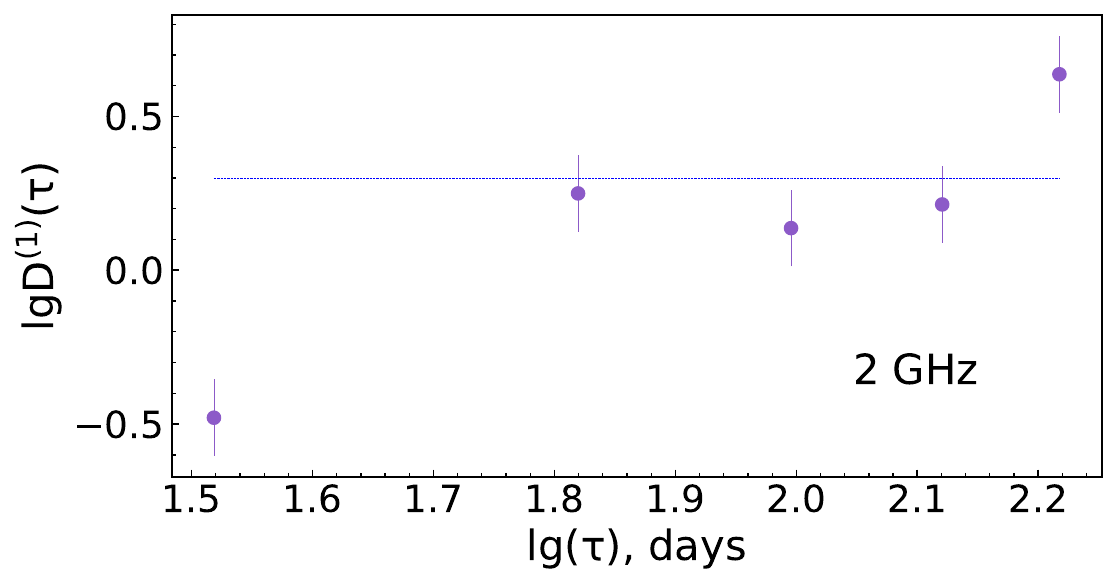}\includegraphics[height=3.2cm]{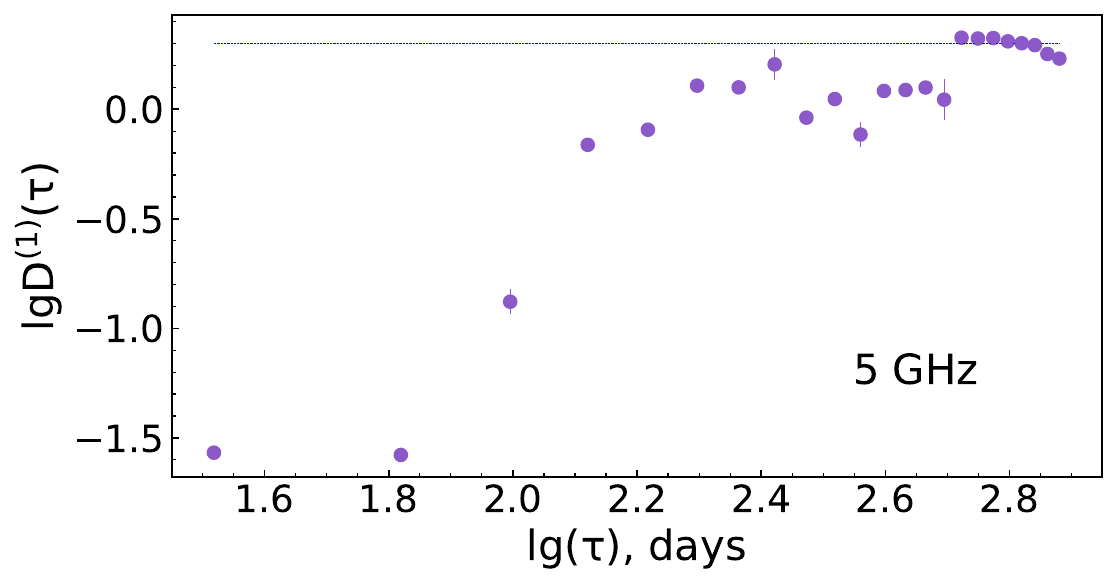}\includegraphics[height=3.2cm]{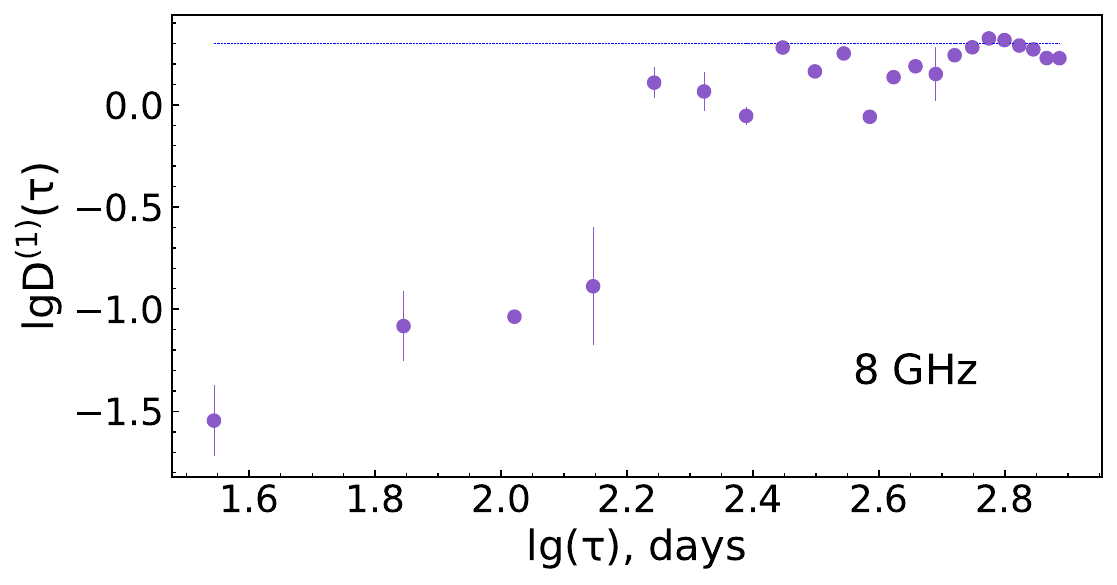}}\centerline{\includegraphics[height=3.2cm]{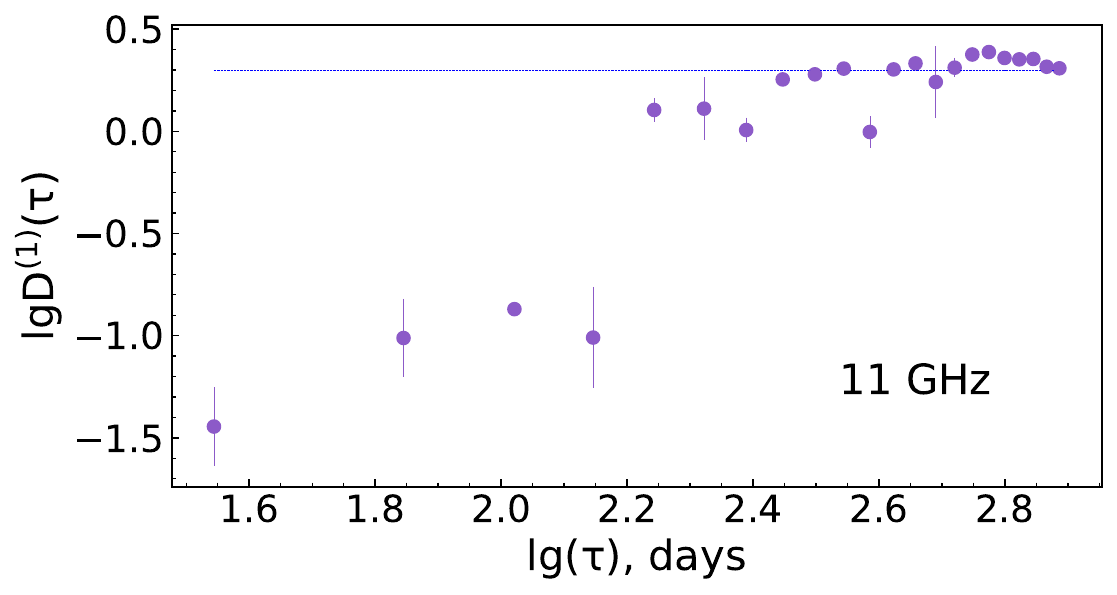}\includegraphics[height=3.2cm]{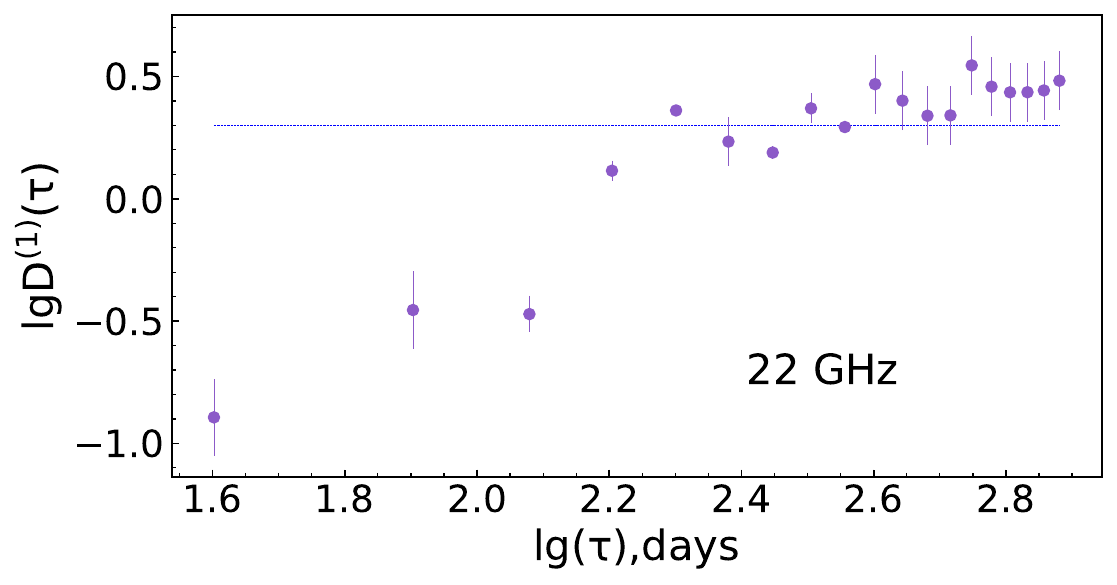}\includegraphics[height=3.2cm]{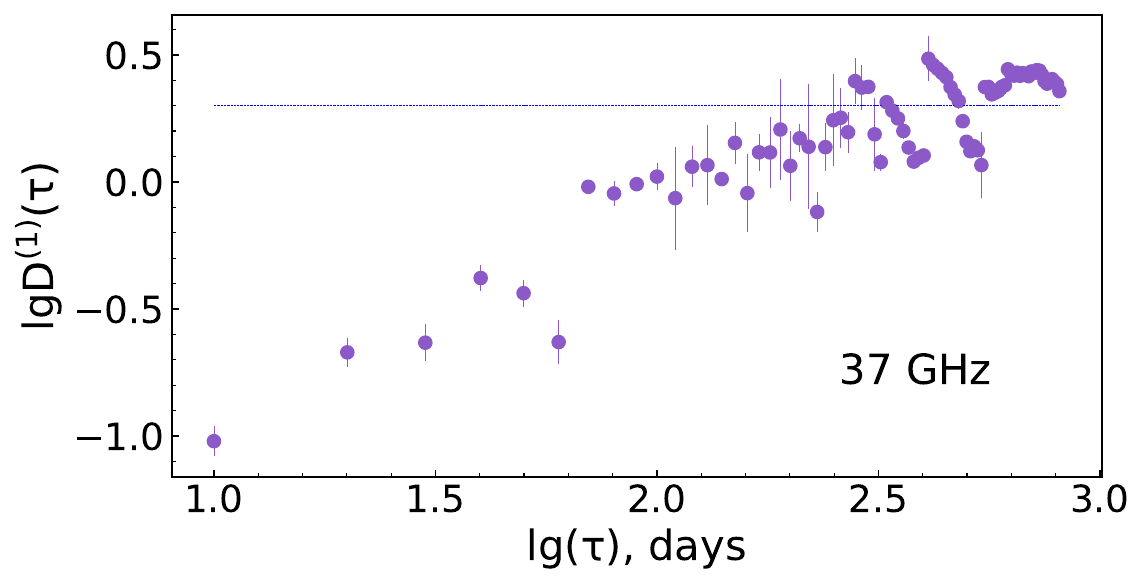}}\centerline{\includegraphics[height=3.2cm]{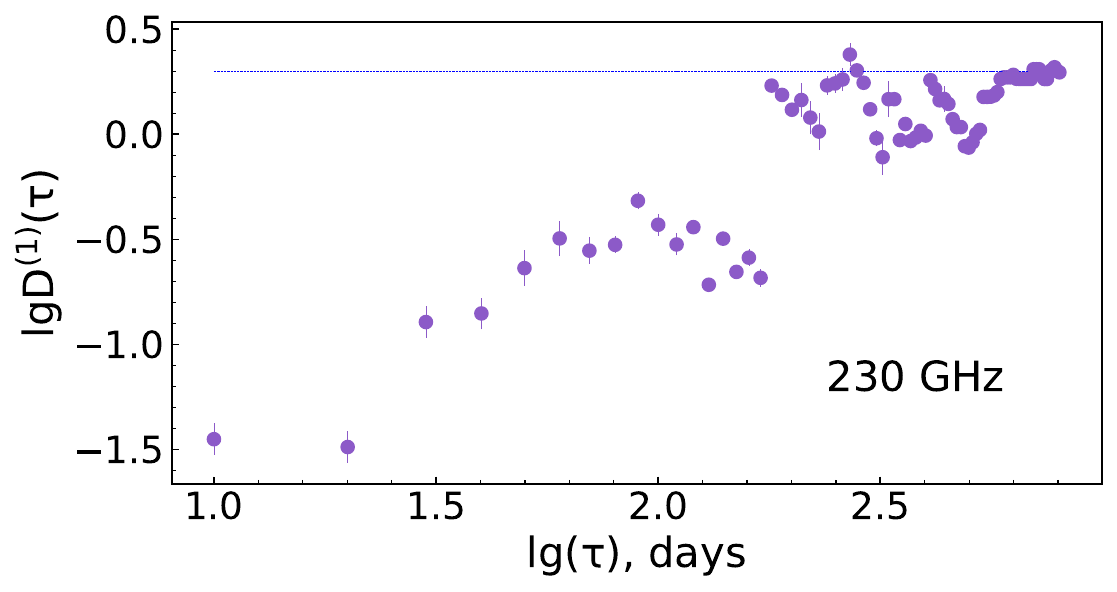}\includegraphics[height=3.2cm]{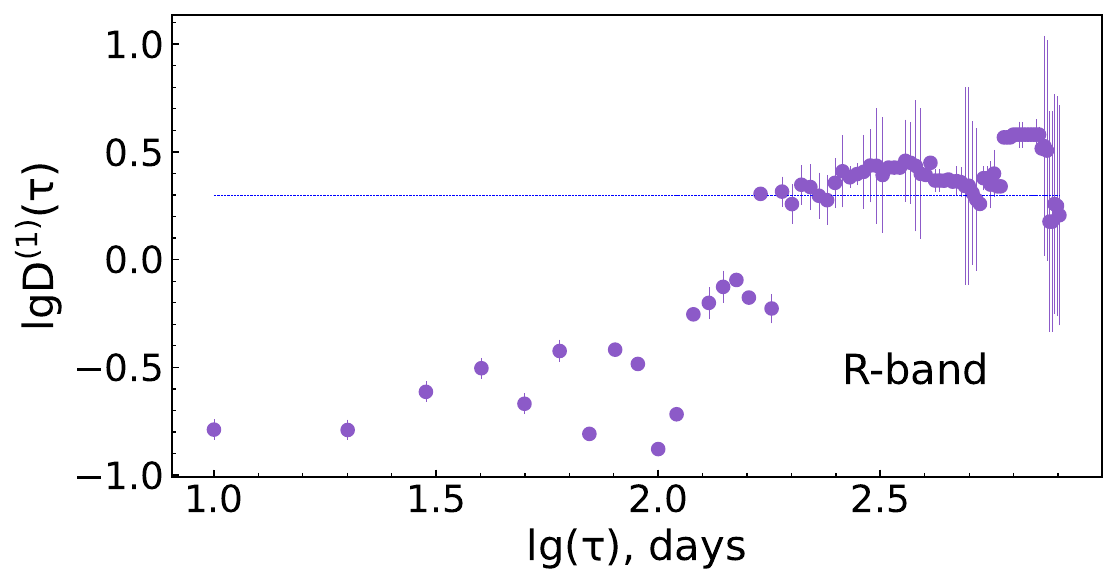}\includegraphics[height=3.2cm]{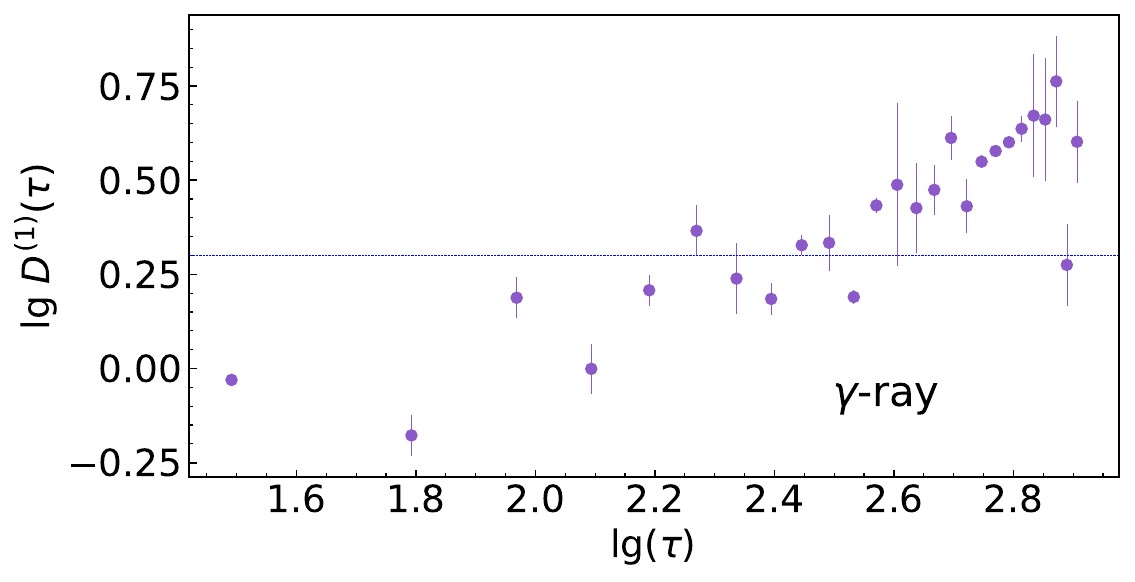}}
\caption{The SFs for the flux density variations during the low activity state in 2009--2014.}
\label{fig:SF}
\end{figure*}


\begin{figure}
\centerline{\includegraphics[width=1.1\columnwidth]{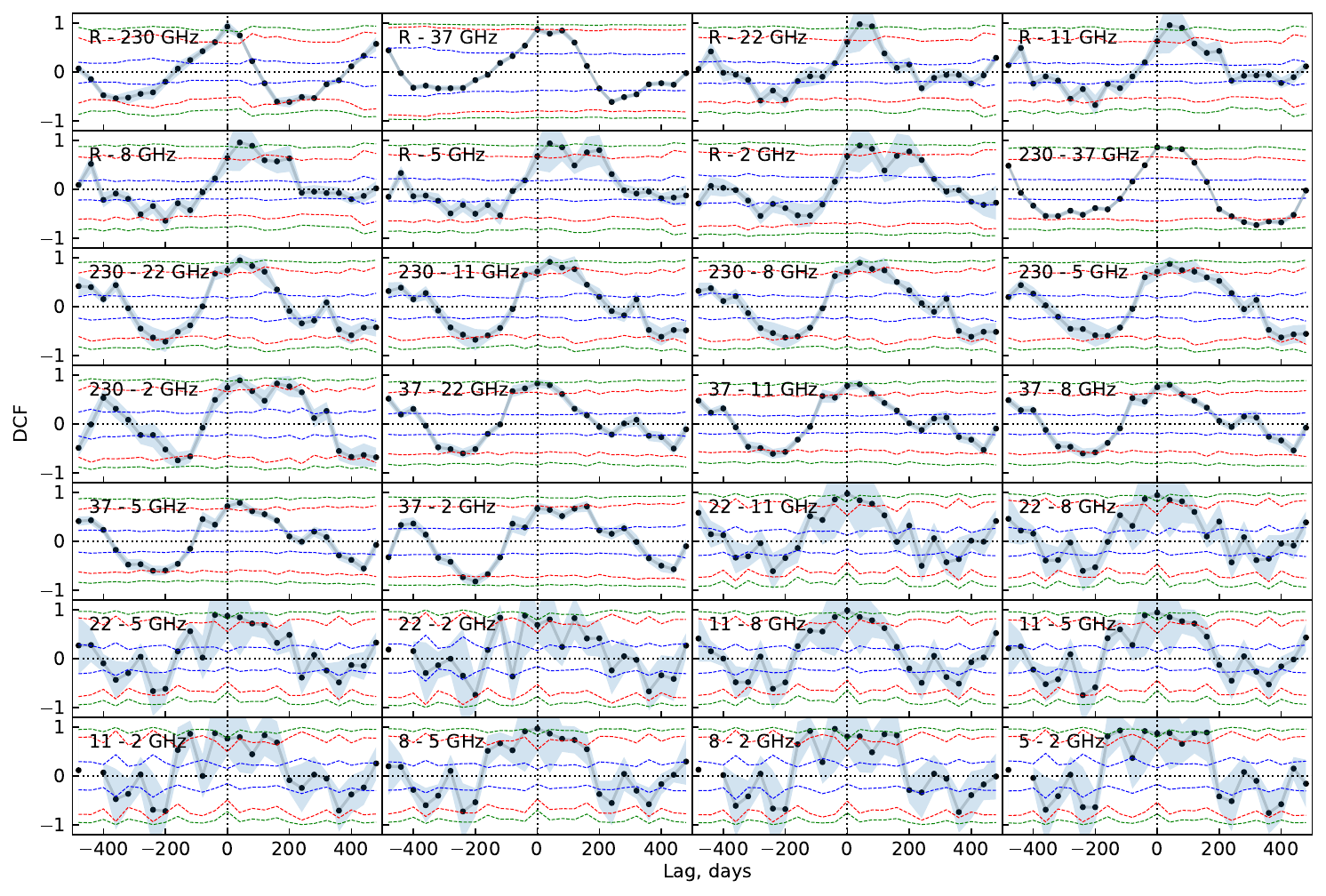}}
\caption{Cross-correlations between different combinations of wavelength bands for epoch 1. The blue, red, and green
lines represent the $1\sigma$, $2\sigma$, and $3\sigma$ significance levels, respectively.}
\label{fig:dcf1}
\end{figure}

\begin{figure}
\centerline{\includegraphics[width=1.0\columnwidth]{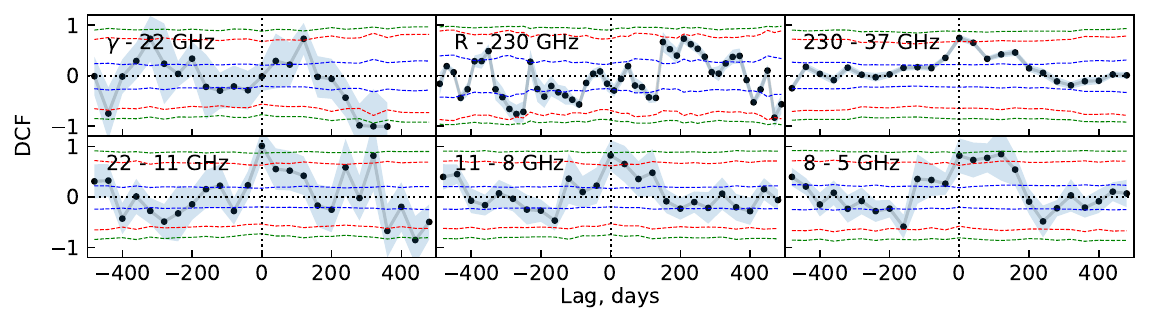}}
\caption{Cross-correlations between different combinations of wavelength bands for the low state in 2009--2014 (epoch 2). The blue, red, and green lines represent the $1\sigma$, $2\sigma$, and $3\sigma$ significance levels, respectively.}
\label{fig:dcf2}
\end{figure}

\begin{figure}
\centerline{\includegraphics[width=1.3\columnwidth]{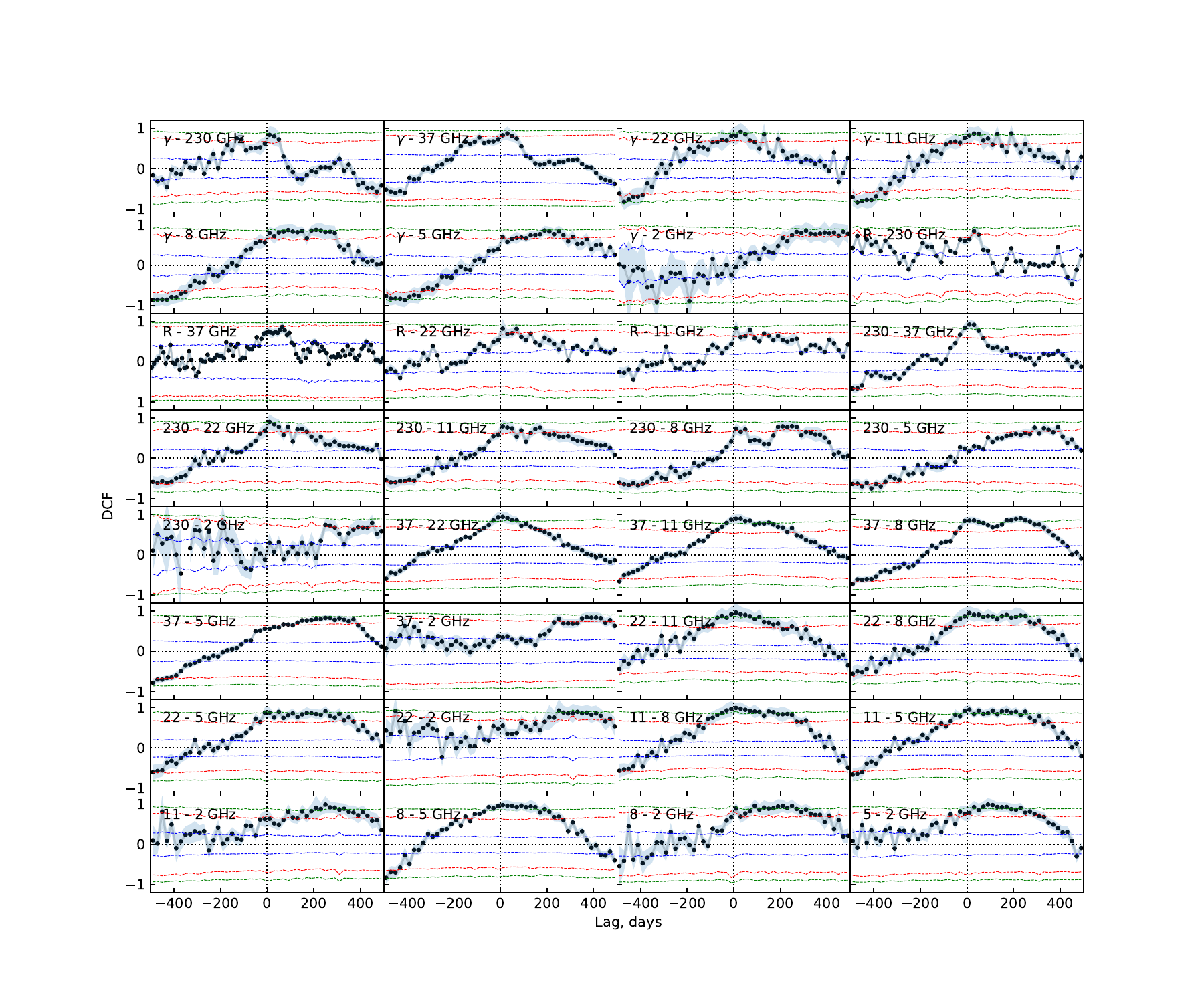}}
\caption{Cross-correlations between different combinations of wavelength bands for epoch 3. The blue, red, and green lines
represent the $1\sigma$, $2\sigma$, and $3\sigma$ significance levels, respectively.}
\label{fig:dcf3}
\end{figure}

\begin{figure}
\centerline{\includegraphics[width=1.3\columnwidth]{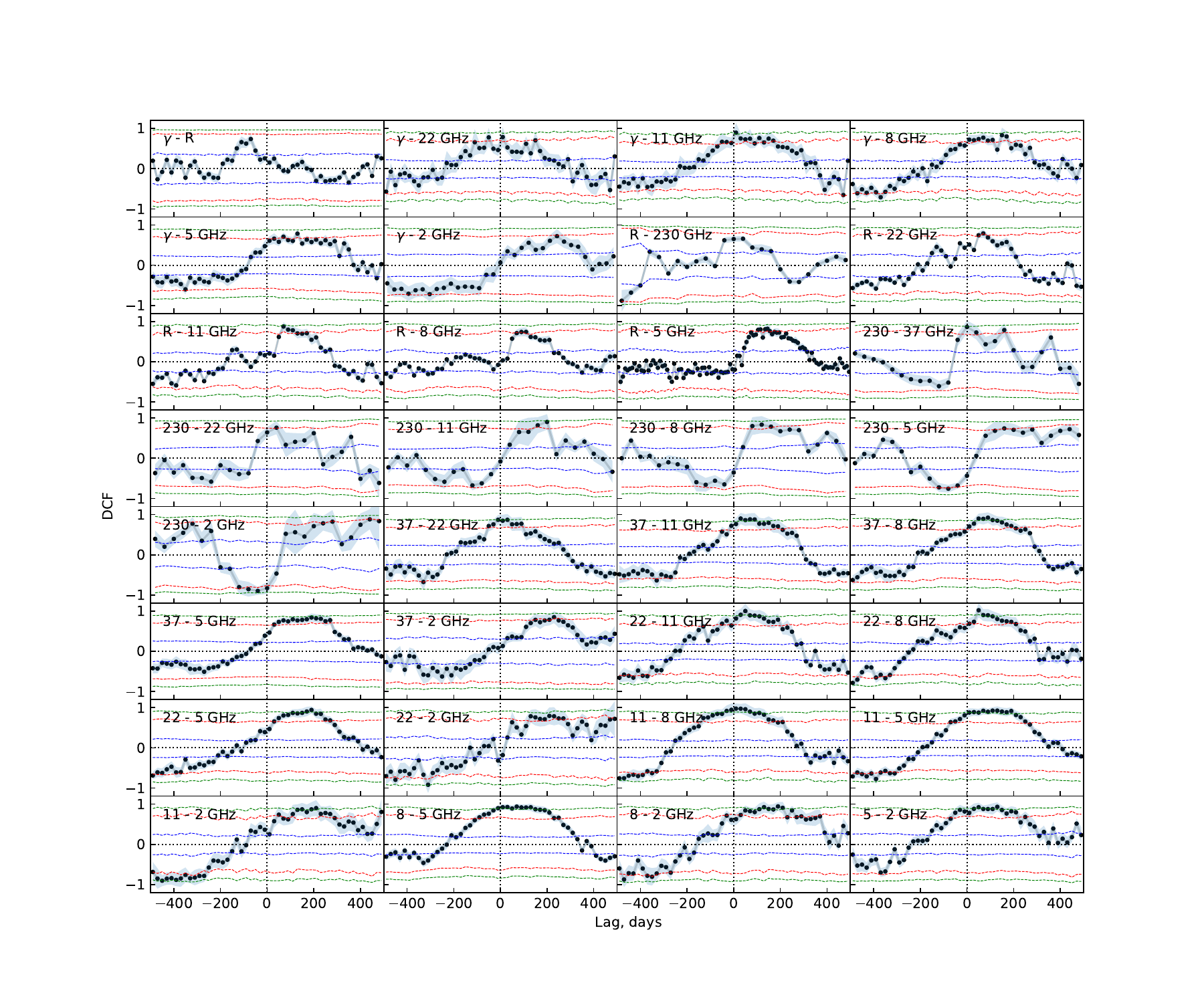}}
\caption{Cross-correlations between different combinations of wavelength bands for epoch 4. The blue, red, and green lines
represent the $1\sigma$, $2\sigma$, and $3\sigma$ significance levels, respectively.}
\label{fig:dcf4}
\end{figure}


\begin{figure*}
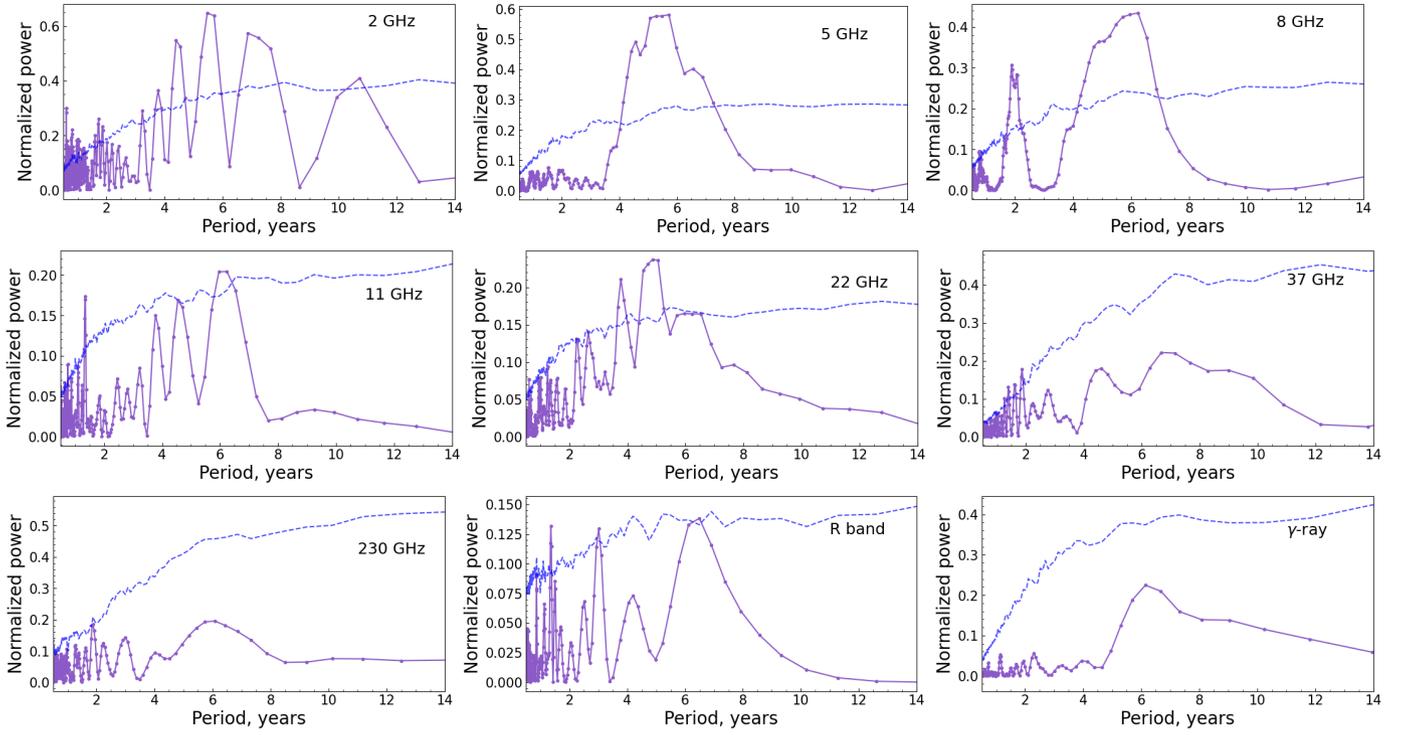

\centerline{\includegraphics[height=3.2cm]{FigA10a.pdf}\includegraphics[height=3.2cm]{FigA10b.pdf}\includegraphics[height=3.2cm]{FigA10c.pdf}}\centerline{\includegraphics[height=3.2cm]{FigA10d.pdf}\includegraphics[height=3.2cm]{FigA10e.pdf}\includegraphics[height=3.2cm]{FigA10f.pdf}}\centerline{\includegraphics[height=3.2cm]
{FigA10g.pdf}\includegraphics[height=3.2cm]{FigA10h.pdf}\includegraphics[height=3.2cm]{FigA10i.pdf}}
\caption{The L--S periodograms for the total flux variations at 2, 5, 8, 11, 22, 37, 230 GHz, in the optical $R$-band, and in
$\gamma$-rays for all the light curves. The dashed lines show
the level of ${\rm FAP} =1$ per cent (false alarm probability).}
\label{fig:LS1}
\end{figure*}

\begin{figure*}
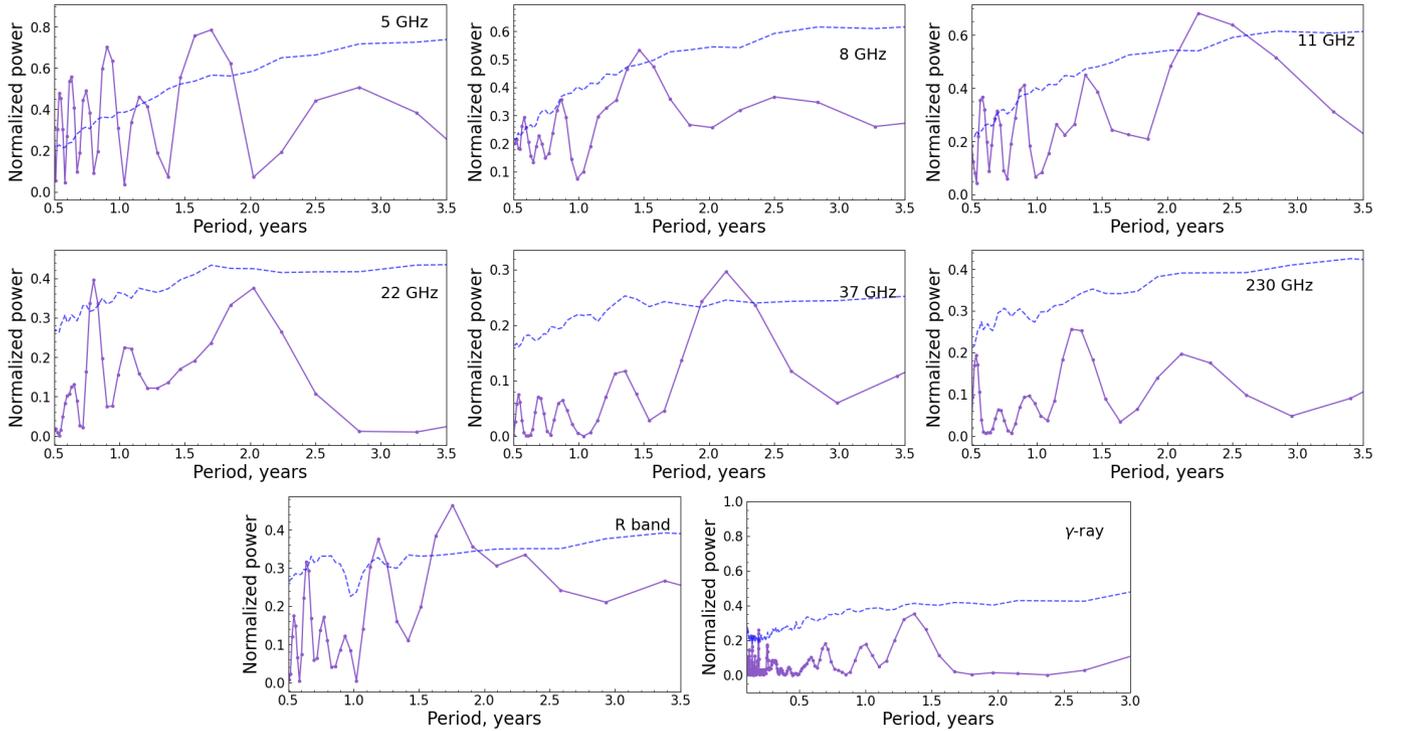

\centerline{\includegraphics[height=3.2cm]{FigA11a.pdf}\includegraphics[height=3.2cm]{FigA11b.pdf}\includegraphics[height=3.2cm]{FigA11c.pdf}}
\centerline{\includegraphics[height=3.2cm]{FigA11d.pdf}\includegraphics[height=3.2cm]{FigA11e.pdf}\includegraphics[height=3.2cm]{FigA11f.pdf}}
\centerline{\includegraphics[height=3.2cm]{FigA11g.pdf}\includegraphics[height=3.2cm]
{FigA11h.pdf}}
\caption{The L--S periodograms for the total flux variations at 5, 8, 11, 22, 37, 230 GHz, in the optical $R$-band , and in
$\gamma$-rays during the low activity state in 2009--2014. The dashed lines show
the ${\rm FAP} =1$ per cent (false alarm probability) level.}
\label{fig:LS2}
\end{figure*}

\clearpage

\bsp
\label{lastpage}

Full list of all authors with their affiliations and ORCID codes

1. V.V. Vlasyuk, Special Astrophysical Observatory of Rus.Ac.Sci., Nizhny Arkhyz, 369167, Russia, 0009-0002-6596-7274

2. Yu.V. Sotnikova,	Special Astrophysical Observatory of Rus.Ac.Sci., Nizhny Arkhyz, 369167, Russia

Kazan Federal University, 18 Kremlyovskaya St, Kazan 420008, Russia, 0000-0001-9172-7237

3. T. V. Mufakharov, Special Astrophysical Observatory of Rus.Ac.Sci., Nizhny Arkhyz, 369167, Russia

Kazan Federal University, 18 Kremlyovskaya St, Kazan 420008, Russia, 0000-0001-9984-127X

4. A.E. Volvach, Crimean Astrophysical Observatory of Rus.Ac.Sci., 	298409, Nauchny, Russia, 0000-0002-3839-3466

5. Yu.A. Kovalev,	Astro Space Center, Lebedev Physical Institute, Russian Academy of Sciences, 117997, Moscow, Russia

Institute for Nuclear Research, Russian Academy of Sciences,
60th October Anniversary Prospect 7a, Moscow 117312, Russia,	 0000-0002-8017-5665

6. O.I. Spiridonova,	Special Astrophysical Observatory of Rus.Ac.Sci.,	Nizhny Arkhyz, 369167, Russia,	0009-0007-7315-3090

7. M.L. Khabibullina,	Special Astrophysical Observatory of Rus.Ac.Sci.,	Nizhny Arkhyz, 369167, Russia,	0000-0001-9515-4552

8. Y.Y. Kovalev,	Max-Planck-Institut fur Radioastronomie, 	Auf dem Hugel 69, Bonn 53121, Germany,	0000-0001-9303-3263

9. A.G. Mikhailov,	Special Astrophysical Observatory of Rus.Ac.Sci.,	Nizhny Arkhyz, 369167, Russia,	0000-0002-0279-0777

10. V.A. Stolyarov,	Special Astrophysical Observatory of Rus.Ac.Sci.,
Nizhny Arkhyz, 369167, Russia

Astrophysics Group, Cavendish Laboratory, University of Cambridge,
Cambridge, CB3 0HE, UK,	0000-0001-8151-828X

11. D.O. Kudryavtsev,	Special Astrophysical Observatory of Rus.Ac.Sci.,	Nizhny Arkhyz, 369167, Russia,	0000-0001-8151-828X

12. M.G. Mingaliev,	Special Astrophysical Observatory of Rus.Ac.Sci.,
Nizhny Arkhyz, 369167, Russia

Kazan Federal University, 18 Kremlyovskaya St, Kazan 420008, Russia

Institute of Applied Astronomy of Rus.Ac.Sci.,
Kutuzova Emb. 10, St. Petersburg 191187, Russia,	0000-0001-8585-1186

13. S. Razzaque, Centre for Astro-Particle Physics and Department of Physics, University of Johannesburg, Auckland Park, 2006, South Africa, 0000-0002-0130-2460

14. T.A. Semenova,	Special Astrophysical Observatory of Rus.Ac.Sci.,	Nizhny Arkhyz, 369167, Russia,	0000-0002-2902-5426

15. A.K. Kudryashova,	Special Astrophysical Observatory of Rus.Ac.Sci.,	Nizhny Arkhyz, 369167, Russia,	0000-0002-3797-3995

16. N.N. Bursov,	Special Astrophysical Observatory of Rus.Ac.Sci., Nizhny Arkhyz, 369167, Russia,	0000-0002-2602-5703

17. S.A. Trushkin,	Special Astrophysical Observatory of Rus.Ac.Sci.,	Nizhny Arkhyz, 369167, Russia,	0000-0002-7586-5856

18. A.V. Popkov,	Moscow Institute of Physics and Technology, Institutsky per. 9, Dolgoprudny 141700, Russia

Astro Space Center, Lebedev Physical Institute Rus.Ac.Sci., 117997, Moscow, Russia, 0000-0002-0739-700X

19. A.K. Erkenov,	Special Astrophysical Observatory of Rus.Ac.Sci., Nizhny Arkhyz, 369167, Russia,	0000-0002-6086-9299

20. I.A. Rakhimov,	Institute of Applied Astronomy of Rus.Ac.Sci.,	Kutuzova Emb. 10, St. Petersburg 191187, Russia,	0000-0002-9185-6239

21. M.A. Kharinov,	Institute of Applied Astronomy of Rus.Ac.Sci.,	Kutuzova Emb. 10, St. Petersburg 191187, Russia,	0000-0002-0321-8588

22. M.A~Gurwell, Center for Astrophysics, Harvard \& Smithsonian, 60 Garden Street, Cambridge, MA 02138, USA

23. P.G. Tsybulev,	Special Astrophysical Observatory of Rus.Ac.Sci., Nizhny Arkhyz, 369167, Russia,	0000-0001-5600-8018

24. A.S. Moskvitin,	Special Astrophysical Observatory of Rus.Ac.Sci., Nizhny Arkhyz, 369167, Russia,	0000-0003-3244-6616

25. T.A. Fatkhullin, Special Astrophysical Observatory of Rus.Ac.Sci., Nizhny Arkhyz, 369167, Russia

26. E.V. Emelianov,	Special Astrophysical Observatory of Rus.Ac.Sci., Nizhny Arkhyz, 369167, Russia,	0000-0003-4845-5955

27. A. Arshinova,	St. Petersburg State University, 	7/9 Universitetskaya Emb., St Petersburg 199034, Russia,	0009-0002-9297-8809

28. K.V. Iuzhanina,	Special Astrophysical Observatory of Rus.Ac.Sci., Nizhny Arkhyz, 369167, Russia

Kazan Federal University, 18 Kremlyovskaya St, Kazan 420008, Russia,	0009-0008-7311-0709

29. T.S. Andreeva, Institute of Applied Astronomy of Rus.Ac.Sci., Kutuzova Emb. 10, St. Petersburg 191187, Russia, 0000-0003-3613-6252

30. L.N. Volvach, Crimean Astrophysical Observatory of Rus.Ac.Sci., 298409, Nauchny, Russia, 0000-0001-6157-003X

31. A. Gosh, Centre for Astro-Particle Physics and Department of Physics, University of Johannesburg, Auckland Park, 2006, South Africa, 0000-0003-2265-0381

\end{document}